\documentclass[11pt]{article}

\newcommand{\preprintpath}{}

\makeatletter
\edef\input@path{{\preprintpath}}
\makeatother

\usepackage{salesforce-preprint}

\usepackage{amsmath,amsfonts,bm}

\def\eqref#1{equation~\ref{#1}}

\def\1{\bm{1}}

\DeclareMathAlphabet{\mathsfit}{\encodingdefault}{\sfdefault}{m}{sl}
\SetMathAlphabet{\mathsfit}{bold}{\encodingdefault}{\sfdefault}{bx}{n}

\newsavebox{\tablewidthbox}
\newenvironment{fitlinewidth}{\begin{lrbox}{\tablewidthbox}}{\end{lrbox}\ifdim\wd\tablewidthbox>\linewidth
    \resizebox{\linewidth}{!}{\usebox{\tablewidthbox}}\else
    \usebox{\tablewidthbox}\fi
}

\DeclareRobustCommand{\modellogo}[1]{\raisebox{-0.18em}{\makebox[1.2em][c]{\includegraphics[width=1em,height=1em,keepaspectratio]{\preprintpath figures/model_logos/#1.png}}}\hspace{0.3em}}

\usepackage{url}
\usepackage{subcaption}
\usepackage{wrapfig}
\usepackage{booktabs}
\usepackage{array}
\usepackage{colortbl}
\usepackage{appendix}
\usepackage{titletoc}
\usepackage{float}
\makeatletter
\let\sf@default@addtocurcol\@addtocurcol
\makeatother
\usepackage{flafter}
\makeatletter
\let\sf@main@addtocurcol\@addtocurcol
\let\@addtocurcol\sf@default@addtocurcol
\newcommand{\maintextfloatlayout}{\let\@addtocurcol\sf@main@addtocurcol
  \@clubpenalty=10000\relax}
\makeatother
\usepackage{tikz}
\usetikzlibrary{arrows.meta,positioning,calc}

\definecolor{demoInfo}{HTML}{536A8A}
\definecolor{demoSuccess}{HTML}{278A78}
\definecolor{demoFailure}{HTML}{B84A4A}
\definecolor{demoTimeout}{HTML}{C47A20}
\definecolor{demoSafe}{HTML}{386FA4}
\newcommand{\demotag}[2]{\begingroup\setlength{\fboxsep}{3pt}\colorbox{#1!15}{\textcolor{#1!80!black}{\sffamily\bfseries\scriptsize #2}}\endgroup}
\newtcolorbox{demoatglance}[2]{
  enhanced,
  colback=#1!4,
  colframe=#1!72!black,
  colbacktitle=#1!14,
  coltitle=black,
  title={#2},
  fonttitle=\bfseries,
  boxrule=0.55pt,
  leftrule=2.5pt,
  arc=1.5mm,
  left=6pt,right=6pt,top=4pt,bottom=4pt,
  before skip=6pt,after skip=7pt
}

\title{CRM-Agent-Bench: Benchmarking Tool-Using Agents in Stateful Enterprise Environments}
\author{Monojit Banerjee\textsuperscript{*},\quad
  Zhujun Lan\textsuperscript{*},\quad
  Defu Cao\textsuperscript{*\textdagger}\\[\preprintauthorsep]\normalfont
  {\protect\sffamily\protect\bfseries\protect\large
    \textcolor{sfnavy}{Salesforce, Inc.}}\\
  {\protect\small
    \href{mailto:mbanerjee@salesforce.com}{mbanerjee@salesforce.com}\quad
    \href{mailto:zhujun.lan@salesforce.com}{zhujun.lan@salesforce.com}\quad
    \href{mailto:defu.cao@salesforce.com}{defu.cao@salesforce.com}}}

\newcommand{\preprintauthornotes}{\begingroup
  \renewcommand{\thefootnote}{\fnsymbol{footnote}}\begin{NoHyper}
    \footnotetext[1]{Equal contribution\quad
      \textsuperscript{\textdagger}Corresponding author}\end{NoHyper}
  \endgroup
}

\date{December 2, 2026}
\preprintgithub{\url{https://github.com/salesforce-misc/crm-agent-bench}}
\preprintprojectpage{\url{https://crm-agent-bench.github.io/}}
\hypersetup{
  pdftitle={CRM-Agent-Bench: Benchmarking Tool-Using Agents in Stateful Enterprise Environments},
  pdfauthor={Monojit Banerjee, Zhujun Lan, Defu Cao},
  pdfsubject={Salesforce preprint edition}
}

\begin{document}
\maketitle
\preprintauthornotes
\begingroup
\maintextfloatlayout
\setstretch{1.15}
\renewcommand{\bottomtitlespace}{5\baselineskip}
\clubpenalty=10000
\widowpenalty=10000
\displaywidowpenalty=10000
\captionsetup{font={small,stretch=1.1},labelfont=bf}
\begin{abstract}

Deploying language agents in real applications demands more than the right answer: an agent must execute a multi-step business process correctly, follow domain-specific rules, and interact with a user across a dialogue, where the gap between saying and doing is the gap between a satisfied customer and a real loss. We argue that a deployable-agent benchmark should jointly assess four facets: the outcome an agent reaches, the trajectory it takes, its operational viability, and its policy compliance. Customer relationship management (CRM), where mutable records meet operational rules and consequential effects, exercises all four. We introduce CRM-Agent-Bench, an execution-grounded benchmark of 555 stateful, multi-turn tasks across nine business domains and 36 capability tags. Each task supplies a task-specific executable contract drawn from checks over actions, arguments, order, response, target state, and protected state; success is the strict conjunction of the checks declared for that task. {We evaluate the complete agent system through these contracts and report scored coverage, latency, and cost.} Across sixteen systems evaluated on the audited final release, strict success spans 64.32\% to 83.60\%; the closely grouped leading systems nonetheless diverge sharply in cost and per-domain reliability. CRM-Agent-Bench offers an auditable framework for measuring contract-grounded business-transaction integrity and adherence to task-declared benchmark
policies, with a portable structure designed for extension as models improve.
\end{abstract}

\section{Introduction}
\label{sec:introduction}

\textbf{From answers to delegated execution.}
Language-model agents increasingly combine reasoning with actions in an external environment~\citep{yao2023react,liu2024agentbench}. This changes the object of evaluation. Once a system receives authority to update a system of record, its failures no longer end with an incorrect answer. It may complete the wrong transaction, continue after a backend rejection, or describe an operation that never succeeded. The gap between saying and doing then becomes a real business loss: an agent that claims it issued a refund but never calls the tool, or that applies a coupon to the wrong account, causes damage that a sounds-helpful score never catches. Customer relationship management (CRM) is a particularly demanding setting for this transition. Customer-facing work ties mutable records to operational rules whose effects propagate into downstream business processes. Even a simple request can require the agent to identify the right entity, follow a prescribed procedure, react to state-dependent feedback, and preserve every record outside the user's intent.
Recent CRM benchmarks bring realistic business data and multi-turn interaction~\citep{huang-etal-2025-crmarena,huang2026crmarena-pro}, yet they largely evaluate retrieval and answer correctness rather than whether an agent carries out the underlying state change correctly and within policy. How, then, should we measure whether the whole system does the right thing rather than merely says it?

\textbf{{Evaluating enterprise execution.}}
Prior benchmarks each cover part of this problem: function calling and argument
generation~\citep{patil2025bfcl}, on-policy dialogue over a mutable
database~\citep{yao2025taubench}, stateful tool use with milestones and
minefields~\citep{lu-etal-2025-toolsandbox}, programmatic state and
collateral-damage tests~\citep{trivedi-etal-2024-appworld}, compositional
enterprise web work~\citep{boisvert2024workarena}, and CRM
realism~\citep{huang-etal-2025-crmarena,huang2026crmarena-pro}; we survey them
in Table~\ref{tab:benchmark-comparison}. Yet each reports only a subset of what
delegated execution demands. A complete assessment must resolve four facets at
once: the \emph{outcome} (the final state reached and the result communicated),
the \emph{trajectory} (the right actions, with the right arguments, in the right
order), \emph{operational} viability (latency and cost), and
\emph{policy compliance} (no forbidden action, no change to protected state,
and no improper disclosure). When a benchmark reports only one facet, a system
can score well while silently failing another the metric never inspects.
Figure~\ref{fig:one-trajectory-four-verdicts} shows exactly this: a real
trajectory whose response, action, and ordering checks all pass, yet whose
duplicated checkout is caught only by the state check, so response-only or
action-only grading would score a failed transaction as a success.
CRM is a revealing lens for all four facets, because
one workflow couples language understanding, process compliance, state
mutation, and asymmetric business risk. We target this execution layer, not
visual perception or low-level control; this is complementary to computer-use
agent benchmarks~\citep{xie2024osworld,kuntz2025osharm}, which add a perception
and GUI-grounding layer in front of similarly stateful execution problems. By
holding the structured action interface fixed, CRM-Agent-Bench isolates whether
an agent selects a valid workflow, respects capability boundaries, and
preserves protected state after an action has been grounded.

\begin{figure}[t]
    \centering
    \includegraphics[width=\linewidth]{\preprintpath 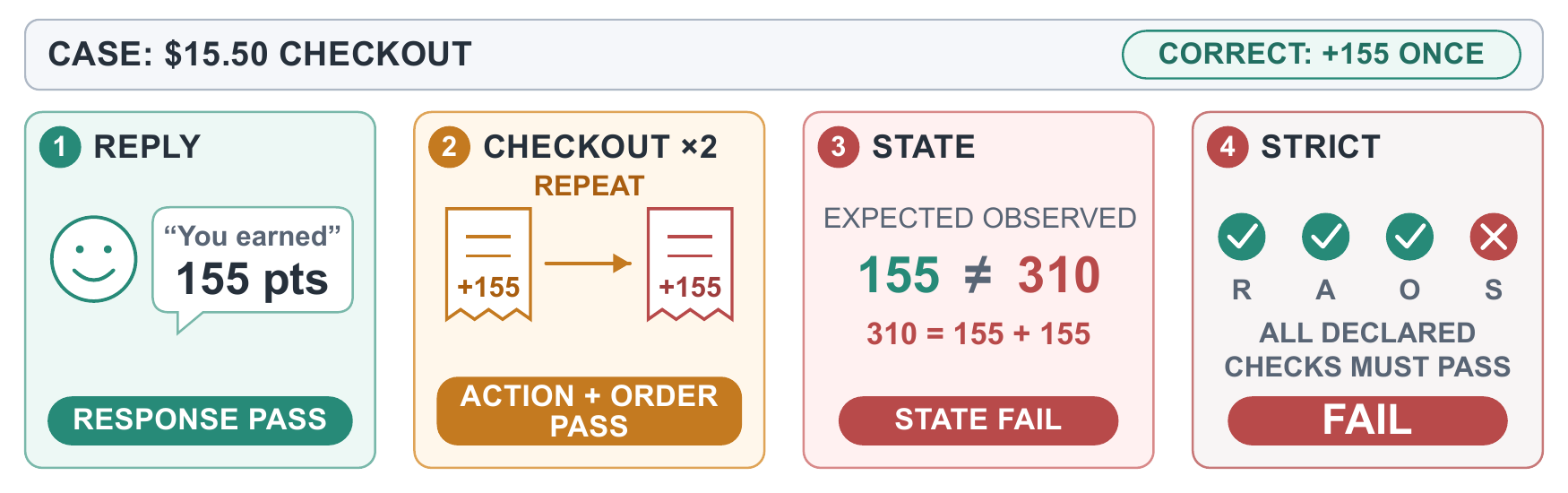}
    \caption{A duplicated \$15.50 checkout: the response, action, and order
    checks all pass, yet the state check catches the doubled loyalty balance
    (310 instead of 155) and the strict contract fails
    (Appendix~\ref{app:case-failure}).}
    \label{fig:one-trajectory-four-verdicts}
\end{figure}
\textbf{CRM-Agent-Bench.} We introduce an execution-grounded benchmark with 555 task instances across nine
stateful business domains and 36 capability tags. Each task defines a user goal,
an initial world state, and a task-specific executable contract, and each trial
starts from a freshly reset environment and unfolds through an on-policy
simulated user that generates every later turn from the preceding user-visible
agent reply rather than replaying a fixed transcript. A contract draws on up to
six check families, and three of the four facets are graded directly on the
trace: required calls, argument constraints, and subsequence or exact ordering
grade the \emph{trajectory}; response-content and target-state checks grade the
\emph{outcome}; and \emph{policy compliance} is enforced by three negative
checks that mirror its definition above: a forbidden-action check (a prohibited
tool must never be called), a protected-state check (records the task did not
authorize must remain unchanged), and a forbidden-argument check (no call may
carry another party's identifiers, so the agent cannot act on or expose records
it was not authorized to touch). The headline success signal is the strict
conjunction of every check a task declares, so all declared requirements must
hold simultaneously; undeclared dimensions are absent rather than counted as
passes. Component scores retain the underlying action, response, and state
evidence for diagnosis. The fourth facet,
\emph{operational} viability, is reported alongside every result as latency and
cost rather than folded into the pass criterion.

\textbf{{Empirical evaluation.}}
{We compare proprietary and open-weight systems using the first-round
reference protocol and quantify how adding task-declared checks changes
suite-wide success.}
We evaluate sixteen systems on 555 audited tasks, including Gemini 3.8 Flash,
Claude Opus 5, Muse 30B, and GPT 5.6 Terra. Strict success
ranges from 64.32\% to 83.60\%.
On the tasks with explicit policy
constraints, systems pass the applicable negative checks on 91.2\% of trials
but satisfy the full contract on only 66.8\%, compared with 82.1\% on tasks
without negative assertions. These results show that avoiding prohibited
behavior and completing the permitted workflow are distinct evaluation dimensions.

\textbf{Our Contribution.} We contribute (i) an execution-grounded CRM benchmark of 555 stateful
multi-turn tasks over nine business domains and 36 capability tags, runnable
without a live CRM or vendor backend; (ii)
task-specific conjunctive contracts that grade the applicable outcome,
trajectory, and policy-compliance requirements while reporting operational
latency and cost; and (iii) an
agent-agnostic evaluation protocol that reports the facets separably rather than
as one score, and audits capability exposure by checking whether each forbidden
tool is even reachable in its domain graph.

\begin{figure}[t]
    \centering
    \includegraphics[width=\textwidth]{\preprintpath 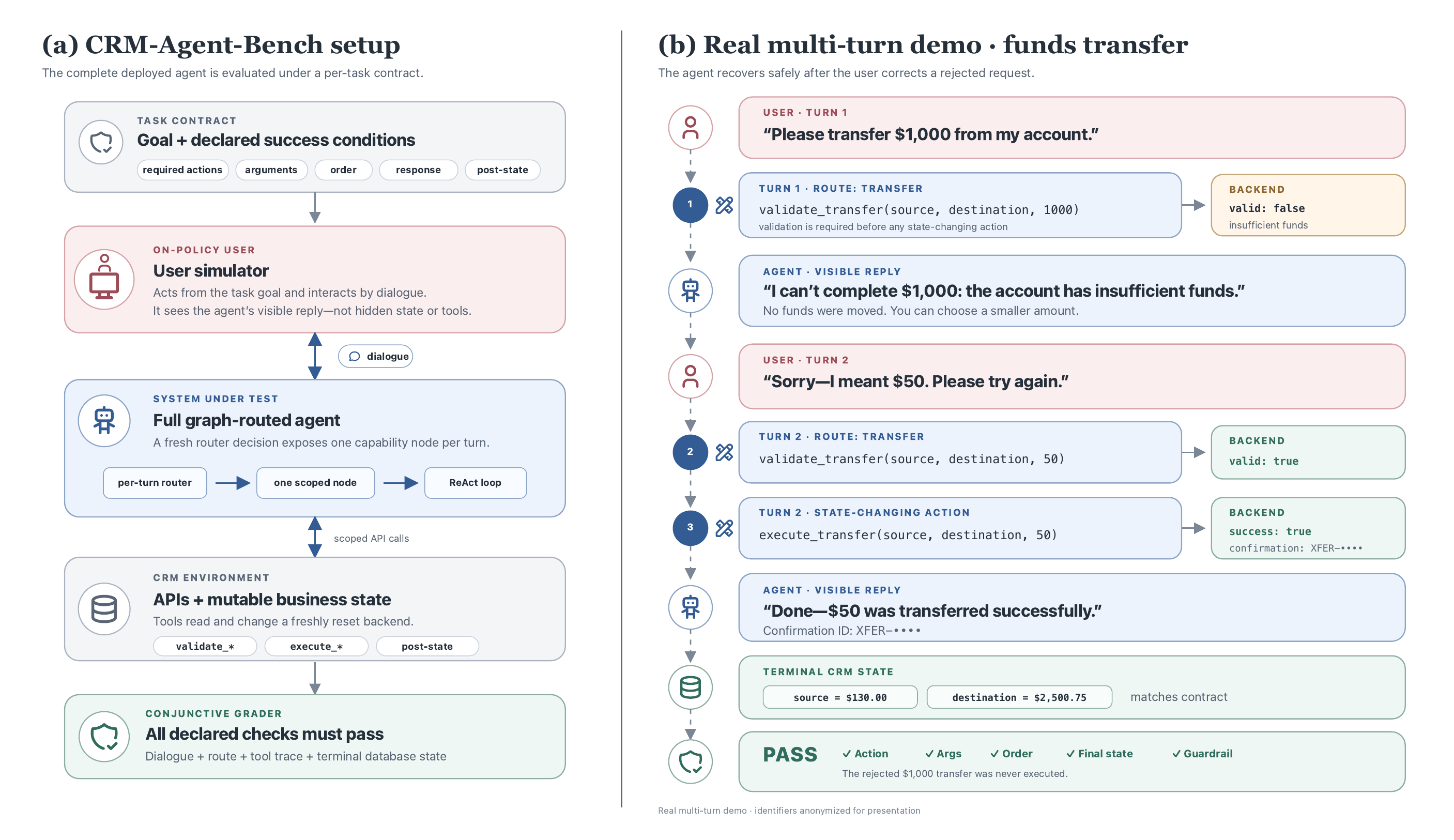}
    \caption{CRM-Agent-Bench setup and a multi-turn demo. (a) An on-policy
    user interacts with the full routed agent; each turn exposes one scoped
    capability over a freshly reset CRM backend. Recorded dialogue, routes,
    tool calls, and terminal state are checked conjunctively against the task
    contract. (b) In an observed synthetic benchmark trajectory, the agent rejects an invalid
    transfer without changing state, then reselects the transfer node, validates, and executes the
    user's corrected request. All declared action, argument, order, state, and
    guardrail checks pass.}
    \label{fig:benchmark-pipeline}
\end{figure}

\section{The CRM-Agent-Bench Benchmark}
\label{sec:method}

\subsection{Evaluation Target}
CRM-Agent-Bench evaluates an \emph{agent system}: a language model embedded in
an orchestration policy, operating through tools against mutable enterprise
state. The benchmark artifact (tasks, initial states, tools, policies, and
graders) is held separate from the system under test, which matters for routed
agents because the same base model can behave differently under a different
tool inventory or routing policy.
Specifically, CRM-Agent-Bench scores every trial along the four facets: \emph{outcome}, \emph{trajectory},
\emph{operational viability}, and \emph{policy compliance}. {Each task's contract specifies its applicable checks, and strict success
requires all of them to pass.} Four measured
components make them observable (Figure~\ref{fig:benchmark-pipeline}): a
resettable domain environment with typed tools, an on-policy user simulator, a
graph-routed reference agent as the system under test, and a task-specific
executable contract that pairs a user goal with its declared success conditions,
adjudicated by programmatic checks and a fixed semantic judge.

\subsection{Tasks and Stateful Environments}
Each task specification
$\mathcal{T}_i=(u_i,s_i^0,\mathcal{C}_i,d_i)$ induces a finite horizon
partially observable Markov decision process (POMDP).
The domain $d_i$ provides the default CRM state and typed tools, while the
optional overlay $s_i^0$ initializes task specific information. At time $t$,
the latent state contains the mutable CRM state $x_t$, the private user
instruction $u_i$, the accumulated interaction history, and the remaining
simulator and executor state. The agent observes only the visible dialogue,
returned tool results, and currently exposed instructions and tool schemas. It
does not observe the raw CRM state, the private user instruction, or the
evaluation contract.
The agent acts through typed tool calls and user visible messages. Transitions
combine programmatic, seed controlled tool execution with an on policy user
simulator, so later observations depend on earlier agent replies rather than a
fixed transcript~\citep{yao2025taubench}. No intermediate reward is exposed.
When the interaction ends, $\mathcal{C}_i$ evaluates the complete trajectory
$\tau_i$ and final CRM state $x_T$, and success requires every declared check
to pass. The access policy determines which tools are available without
changing the task itself, and the routed reference agent is one concrete policy
within this process~\citep{yao2023react}. \subsection{Trial Execution and Grading}

\textbf{On-policy user simulator.}
Following $\tau$-bench and ToolSandbox
~\citep{yao2025taubench,lu-etal-2025-toolsandbox}, evaluation is on-policy:
the simulated user observes the agent's user-visible reply and generates the
next turn conditioned on the task goal and conversation so far. It does not
receive the agent's private reasoning, tool calls, tool results, or database
state. {The simulator is therefore conditioned on user-visible evidence.}
The simulator emits a stop marker when it judges the interaction complete;
otherwise interaction continues until the 30-turn cap or the 1,200-second trial
timeout. {The reference protocol uses \texttt{gpt-5-mini} as
its user simulator, with settings given in Table~\ref{tab:frozen-protocol}.
Appendix~\ref{app:simulator-sensitivity} compares six systems under GPT-6
Luna and Claude Haiku 5.5 as simulators over three paired rounds.}

\textbf{Executable contract and semantic judge.}
Each task declares a contract $\mathcal{C}_i$ of only the checks relevant to it,
drawn from six families: expected actions, argument constraints, ordering,
response content, target state, and forbidden actions, arguments, or state.
Writing $D_i$ for the declared checks and $c(\tau_i,x_T)\in\{0,1\}$ for one
check on trace $\tau_i$ and final state, task success is their strict conjunction
$Y_i=\prod_{c\in D_i} c(\tau_i,x_T)$ (Figure~\ref{fig:contract-anatomy}),
so a persuasive reply cannot offset a missing transaction and a correct endpoint
cannot erase a forbidden intermediate action; undeclared checks are absent rather
than passed, and per-check components are retained to localize failure. These six
families instantiate three of the four facets of Section~\ref{sec:introduction}:
\emph{outcome} (target state and response), \emph{trajectory} (actions,
arguments, and ordering), and \emph{policy compliance} (forbidden actions,
forbidden or collateral state, and identity-gated disclosure); \emph{operational
viability} is reported alongside rather than folded into the conjunction. Only
the response-content criteria are sent to a fixed LLM judge, which sees just the
user instruction and the visible replies and cannot override the programmatic
checks; the per-family check semantics and the full judge protocol are given in
Appendix~\ref{app:protocol-details}. Appendix~\ref{app:human-judge-agreement}
reports human agreement with this judge on 108 saved response decisions.
{Expected-action checks match successfully executed calls;
forbidden-action checks reject prohibited attempts, including calls rejected
by the tool.}

\textbf{Reporting.}
{The static capability-exposure audit measures graph-wide reachability
of forbidden tools; turn-level exposure depends on routing.}
We report \emph{scored coverage}, the fraction of scheduled tasks yielding a
valid grade, alongside conditional success\begin{equation}
    \mathrm{Pass}^{1}_{\mathrm{scored}}
    = \frac{\sum_{i\in G}Y_i}{|G|},
\end{equation}
where $G$ excludes infrastructure failures but retains model-attributable
failures such as hallucinated tools and human-verified non-convergence at
timeout. {Success counts are paired with $|G|/555$ and failure-category counts;
latency and cost cover the tested agent model.}
We report along these four facets rather than collapsing them into a composite
index: strict success and its per-check components for outcome and trajectory,
latency and cost for operational viability, and the forbidden-action,
forbidden-argument, and collateral-state checks for policy compliance. A system
can look competitive on a single headline rate while differing sharply in its
per-component pass rates or in cost, so tables and figures keep the component
evidence and cost behind any headline rate.

\section{Benchmark Construction}
\label{sec:construction}

\subsection{Task and Reference-Agent Construction}
\label{sec:task-agent-construction}

As shown in Figure~\ref{fig:benchmark-pipeline}, CRM-Agent-Bench is built around business workflows rather than isolated
utterances or tool schemas. Each source workflow specifies an actor, a business
goal, a stopping condition, the information needed to complete it, and both
intended and prohibited effects, and is ported through a common pipeline into
three separable artifacts: (i) a resettable domain environment of synthetic
records, business policies, and deterministic tools, including structured
business rejections; (ii) an architecture-neutral task definition binding a user
goal and an optional initial-state overlay to an executable contract; and (iii)
a graph-structured reference agent whose capability-specific nodes hold scoped
instructions and tools. The reference graph is one fixed orchestration
scaffold; the task and its contract do not require a tested agent to adopt it.
Because all records and transactions are synthetic, the benchmark runs without a
live vendor CRM system.

CRM-Agent-Bench contains 555 tasks drawn from two complementary sources. The
first derives from five CRM applications developed and reviewed with
practitioners and domain experts. The second adapts 22 tool using workflow
families from a public, versioned collection of agent recipes
~\citep{salesforce2026agentscriptrecipes}.
The resulting tasks are organized into nine business domains and cover routine
workflows, capability routing, multi step discovery, ambiguous requests, policy
refusal, adversarial behavior, and collateral damage. Every task undergoes the
same schema, tool reference, and fresh environment gold path validation.
Here, schema refers to the benchmark task format rather than the agent's tool
interface. Validation confirms that the task is well formed, every referenced
tool exists, and a reference solution can reach the contracted state from a
fresh environment.

\subsection{Workflow Taxonomy and Coverage}
Within a domain, coverage is organized at four levels: \emph{domain} (a shared
mutable world and tool registry) $\rightarrow$ \emph{capability} (a coverage
tag, not an oracle route and not revealed to the agent) $\rightarrow$
\emph{scenario} (a persona goal and stopping condition over seed records)
$\rightarrow$ \emph{executable contract}. Because a multi-turn task may traverse
several nodes, this hierarchy describes coverage without predetermining the
agent's route, and grading is execution-first: every declared action resolves to
a tool in the task's domain, every state assertion addresses the runtime
environment, and the grader consumes the recorded trace and final state rather
than a post-hoc label.

\begin{figure}[!htbp]
    \centering
    \includegraphics[width=0.8\linewidth]{\preprintpath 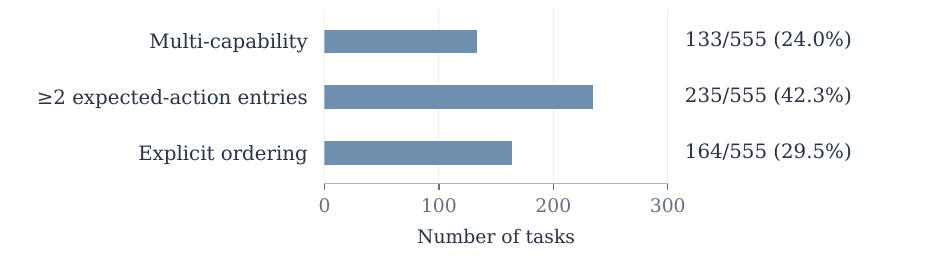}
    \caption{{Multi-capability coverage and execution requirements in
    the 555-task evaluation set. Each bar counts tasks with the indicated
    property; percentages use all 555 tasks. The categories overlap and
    describe declared task requirements.}}
    \label{fig:task-structure-overview}
\end{figure}

The nine domains carry 36 capability tags, and each task is labeled easy,
medium, or hard by its authors as a descriptive stratum rather than an input to
the agent or grader. The benchmark spans a per-domain
distribution across difficulty together with a count of tasks carrying a
negative (forbidden action, argument, or state) assertion. Coverage spans direct
workflows, cross-capability routing, multi-hop discovery, ambiguity, policy
refusal, adversarial requests, and collateral-damage cases, so a single contract
may combine a valid update, an ordering constraint, a customer-facing
explanation, and preservation of a protected record. We freeze task identifiers
and configuration before model comparison so that all systems face the same
workload and evaluator, following the matched-protocol principle of stateful
interactive benchmarks~\citep{yao2025taubench,barres2025tau2bench}.
{Figure~\ref{fig:task-structure-overview} summarizes three overlapping
properties of the task contracts: 133 tasks carry multiple capability tags,
235 declare at least two expected-action entries, and 164 impose explicit
ordering constraints.}

\section{Results and Analysis}
\label{sec:experiments}

We evaluate sixteen systems spanning the Gemini, Claude, GPT, and open-weight
families on the 555-task canonical set.
{
Each run yields valid grades for all 555 tasks, with zero infrastructure
exclusions, so {strict accuracy} (passed\,/\,555) is directly comparable
across models. The main leaderboard uses
one execution per task, so we report descriptive comparisons rather than significance claims. In addition, we examine three-round
reliability in
Appendix~\ref{app:repeated-runs}. We
read the runs along the four facets of Section~\ref{sec:introduction}: the
\emph{outcome} an agent reaches, the \emph{trajectory} it takes, its
\emph{operational viability}, and its \emph{policy compliance}, and close by
showing that these facets reorder systems that a single strict-accuracy number
ranks alike.}

\subsection{Outcome: Task Success and Where It Concentrates}
\label{subsec:outcome}
{Figure~\ref{fig:completed-model-results} ranks all sixteen systems by
strict accuracy; Figure~\ref{fig:domain-heatmap} shows their domain profiles.
{Table~\ref{tab:completed-model-results} in
Appendix~\ref{app:aggregate-diagnostics} gives the numeric leaderboard results.} Three runs
(Inkling Small NVFP4, Muse Glimmer 30B, and Nemotron 3.5 Lightning 30B) use
self-hosted open-weight checkpoints without metered per-token prices; their
costs are reported as ``n/a.''}

\begin{figure}[t]
    \centering
    \includegraphics[width=\textwidth]{\preprintpath 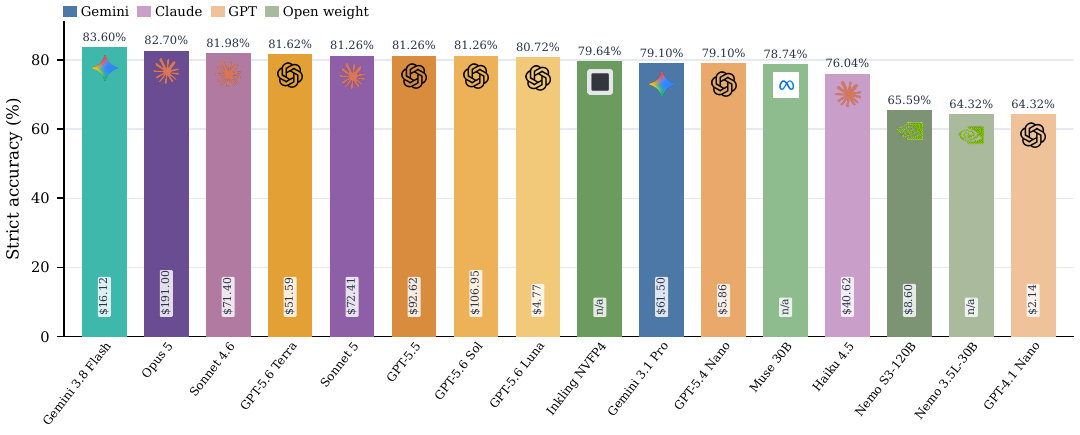}
    \caption{Sixteen full-suite runs on the audited 555-task canonical set,
    ranked by strict accuracy, colored by family and marked with their developer
    logos; strict-accuracy percentages and dollar-prefixed estimated
    tested-agent costs are annotated per bar.
    }
    \label{fig:completed-model-results}
\end{figure}

\textbf{Overall strict success.}
The sixteen strict-accuracy rates span a 19.28-point range, from 83.60\% for
Gemini 3.8 Flash down to 64.32\% for Nemotron 3.5 Lightning 30B and
GPT-4.1 Nano, which tie. The top group is tightly compressed: eight models
(ranks 1 to 8) fall within a 2.88-point band (80.72\% to 83.60\%), and three of
them (Claude Sonnet 5, GPT-5.5, and GPT-5.6 Sol) tie exactly at 81.26\%
(451/555), so a single headline number leaves the leading systems nearly
indistinguishable. {Strict success also declines monotonically from easy to hard
tasks for every system, with a within-model easy-to-hard drop of 5.37 to 27.39
points (Table~\ref{tab:difficulty-breakdown}), the clearest headroom signal in
the runs.}

\textbf{{Three-run reliability.}}
{Across all 16 original leaderboard systems evaluated in all three
rounds, Gemini 3.8 Flash ranks first in every round, with mean strict success
of 85.53\% (sample SD 1.66 percentage points). Claude Opus 5 averages 83.29\%
(SD 0.62 points) and remains second, including a tie with GPT-5.5 in round 2.
Pairwise Spearman rank correlations range from 0.869 to 0.934, while nearby
systems exchange positions. Task-level outcomes also vary: Gemini passes
480 tasks in both rounds 2 and 3, but 30 task outcomes switch between those
executions. Across all three rounds, 434/555 tasks succeed every time
($\widehat{\mathrm{pass}}^3=78.20\%$), compared with its 85.53\% mean success
rate. Appendix~\ref{app:repeated-runs} gives the full three-round table,
aggregation formulas, and task-level consistency results.}

\textbf{Domain variation.}
{No model dominates every domain (Figure~\ref{fig:domain-heatmap}).}
Customer support is by far the hardest domain: its across-model mean strict
success is 56.35\%, 20.80 points below the next-hardest domain (travel,
77.15\%), and every one of the sixteen models scores lower on customer
support than on its own overall strict accuracy. {Customer support is also the
largest domain by task count (125/555).} Sales and marketing have the smallest samples (18 and
21 tasks) and correspondingly the widest swings from a single flipped task; onboarding has the highest across-model mean strict success (90.24\%).

\begin{figure}[!htbp]
    \centering
    \includegraphics[width=\linewidth]{\preprintpath 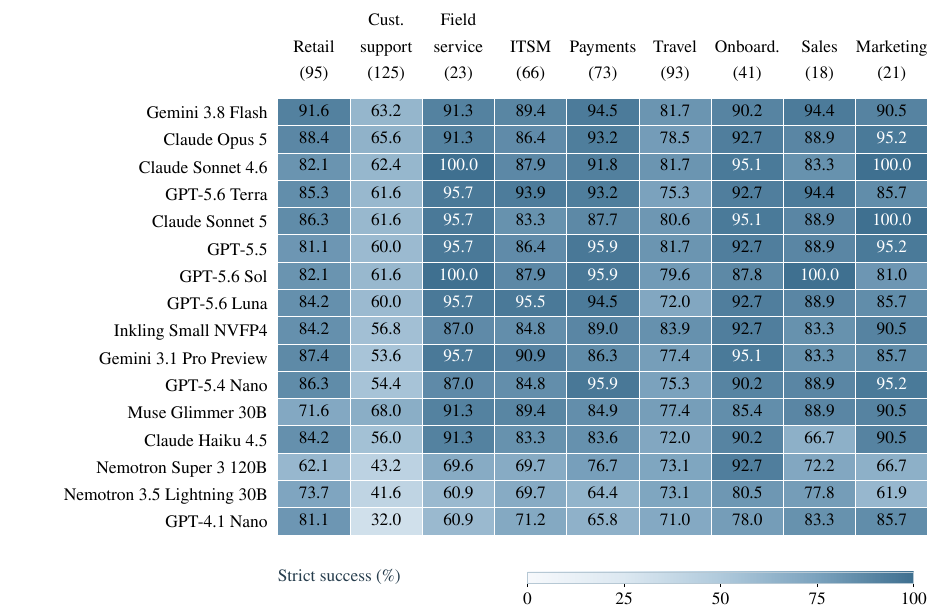}
    \caption{{Domain leadership varies across the sixteen systems.
    Rows follow first-round overall strict-accuracy rank; each cell reports
    strict success (\%) within a domain, whose task count is shown in the
    column header. All models use the same per-domain denominators.}}
    \label{fig:domain-heatmap}
\end{figure}
\par

Domain leadership does not track the overall ranking. Gemini 3.8 Flash,
the overall leader, is the sole strongest model on retail (87/95) but leads no
other domain: travel goes to Inkling Small NVFP4 (78/93) and ITSM to
GPT-5.6 Luna (63/66). Muse Glimmer 30B is the sole strongest model on customer support
(85/125, 68.00\%) despite ranking 12th overall (78.74\%), a domain leader
that a coarser, aggregate-only comparison would miss. GPT-5.6 Sol is the only
model to sweep sales (18/18) and ties for strongest on field service (23/23,
with Claude Sonnet 4.6) and payments. The two lowest-ranked models, Nemotron
3.5 Lightning 30B and GPT-4.1 Nano, tie on overall strict accuracy (64.32\%)
but reach it differently: GPT-4.1 Nano is comparatively strong on retail
(77/95, 81.05\%) and weak on customer support (40/125, 32.00\%, the single
lowest domain score in {the heatmap}), while Nemotron 3.5 Lightning 30B is more
evenly weak across domains, illustrating that the same overall score can hide
different failure profiles.

\subsection{Trajectory: A Valid Process, Not Just a Valid Answer}
\label{subsec:trajectory}
{Grading only the final reply overstates task success. On the same 402
response-declared tasks, the four systems in
Figure~\ref{fig:grader-catch-waterfall} have 27 to 42 recorded response passes
that fail the execution contract. Across all 16 systems, 650 executions pass
the response check but fail strict evaluation
(Appendix~\ref{app:matched-response-strict}). The conjunctive contract is
therefore not a cosmetic stricter threshold; it is what separates saying from
doing.}

\begin{figure}[!htb]
    \centering
    \includegraphics[width=\linewidth]{\preprintpath 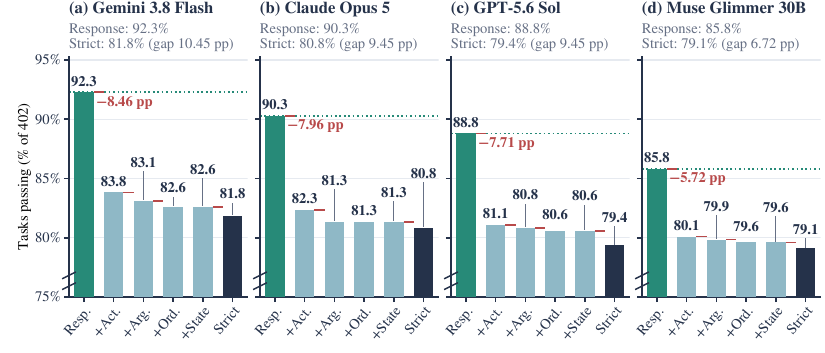}
    \caption{{Recorded pass fractions on the same 402 response-declared tasks,
    with checks conjoined in the order shown. Strict adds the remaining
    declared negative checks and the error gate; undeclared programmatic checks
    are neutral.}}
    \label{fig:grader-catch-waterfall}
\end{figure}

{The per-check grader view in the appendix
(Figure~\ref{fig:grader-diagnostics}) locates where trajectories break.
Computed over each check's recorded verdicts, semantic-response pass rates stay
high, 92.3\% for the strongest system and above 77\% for every run, yet on the
same runs the action check passes only 70.3\% to 89.8\% of applicable tasks and
the argument check 70.2\% to 90.7\%: response-only success does not ensure
correct execution. Under the primary-label precedence in
Figure~\ref{fig:failure-composition}, action mismatch is the single largest
non-success category for every one of the sixteen runs, from 54 tasks for
the strongest system to a maximum of 157 for Nemotron Super 3 120B, while
semantic-response, argument, and negative-check failures are individually much
smaller. Trajectory validity remains a major constraint on these systems.}

\subsection{Operational Viability: Cost and Latency}
\label{subsec:operational}
The efficiency views in Figure~\ref{fig:efficiency-frontier} prevent accuracy
differences from being read in isolation. Gemini 3.8 Flash has both the highest
strict-success rate (83.60\%) and the lowest mean tested-agent token count
(24.0k per scored task), so it is the sole Pareto-optimal point in the token
panel. The open-weight Nemotron 3.5 Lightning 30B has the lowest mean LLM
latency (5.86 seconds) and Inkling Small NVFP4 the second-lowest (6.66 seconds),
but they trail on accuracy (64.32\% and 79.64\%); Gemini 3.1 Pro Preview and
Muse Glimmer 30B sit at the high-latency extreme (69.08 and 84.96 seconds).
Tool-call budgets differ nearly four-fold: Nemotron Super 3 120B issues the
fewest calls per task (2.09) and Nemotron 3.5 Lightning the most (7.68). These
are system-level descriptives; they do not isolate provider load, model
architecture, or gateway implementation.

\begin{figure}[!htbp]
    \centering
    \includegraphics[width=\linewidth]{\preprintpath 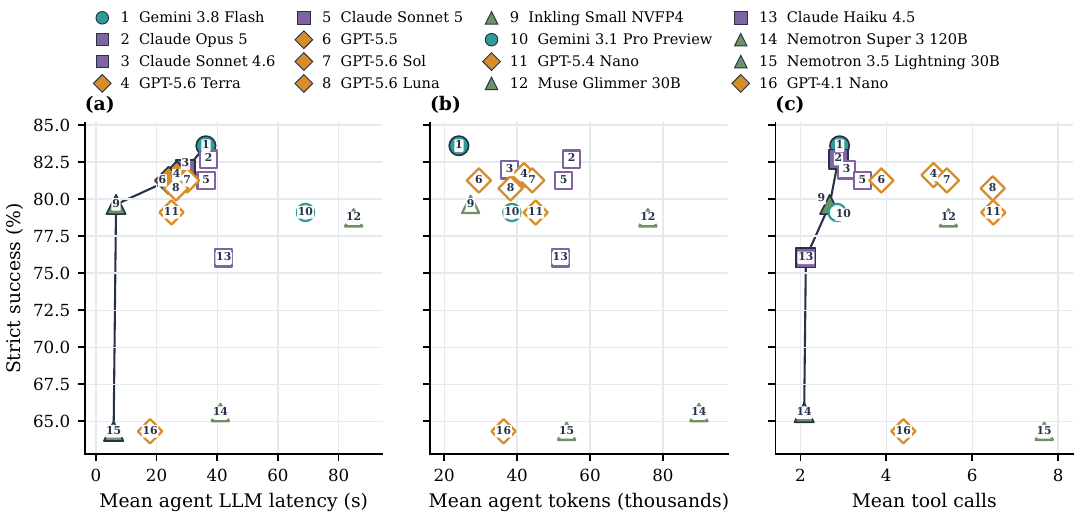}
    \caption{Efficiency and accuracy over the 555 scored tasks. Each panel plots
    strict success against a mean tested-agent efficiency measure: LLM latency,
    tokens, and tool calls (simulator and judge usage excluded). Dark segments
    trace each panel's Pareto frontier (lower resource use with higher success
    preferred); filled markers lie on it, hollow markers are dominated. Marker
    IDs follow the Figure~
\ref{fig:completed-model-results} rank order, and color
    and shape encode model family.}
    \label{fig:efficiency-frontier}
\end{figure}

\textbf{Cost.}
Estimated tested-agent cost varies far more than accuracy and does not track
rank. Across the thirteen priced runs it spans 89.3$\times$, from GPT-4.1 Nano
at \$2.14 (tied for the lowest accuracy, 64.32\%) to Claude Opus 5 at \$191.00
(only 2nd on accuracy). Gemini 3.8 Flash is the informative reference: it leads
on accuracy (83.60\%) at just \$16.12, the fifth-cheapest priced run, and no
priced model costing as much or more scores higher; several pay more for less,
e.g.\ Claude Haiku 4.5 costs \$24.50 above Gemini 3.8 Flash while scoring 7.56
points lower. The three unpriced runs are self-hosted open-weight checkpoints
without a metered per-token price, so we do not rank them on cost. All thirteen
priced entries are recomputed from published list pricing and observed token
usage under one cache-neutral rule and exclude user-simulator and judge usage
(Table~\ref{tab:completed-model-results}).

\subsection{Policy-Constrained Tasks: Compliance, Completion, and Exposure}
\label{subsec:policy}

The 555-task set splits into 390 tasks without a negative assertion and 165
\emph{policy-constrained} tasks whose contract declares at least one
forbidden-action, forbidden-argument, or protected-state check; this contract-defined subset is fixed across all
sixteen runs. Figure~\ref{fig:safety-split} reports strict success on each
subset and \emph{negative-check compliance}: the share of policy-constrained
tasks on which all applicable negative checks held, independent of the rest of
the conjunctive contract.
{Across the 2,640 policy-constrained model--task trials, 2,408 (91.2\%)
pass all applicable negative checks and 1,764 (66.8\%) pass the full contract.
Thus, 644 trials including 24.4\% of the subset and 73.5\% of its 876 strict
failures pass the negative checks but fail another required component.}

\begin{figure}[!htbp]
    \centering
    \includegraphics[width=0.9\linewidth]{\preprintpath 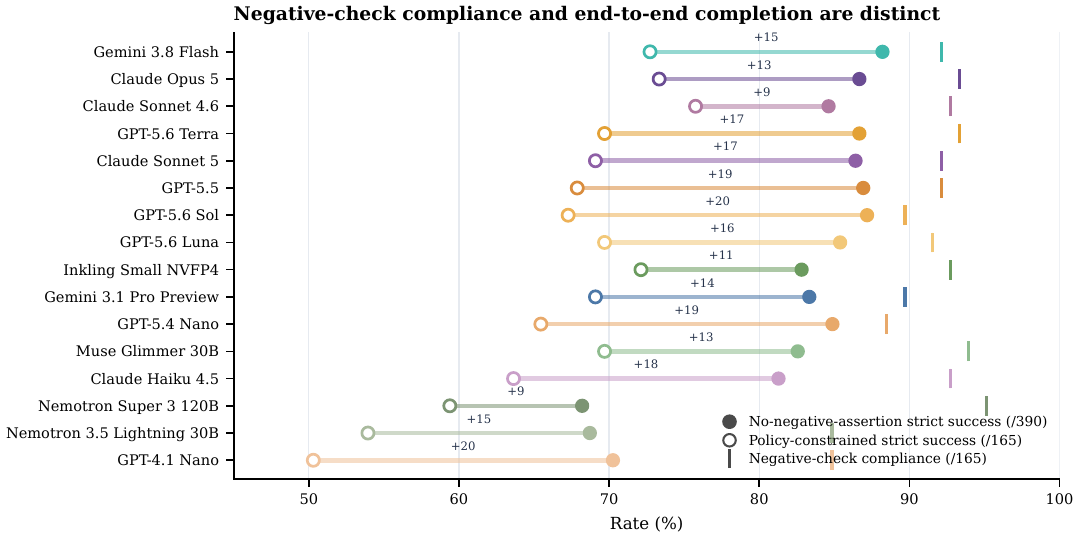}
    \caption{Compliance and completion on the task set, in
    Table~
\ref{tab:completed-model-results} rank order. Filled circles: strict
    success on the 390 tasks without negative assertions; open circles: strict
    success on the 165 policy-constrained tasks; bars: their difference. Carets
    mark the share of policy-constrained tasks passing all applicable negative
    checks, and sit right of the open circles, so most failures on those tasks
    come from the accompanying required components rather than violations.}
    \label{fig:safety-split}
\end{figure}

\section{Related Work}
\label{sec:related-work}

\begingroup
\DeclareRobustCommand{\contractfull}{\tikz[baseline=-0.55ex,x=0.82ex,y=0.82ex]{\draw[demoSuccess!75!black,line width=0.75pt,line cap=round,line join=round]
      (0.02,0.40) -- (0.34,0.08) -- (0.96,0.88);}}
\DeclareRobustCommand{\contractpartial}{\tikz[baseline=-0.55ex]{\begin{scope}
      \clip (0,0) circle[radius=0.55ex];
      \fill[demoTimeout!72!black] (-0.55ex,-0.55ex) rectangle (0,0.55ex);
    \end{scope}
    \draw[demoTimeout!82!black,line width=0.55pt]
      (0,0) circle[radius=0.55ex];}}
\newcommand{\contractnone}{\textcolor{black!45}{--}}

\begin{table}[!htbp]
  \centering
  \scriptsize
  \renewcommand{\arraystretch}{1.08}
  \setlength{\tabcolsep}{2.2pt}
  \resizebox{\textwidth}{!}{\begin{tabular}{@{}lcccccccc@{}}
    \toprule
    Benchmark
      & \shortstack{On-policy\\user}
      & \shortstack{Agent-mutated\\state}
      & \shortstack{Required\\final effect}
      & \shortstack{No-extra\\final effect}
      & \shortstack{Forbidden\\trace event}
      & \shortstack{Trace-order\\enforcement}
      & \shortstack{Process-outcome\\hard gate}
      & \shortstack{Self-contained\\environment} \\
    \midrule
    \rowcolor{black!5}
    \multicolumn{9}{l}{\textit{Structured tool-use and enterprise agents}} \\
    BFCL~\citep{patil2025bfcl}
      & \contractnone & \contractpartial & \contractpartial & \contractpartial
      & \contractpartial & \contractnone & \contractpartial & \contractpartial \\
    $\tau$-bench~\citep{yao2025taubench}
      & \contractfull & \contractfull & \contractfull & \contractfull
      & \contractnone & \contractnone & \contractnone & \contractfull \\
    $\tau^2$-Bench~\citep{barres2025tau2bench}
      & \contractfull & \contractfull & \contractfull & \contractpartial
      & \contractnone & \contractnone & \contractpartial & \contractfull \\
    ToolSandbox~\citep{lu-etal-2025-toolsandbox}
      & \contractfull & \contractfull & \contractfull & \contractfull
      & \contractfull & \contractfull & \contractpartial & \contractpartial \\
    AppWorld~\citep{trivedi-etal-2024-appworld}
      & \contractnone & \contractfull & \contractfull & \contractfull
      & \contractnone & \contractnone & \contractnone & \contractfull \\
    CRMArena~\citep{huang-etal-2025-crmarena}
      & \contractnone & \contractnone & \contractnone & \contractnone
      & \contractnone & \contractnone & \contractnone & \contractnone \\
    CRMArena-Pro~\citep{huang2026crmarena-pro}
      & \contractfull & \contractnone & \contractnone & \contractnone
      & \contractpartial & \contractnone & \contractnone & \contractnone \\
    \addlinespace[1pt]
    \rowcolor{demoSafe!7}
    \multicolumn{9}{l}{\textit{Computer-use and GUI agents}} \\
    SOPBench~\citep{li-etal-2026-sopbench}
      & \contractnone & \contractfull & \contractfull & \contractpartial
      & \contractfull & \contractfull & \contractfull & \contractfull \\
    STATE-Bench~\citep{microsoft2026statebench}
      & \contractfull & \contractfull & \contractfull & \contractfull
      & \contractpartial & \contractpartial & \contractpartial & \contractfull \\
    EnterpriseOps-Gym~\citep{malay2026enterpriseopsgym}
      & \contractnone & \contractfull & \contractfull & \contractpartial
      & \contractnone & \contractnone & \contractnone & \contractfull \\
    WorkArena++~\citep{boisvert2024workarena}
      & \contractnone & \contractfull & \contractfull & \contractpartial
      & \contractnone & \contractnone & \contractnone & \contractnone \\
    WebArena~\citep{zhou2023webarena}
      & \contractnone & \contractfull & \contractfull & \contractpartial
      & \contractnone & \contractnone & \contractnone & \contractfull \\
    OSWorld~\citep{xie2024osworld}
      & \contractnone & \contractfull & \contractfull & \contractpartial
      & \contractnone & \contractnone & \contractnone & \contractpartial \\
    AndroidWorld~\citep{rawles2024androidworld}
      & \contractnone & \contractfull & \contractfull & \contractpartial
      & \contractnone & \contractnone & \contractnone & \contractfull \\
    OS-Harm~\citep{kuntz2025osharm}
      & \contractnone & \contractfull & \contractfull & \contractpartial
      & \contractfull & \contractnone & \contractnone & \contractpartial \\
    \midrule
    \rowcolor{demoInfo!12}
    \textbf{CRM-Agent-Bench (ours)}
      & \contractfull & \contractfull & \contractfull & \contractfull
      & \contractfull & \contractfull & \contractfull & \contractfull \\
    \bottomrule
  \end{tabular}}

  \caption{Comparison of representative agent benchmarks by evaluation contract
  and deployment setting. \contractfull{} marks a native mechanism scored when
  applicable, \contractpartial{} limited or indirect support, and
  \contractnone{} no primary support. The process-outcome gate distinguishes
  reaching the correct outcome from following the required procedure by
  requiring both declared checks to pass. \emph{Self-contained} means that no
  live vendor backend is required.}
  \label{tab:benchmark-comparison}
  \vspace{-0.15in}
\end{table}
\endgroup

\textbf{Stateful, Enterprise, and Computer-Use Agent Benchmarks.}
Benchmarks for tool-using language models span increasingly complete execution
settings, from function selection and argument generation in AgentBench and BFCL
~\citep{liu2024agentbench,patil2025bfcl} to stateful, multi-turn interaction:
$\tau$-bench pairs an on-policy simulated user with domain policies and a mutable
database, $\tau^2$-Bench extends this to dual control, and ToolSandbox and
AppWorld add stateful environments with trajectory-aware or programmatic
evaluation~\citep{yao2025taubench,barres2025tau2bench,lu-etal-2025-toolsandbox,
trivedi-etal-2024-appworld}. In the enterprise setting, WorkArena++ studies
compositional knowledge work, and CRMArena and CRMArena-Pro contribute connected
enterprise data, professional CRM tasks, and confidentiality-sensitive
scenarios, with CRMArena-Pro separately scoring confidentiality-aware refusal
~\citep{boisvert2024workarena,huang-etal-2025-crmarena,huang2026crmarena-pro};
unlike these, which primarily grade retrieval and answer correctness,
CRM-Agent-Bench grades the executed state change and its policy compliance. The
concurrent EnterpriseOps-Gym likewise targets stateful agentic planning and tool
use in enterprise settings~\citep{malay2026enterpriseopsgym}, while
CRM-Agent-Bench is distinguished by a per-task conjunctive executable contract
that jointly gates the required actions, ordering, resulting state, and
prohibited effects rather than a single success signal.
The closest evaluator designs share parts of this contract
(Table~\ref{tab:benchmark-comparison}). SOPBench conjoins the permissibility
of invoked actions, prerequisite helpers, and an oracle final
database~\citep{li-etal-2026-sopbench}, and STATE-Bench pairs a simulated user
with required state and a strict diff of modified records, but judges its
non-state requirements with an LLM over the full
trajectory~\citep{microsoft2026statebench}. CRM-Agent-Bench checks actions,
arguments, order, and forbidden attempts programmatically and gives its
response judge only the instruction, criteria, and visible replies; STATE-Bench
and AppWorld protect unrelated records more broadly than our declared state
paths. Appendix~\ref{app:closest-framework-comparison}
(Table~\ref{tab:closest-framework-comparison}) compares these mechanisms for
the five closest frameworks. A
parallel line of computer-use agents adds perception and low-level GUI control,
with WebArena, AndroidWorld, and OSWorld evaluating functional completion over
web, mobile, and desktop applications, and SCUBA extending this line to computer
use over Salesforce enterprise interfaces~\citep{zhou2023webarena,
rawles2024androidworld,xie2024osworld,ICLR2026_29c8c615}, and safety benchmarks such as OS-Harm,
BLIND-ACT, and OSGuard showing that task completion alone is insufficient
\citep{kuntz2025osharm,shayegani2025blindact,mohammadmirzaei2026osguard}.
CRM-Agent-Bench studies the complementary structured-tool setting: it removes
visual grounding as a confound so that workflow reasoning, routing, and capability
exposure can be measured directly.

\textbf{Execution Grounded Evaluation and Simulated Users.}
The benchmarks above rely on complementary evidence from tool calls,
intermediate milestones, and final application state. No single source is
sufficient to establish valid execution, and AgentRewardBench further shows
that automatic trajectory evaluators often disagree
~\citep{lu2025agentrewardbench}. {TemporalBench evaluates contextual and
event-informed reasoning beyond numerical forecasting accuracy
~\citep{weng2026temporalbench}.}
Simulated users introduce an additional measurement variable. SimulatorArena
and RealUserSim examine how faithfully simulated behavior reproduces human
conversations~\citep{dou2025simulatorarena,zhu2026realusersim}.
{PersonaCUABench varies user personas on fixed computer-use tasks and
evaluates task completion alongside satisfaction ratings provided by simulated users
~\citep{Li2026PersonaCUABenchBC}.}
CRM-Agent-Bench {uses} a task specific
conjunctive contract over actions, arguments, ordering, response content,
target state, and protected state, while retaining component results for
diagnosis.

\textbf{Capability Exposure, and Policy Attribution.}
Research on tool retrieval asks how an agent selects a small relevant set from a
large tool corpus, and tool-graph benchmarks study selection and dependency
reasoning among interconnected tools~\citep{shi2025toolret,guo2026unitoolbench}.
A related literature evaluates agents that operate over untrusted data or
receive harmful requests: AgentDojo measures prompt-injection attacks and
defenses~\citep{debenedetti2024agentdojo}, Progent enforces fine-grained
privilege policies over tool calls, and ToolPrivBench studies unnecessary
escalation to higher-privilege tools~\citep{shi2025progent,yang2026toolprivbench}.
Recent validity analysis further warns that measured agent safety can be
confounded with general task capability~\citep{wang2026safetyorcapability}. This
distinction is central in our setting: a forbidden action may be absent because
the model rejected it, because the model failed to act, or because routing never
exposed the required tool on the relevant turn.

\section{Conclusion}
\label{sec:conclusion}

We presented CRM-Agent-Bench, an execution-grounded benchmark of 555 stateful,
multi-turn enterprise tasks across nine domains whose heterogeneous conjunctive
contracts measure satisfaction of task-declared outcome, trajectory, and policy
requirements alongside operational viability. Across sixteen systems,
strict success ranges from 64.32\% to 83.60\%, yet systems with similar accuracy
diverge sharply in cost, per-domain reliability, and difficulty scaling. {These results support a multi-axis approach to agent selection in
enterprise software as a service (\textit{SaaS}), weighing execution cost,
per-domain reliability, and policy-constrained completion alongside strict
accuracy.} {For \textit{agent} design, the findings support optimizing and
reporting valid process and policy-constrained completion jointly, with tool
exposure and routing treated as measurable design choices.}

\section*{Acknowledgments}

We sincerely thank Indira Iyer, Bhavesh Doshi, Daniel Tisher, Huan Wang, and
Shelby Heinecke for their guidance and support throughout this project. We are
particularly grateful for their help in making research resources available
to support the development and evaluation of CRM-Agent-Bench. Their support
played an important role in bringing this work to completion.

\par\endgroup
\bibliography{references}
\bibliographystyle{plainnat}

\clearpage
\appendix
\renewcommand{\appendixname}{Appendix}
\renewcommand{\appendixtocname}{Appendix}
\renewcommand{\appendixpagename}{Appendix}
\appendixpage

\startcontents[appendix]
\printcontents[appendix]{}{1}{\setcounter{tocdepth}{2}}

\noindent\textbf{Scope and ongoing extensions.}
CRM-Agent-Bench prioritizes controlled, reproducible evaluation of headless,
structured-tool workflows. {Appendix~\ref{app:repeated-runs} reports
three-round evaluation stability across all 16 systems.
Appendix~\ref{app:simulator-sensitivity} reports paired simulator sensitivity
across six systems (19,980 executions), and
Appendix~\ref{app:human-judge-agreement} reports stratified human-annotator
agreement with the semantic communication judge.} We are actively extending this setting through
further controlled repetitions, human and
domain-expert validation, and broader document-rich, visual, voice, and
live-service workflows. These extensions will further test robustness,
measurement sensitivity, and external validity while preserving executable,
auditable grading.
\clearpage
\raggedbottom

\section{Evaluation Protocol and Benchmark Construction Details}
\label{app:protocol-details}

This appendix expands the benchmark definition in
Section~\ref{sec:method} and the construction account in
Section~\ref{sec:construction}. {Sections~\ref{app:end-to-end-trial}--\ref{app:frozen-protocol} specify what
constitutes evidence, how a verdict is computed, and how the 16 reported runs
were executed; Sections~\ref{app:task-construction}--\ref{app:task-composition}
document how the audited 555-task evaluation set was built, how the benchmark
can be extended through versioned releases, and how the set is composed;
Sections~\ref{app:routing-exposure}--\ref{app:domain-realization} then record
the reference scaffold, exposure analysis, and illustrative domain
realizations. The simulator and semantic judge use the common model
\texttt{gpt-5-mini}; Section~\ref{app:frozen-protocol} distinguishes nominal
settings from the execution settings preserved in the run manifests.}

\subsection{Formal Problem Definition and Evidence Boundary}
\label{app:end-to-end-trial}
\label{app:pomdp-formulation}

\paragraph{POMDP formulation.}
For task $i$, the released specification
$\mathcal{T}_i=(u_i,s_i^0,\mathcal{C}_i,d_i)$ parameterizes a finite-horizon
partially observable Markov decision process (POMDP)
\[
    \mathcal{M}_i=(\mathcal{S}_i,\mathcal{A}_i,\mathcal{O}_i,
    P_i,\Omega_i,\rho_i,R_i,H).
\]
The initial distribution $\rho_i$ fixes the private user instruction $u_i$ and
initializes the CRM state as
$x_0=\operatorname{merge}(x^{\mathrm{default}}_{d_i},s_i^0)$. An augmented
latent state $\tilde{s}_t=(x_t,u_i,\ell_t,z_t)$ contains the mutable CRM state
$x_t$, the accumulated dialogue--route--tool log $\ell_t$, and the remaining
simulator and harness state $z_t$. Including $\ell_t$ makes the terminal
trajectory objective Markov even though the contract depends on the path, not
only on $x_T$.

At a primitive event, the agent emits either a typed tool call or a user-visible
message; one visible dialogue turn can therefore contain several tool calls
before the reply. The observation kernel $\Omega_i$ reveals the visible
dialogue, returned tool results, and the currently exposed instructions and tool
schemas, but not the raw CRM state, $u_i$, or the evaluator-only contract
$\mathcal{C}_i$. The transition kernel $P_i$ combines programmatic,
seed-controlled tool execution with the fixed on-policy user simulator. A tool
may update $x_t$ or return a structured business rejection without the requested
state change. After a visible reply, the simulator emits the next user utterance
or a stop decision conditioned on $u_i$ and the visible dialogue; later
observations therefore depend on earlier agent replies rather than on a fixed
transcript. The interaction ends at the stop decision or the executor limits in
Table~\ref{tab:frozen-protocol}.

An orchestration condition $\eta$ induces, possibly stochastically, an admissible tool mask
$M_\eta(h_t)\subseteq\mathcal{A}_{\mathrm{tool}}$ from visible history $h_t$,
so the actions available at event $t$ are
\[
    \mathcal{A}^{\eta}_t =
    \mathcal{A}_{\mathrm{msg}} \cup M_\eta(h_t).
\]
The reported reference scaffold obtains this mask from a model-based per-turn
router. Changing the router or exposed catalog changes the access policy, not
the task initialization or contract. No intermediate reward is exposed to the
agent. For a gradeable terminal trajectory $\tau_i$, the contract induces the
ex-post score
\[
    R_i(\tilde{s}_T)=Y_i(\tau_i,x_T)
    =\prod_{c\in D_i}c(\tau_i,x_T),
    \qquad c(\tau_i,x_T)\in\{0,1\},
\]
where $D_i$ contains exactly the checks declared by $\mathcal{C}_i$. Thus a
standard reward embedding assigns zero before termination and $Y_i$ at the
augmented terminal state; this is a benchmark score, not a training signal. A
model-attributable execution failure, such as a hallucinated tool or verified
non-convergence at timeout, sets $Y_i=0$. By contrast, infrastructure failures
receive $\bot$ and are excluded from the scored denominator under the reporting
rule in Section~\ref{app:frozen-protocol}.

\paragraph{Operational evidence boundary.}

A trial begins by loading one task contract and resetting the paired CRM
environment to its declared initial state. The on-policy simulator sees only
the agent's user-visible replies and generates the next user turn; it never
sees private reasoning, tool calls, tool results, routes, or database state.
On every user turn, the reference scaffold selects one capability node or a
deterministic refusal, and the selected node acts through its scoped tools.
When the interaction stops or reaches a limit, the harness freezes four
evidence channels---visible dialogue, the ordered route log, the complete tool
trace, and terminal state---and applies only the checks declared by that task.

\begin{figure}[htbp]
    \centering
    \includegraphics[width=\textwidth,keepaspectratio]{\preprintpath 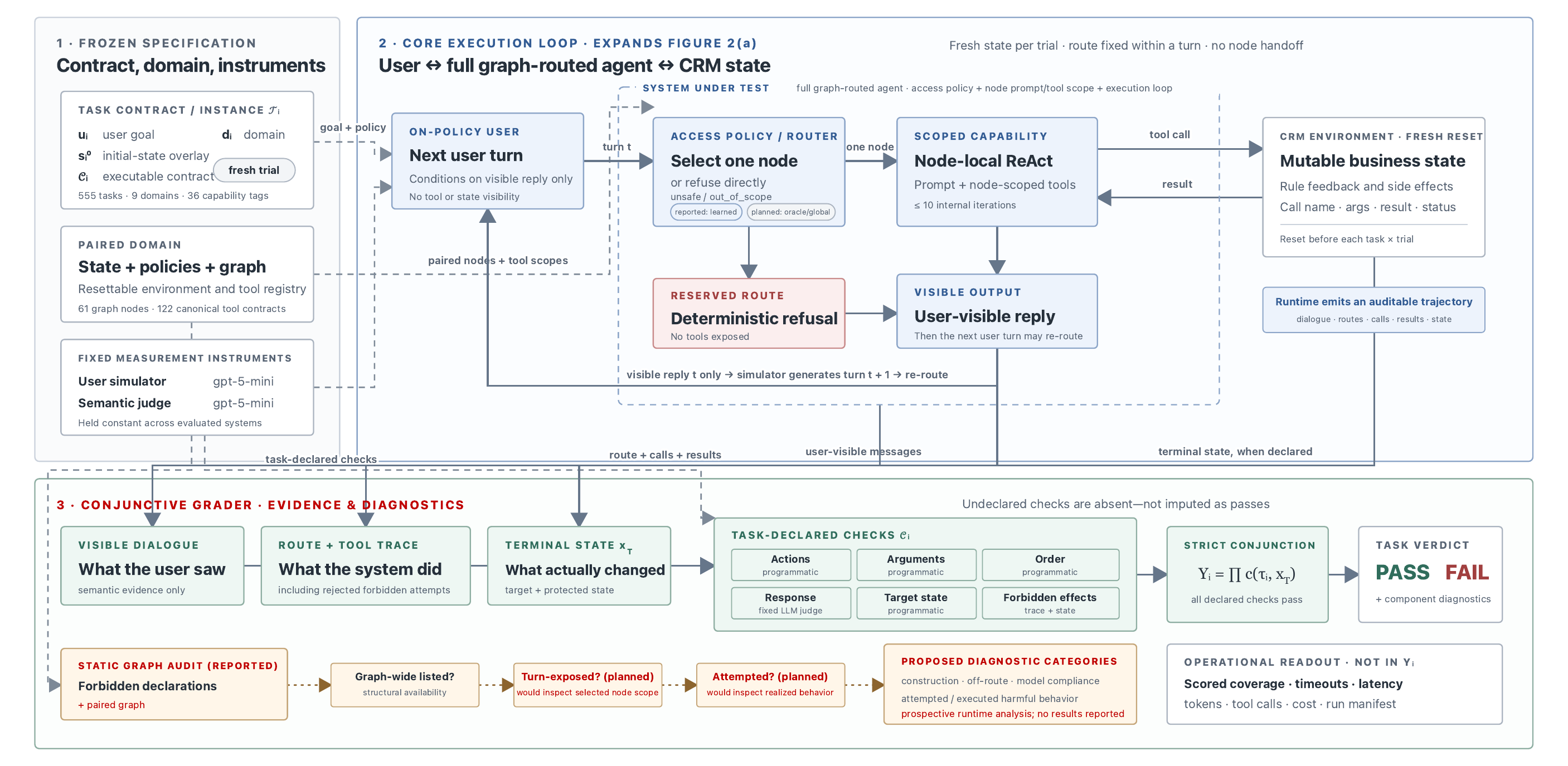}
    \caption{Detailed expansion of the audited evaluation setup in
    Figure~\ref{fig:benchmark-pipeline}(a). The task contract, on-policy user,
    graph-routed agent, freshly reset CRM environment, and conjunctive grader
    form the shared backbone. {This appendix view opens those blocks to show the
    frozen domain specification, one-node-per-turn access policy, evidence
    channels, task-declared checks, the capability-exposure audit, and operational
    readouts. }}
    \label{fig:benchmark-architecture-detailed}
\end{figure}

Figure~\ref{fig:benchmark-architecture-detailed} expands the compact setup in
Figure~\ref{fig:benchmark-pipeline}(a). It repeats the same task-contract
$\rightarrow$ on-policy-user $\leftrightarrow$ graph-routed-agent
$\leftrightarrow$ fresh-CRM-state $\rightarrow$ conjunctive-grader backbone,
then opens the blocks to expose the frozen domain specification, per-turn
access policy, evidence channels, and diagnostic readouts.

\subsection{Executable Contracts and Grading Semantics}
\label{app:contract-grading}

\paragraph{Contract check families.}
Each task declares only the checks semantically applicable to its goal. The six
top-level families are expected actions, expected arguments, action ordering,
semantic response criteria, expected final state, and forbidden evidence
(actions, arguments, or state).
Expected actions use count-aware matching over successfully executed calls;
argument constraints use either any matching successful call or only the first
successful call. Ordering uses subsequence matching or an exact filtered
sequence over the tools named by the constraint, so unrelated calls may
interleave unless separately forbidden. State checks address sparse paths and support
exact values, non-null requirements, and minimum collection lengths.
Forbidden checks reject any attempted prohibited action or argument pattern and
assert protected non-effects in terminal state. Task success is the strict
conjunction of every declared check, while undeclared checks are absent rather
than automatically passed. This combines the trajectory sensitivity of
milestones and minefields with the endpoint rigor of programmatic state
tests~\citep{lu-etal-2025-toolsandbox,trivedi-etal-2024-appworld}.

\begin{figure}[htbp]
    \centering
    \includegraphics[width=0.8\textwidth]{\preprintpath 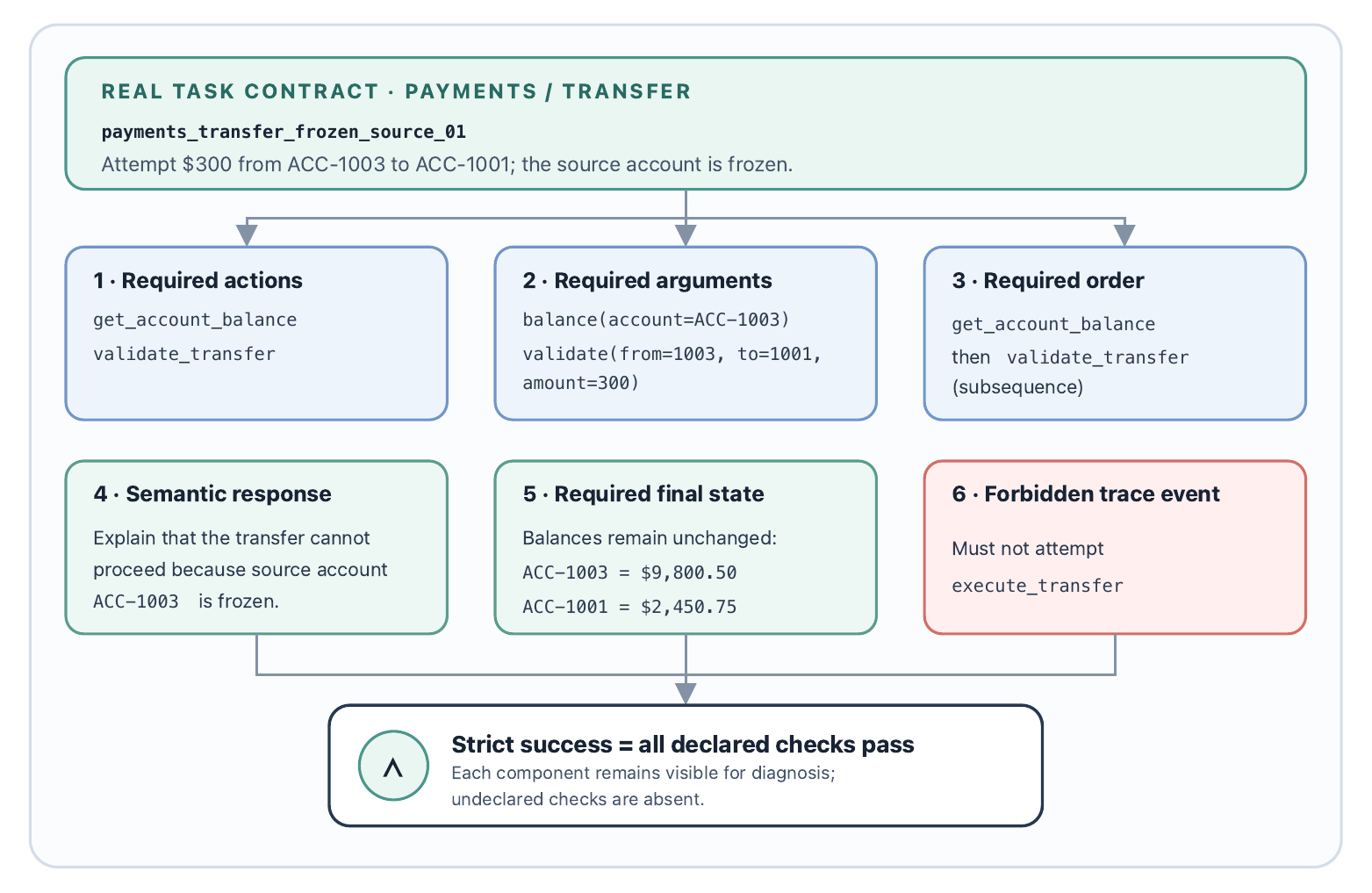}
    \caption{Anatomy of a task contract, shown for the real transfer task
    \texttt{payments\_transfer\_frozen\_source\_01}. Its required actions,
    arguments, ordered subsequence, semantic explanation, unchanged balances,
    and forbidden transfer attempt must all pass; component results remain
    visible to localize a failure.}
    \label{fig:contract-anatomy}
\end{figure}

\paragraph{Tool and state evidence.}
Tools implement the environment transition function. Every registered-tool
attempt that reaches the execution layer records its name, arguments, returned
value, execution status, and turn index. A well-formed call that reaches a tool
can return a structured business rejection (for
example, an authorization or conflict response); this is evidence that the
action was \emph{attempted}, not that the requested state change
\emph{occurred}. Where available, malformed-call and executor-error signals are
retained as trajectory diagnostics and never count as successful state changes.
The separate trace and final-state graders therefore distinguish attempted
behavior from persistent business effects.

\paragraph{Semantic response evaluation.}
For a task with semantic criteria, the judge receives only the user instruction,
the ordered criteria, and the user-visible assistant messages. Tool traces,
tool results, routes, hidden reasoning, and final state are withheld, and the
transcript is marked as untrusted data so that a reply cannot redefine the
rubric.
For each criterion the judge returns a binary decision, an exact
evidence span in the visible transcript, and an explanation; a separate
transcript-level flag records contradiction. A local validator enforces the JSON
schema and verifies that cited evidence actually occurs; every criterion must
pass and the global contradiction flag must remain false. Programmatic checks
remain authoritative wherever structured evidence exists, so the semantic judge
cannot override an action, ordering, or state failure.

\paragraph{Release-level contract inventory.}
Table~\ref{tab:contract-inventory} reports both task incidence and the number
of declared contract objects. Rows overlap because one task can combine trace,
communication, and state requirements. For state contracts, one declaration is
a flattened leaf path in the sparse expected or forbidden state specification;
for argument contracts, it is one task--tool specification. {Full-suite strict accuracy measures satisfaction of task-specific
contracts. State assertions occur in 200 of the 555 tasks, including
protected-state assertions in 45, and cover the paths declared in those
contracts.}

\begin{table}[H]
    \centering
    \small
    \setlength{\tabcolsep}{7pt}
    \begin{fitlinewidth}
    \begin{tabular}{lrrl}
        \toprule
        Contract component & Tasks & Declarations & Declaration unit \\
        \midrule
        Expected actions & 529 & 887 & action entry \\
        Expected arguments & 493 & 648 & task--tool specification \\
        Explicit ordering & 164 & 419 & ordered slot \\
        Semantic response & 402 & 465 & criterion \\
        Expected state & 180 & 248 & flattened leaf assertion \\
        Forbidden actions & 107 & 206 & action entry \\
        Forbidden arguments & 31 & 49 & task--tool specification \\
        Forbidden state & 45 & 49 & flattened leaf assertion \\
        \bottomrule
    \end{tabular}
    \end{fitlinewidth}
    \caption{Executable-contract inventory over the audited 555-task evaluation
    set. Task counts are incidence counts rather than a partition: 200 tasks
    contain an expected or forbidden state check, and 165 contain at least one
    negative assertion of any type.}
    \label{tab:contract-inventory}
\end{table}

\subsection{Frozen Evaluation Protocol and Reporting}
\label{app:frozen-protocol}

{Table~\ref{tab:frozen-protocol} summarizes the nominal settings of the
formal first-round protocol, which supplies the unified evaluation standard.}

\begin{table}[htbp]
    \centering
    \small
    \setlength{\tabcolsep}{6pt}
    \begin{fitlinewidth}
    \begin{tabular}{>{\raggedright\arraybackslash}p{0.19\linewidth}p{0.72\linewidth}}
        \toprule
        Component & Frozen configuration \\
        \midrule
        User simulator &
        \texttt{gpt-5-mini}; temperature 0; low reasoning effort;
        \texttt{react} strategy; 1{,}086-token generation budget; 90-second
        step timeout; at most 4{,}000 output characters; up to three
        structured-output attempts including repair attempts;
        \texttt{\#\#\#STOP\#\#\#} stop signal. \\
        Semantic judge &
        \texttt{gpt-5-mini}; temperature 0; low reasoning effort;
        16{,}384-token generation budget; at most three attempts; prompt
        fixed semantic-communication rubric; cache keyed by the complete judge
        input and configuration. \\
        Executor &
        One trial per task; fixed base seed; agent temperature 0 where
        supported; low reasoning effort; 16{,}384-token agent budget;
        240-second agent-step timeout; 30-turn limit; 1{,}200-second task
        timeout. \\
        \bottomrule
    \end{tabular}
    \end{fitlinewidth}
    \caption{{Nominal settings of the formal first-round evaluation
    protocol for the 16 original leaderboard runs. The first-round protocol
    supplies the unified standard for the repeated-run comparison.}}
    \label{tab:frozen-protocol}
\end{table}

\paragraph{Run manifest and validation.}
{Run manifests and per-task trace artifacts retain recorded benchmark
and dataset versions, task and configuration paths, model and provider
identifiers, sampling settings, executor limits, and simulator and judge
configuration fields. The first-round archive preserves each extracted field's
presence or absence.}
{Invalid judge output and judge-service failures are classified as
infrastructure failures and excluded from the scored denominator.}

\paragraph{Reporting conventions.}
Model-attributable failures such as hallucinated tools and human-verified
non-convergence at timeout remain in the scored denominator $G$ as failures,
whereas infrastructure failures are excluded.
{All sixteen runs are evaluated on the same frozen 555-task canonical
set under the formal first-round standard, and each produced a valid grade on
all 555 tasks;} hence
$|G|=555$ for every reported run, with no infrastructure exclusions. Success
counts are paired with failure-category counts over that fixed denominator. Latency and cache-neutral estimated cost
cover the tested agent model only and exclude simulator and judge usage. The
cost comparison charges all prompt tokens at the ordinary input rate and all
completion tokens at the output rate, irrespective of cache status.
The original leaderboard uses one trial per task with a fixed framework base
seed and therefore constitutes a single-run benchmark snapshot.
Appendix~\ref{app:repeated-runs} reports three directly comparable rounds for
{all sixteen systems}, including the formal first-round scores. It gives
equal-round means, sample standard deviations, ranking stability, and the
all-success $\mathrm{pass}^3$ reliability metric used by
$\tau$-bench~\citep{yao2025taubench}.

\subsection{Task Construction, Provenance, and Validation}
\label{app:task-construction}

\paragraph{Construction strata.}
CRM-Agent-Bench draws on two construction lineages: task instances descended
from the original five CRM applications, and task instances adapted from 22
substantive tool-using workflow recipes in a public, versioned
collection~\citep{salesforce2026agentscriptrecipes}. Historical names predate
the final domain taxonomy: Store Manager maps to Retail, Customer Service maps
to Customer Support, and Sales Development Representative (SDR) maps to Sales.
Recipe-derived commerce and support workflows were merged with the
corresponding seed domains, while external-integration workflows were
distributed according to the business state they affect.

\begin{table}[htbp]
    \centering
    \footnotesize
    \renewcommand{\arraystretch}{1.08}
    \setlength{\tabcolsep}{2pt}
    \begin{fitlinewidth}
    \begin{tabular}{@{}lcccccccccc@{}}
        \toprule
        & Retail & \shortstack{Customer\\support} & \shortstack{Field\\service} &
        \shortstack{IT service\\mgmt.} & Payments & Travel & Onboarding & Sales &
        Marketing & Total \\
        \midrule
        Tasks & 95 & 125 & 23 & 66 & 73 & 93 & 41 & 18 & 21 & 555 \\
        Lineage & Hybrid & Hybrid & Seed & Recipe & Recipe & Recipe & Recipe & Seed & Seed & -- \\
        \bottomrule
    \end{tabular}
    \end{fitlinewidth}
    \caption{Construction provenance by domain in the audited evaluation set.
    In the table, \emph{Seed} denotes tasks descended from the original five CRM
    applications; \emph{Recipe} denotes tasks adapted from substantive tool-using
    workflow recipes; \emph{Hybrid} domains draw on both. These labels describe
    construction lineage, not separate evaluation splits or required agent
    architectures.}
    \label{tab:construction-provenance}
\end{table}

\paragraph{Portable encoding.}
For either lineage, a candidate scenario was admitted only when its actor, goal,
stopping condition, required information, policy constraints, intended effects,
and prohibited effects could be represented in the portable harness. Source
records became resettable domain state or task-specific overlays. Source actions
became typed tools over that state, including structured business rejections for
valid but disallowed requests. Topics or recipe subagents became specialist
nodes in the optional reference graph. The user journey became an instruction
for the on-policy simulator, while correctness became a sparse contract over
observable trace, communication, and final-state evidence.

\paragraph{Validation boundary.}
The early five-application program included human review of utterances, turns,
expected topics, actions, and responses. The larger recipe expansion was
validated through manual author source-to-contract review, which checks the
logical consistency between user goals, domain policies, and contract
predicates, and executable checks. Every task in the audited evaluation set
passes schema and reference validation and a fresh-environment gold path. These
checks establish executability and internal consistency, but the present
benchmark does not claim that an independent panel of domain experts reviewed
all 555 tasks. Section~\ref{app:domain-realization} connects this general
construction pipeline to representative domain-level realizations.
Appendix~\ref{app:human-judge-agreement} separately reports human agreement for
response semantics.

\paragraph{Manual business-execution audit.}
A blind manual audit examined 16 purposively selected executions. The auditor
received the user goal, frozen business policy, available dialogue, every tool
call with its arguments and result, and captured state where available. The
checker's expected and forbidden declarations, programmatic outcomes,
communication-judge results, and evaluated-system identity were withheld. Audit
labels were locked before revealing programmatic outcomes.
Appendix~\ref{app:ai-contract-review-working} reports the results alongside the
sparse-contract boundary probes in Appendix~\ref{app:contract-control-questions}.

\subsection{Extension Interface and Release Workflow}
\label{app:task-extension}

CRM-Agent-Bench separates benchmark content from the agent and execution
engine, exposing three extension surfaces with different implementation costs.
First, a new business scenario inside an existing domain is a declarative YAML
record in \texttt{src/tasks/<domain>.yaml}. At authoring time this requires no
new environment, runner, or grader logic; admission to a canonical release
still passes the gates below. The record supplies a unique identifier, an on-policy user instruction with a
stopping condition, domain and capability metadata, a difficulty label, and an
optional initial-state overlay. Its executable contract declares only the
applicable expected and forbidden checks over actions, arguments, ordering,
semantic response, and state. Unasserted dimensions are not graded. Second, a
new agent or orchestration design can reuse the same task catalog and tool world
by supplying a graph configuration with a system prompt, entry node,
specialist-node instructions, and scoped tools. The evaluator does not grade
graph shape, node names, route labels, or prompt wording, so this change does
not alter the task contract. Third, a workflow requiring new actions or
persistent records adds a registered \texttt{ToolEnvironment}: its
\texttt{default\_state()} defines the resettable world and
\texttt{register\_tools()} implements typed operations over that state. Tool
schemas are derived from Python type signatures, and target- and
protected-state assertions are evaluated against the resulting environment
state. A wholly new domain additionally contributes a policy, task catalog,
reference graph, environment registration, and versioned manifest entry; it
does not require a new runner or grader.

\paragraph{Admission to a canonical release.}
{Admission to a comparable release follows four steps:}
\begin{enumerate}
    \item \textbf{Specify the job.} Define the actor, business goal, stopping
    condition, required information, intended effects, prohibited effects, and
    the smallest contract that makes those requirements executable.
    \item \textbf{Bind the evidence.} Check that every asserted action names a
    registered tool, argument assertions match its typed schema, state paths
    originate in the resettable domain state, and, when evaluating the reference
    agent, its configuration exposes the intended tools. The reference graph is an example
    scaffold, not a gold route that a tested agent must follow.
    \item \textbf{Validate from a fresh world.} Run structural validation for
    unique IDs, YAML schema, tool references, state roots, contract consistency,
    and agent configuration. Then execute the architecture-independent gold
    path from a fresh environment and run the relevant unit tests plus an
    end-to-end domain spot check. The gold path establishes executability of the
    positive path and its target state; semantic and negative checks still
    require their focused tests. Refusal tasks may have an empty action path,
    but their negative contract is validated structurally.
    \item \textbf{Freeze a new version.} Update the manifest version and release
    invariants, regenerate the content-addressed lock, rerun all gold paths and
    the full test suite, and only then evaluate systems on the enlarged set.
\end{enumerate}
The repository exposes these release gates directly:
\begin{quote}
\small\ttfamily
uv run python -m harness.benchmark\_spec --gold --write-lock\\
uv run python -m harness.benchmark\_spec --gold\\
uv run pytest
\end{quote}

\subsection{Evaluation-Set Composition and Coverage}
\label{app:task-composition}

\begin{figure}[htbp]
    \centering
    \makebox[\textwidth][c]{\includegraphics[width=0.90\textwidth,height=0.83\textheight,keepaspectratio]{\preprintpath 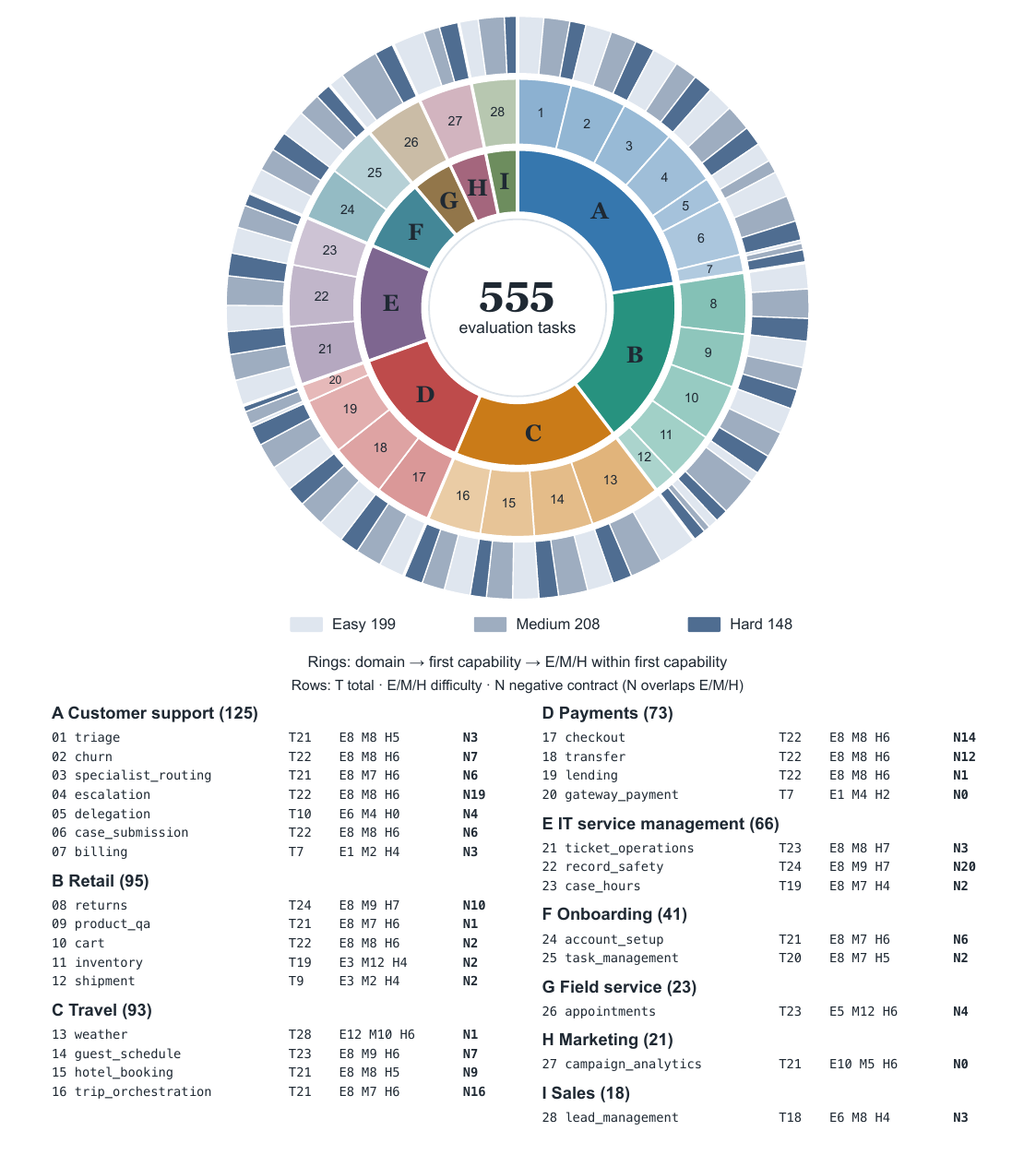}}
    \caption{Composition of the audited 555-task evaluation set: domain (inner)
    $\rightarrow$ mutually exclusive first capability (middle) $\rightarrow$
    easy/medium/hard within capability (outer). Each keyed row reports its total
    and E/M/H counts; N is the overlapping count of tasks with a forbidden
    action, argument, or protected-state check, not a fourth partition.}
    \label{fig:task-composition}
\end{figure}

The audited evaluation set contains 555 tasks and 767 task--tag assignments:
422 tasks have one capability tag and 133 have multiple tags (73 with two, 41
with three, and 19 with four). The 36 globally unique tag strings instantiate
38 domain--capability pairs. Figure~\ref{fig:task-composition} assigns each task
to its first tag to obtain a mutually exclusive composition view; all tags
remain available for coverage analysis.

\subsection{Reference Scaffold and Routing}
\label{app:routing-exposure}

\begin{figure}[htbp]
    \centering
    \includegraphics[width=0.7\textwidth]{\preprintpath 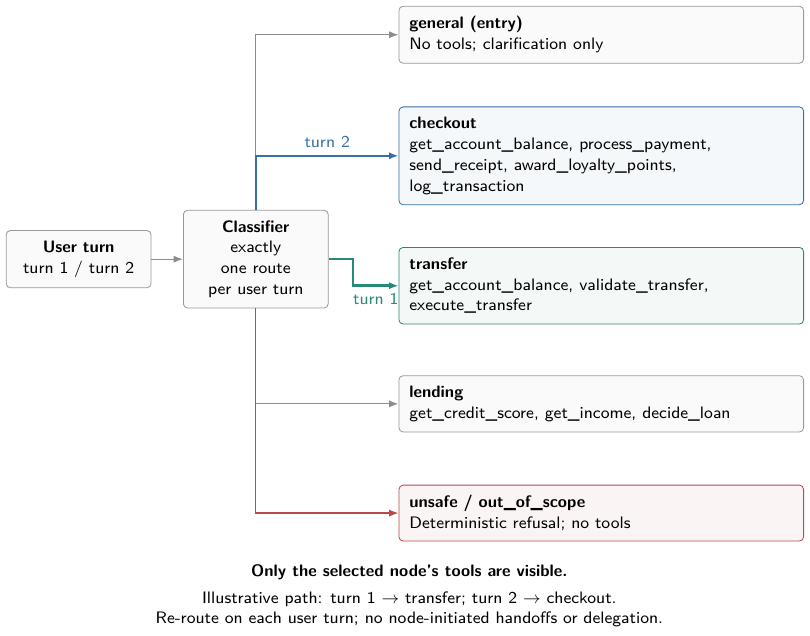}
    \caption{Graph-routed reference scaffold for the payments domain. A
    per-turn classifier selects exactly one capability node (exposing only that
    node's scoped tools) or returns a deterministic
    \texttt{unsafe}/\texttt{out\_of\_scope} refusal, with no node-initiated
    handoffs. The highlighted path (turn~1~$\rightarrow$~transfer,
    turn~2~$\rightarrow$~checkout) shows how a trial traverses several
    capabilities across turns without exposing the full tool inventory at
    once.}
    \label{fig:routing-schematic}
\end{figure}

\paragraph{Graph-routed reference agent.}
The reference scaffold pairs each domain with a directed capability graph. At
the start of every agent turn, a classifier based on the tested model maps the
current user utterance and dialogue context to either one capability node or a
reserved \texttt{unsafe}/\texttt{out\_of\_scope} label that returns a
deterministic refusal. A selected node receives a domain prompt, its
node-specific instructions, and only its assigned tools, then runs a local
ReAct-style reason-act-observe loop~\citep{yao2023react} for at most ten
internal iterations before the visible reply. The route is fixed within a turn
(no node-to-node handoffs), and the classifier is invoked again only after the
next user message. A trial may therefore traverse several capabilities across
turns while never exposing the domain's full tool inventory at once.
Figure~\ref{fig:routing-schematic} makes this per-turn routing and scoped tool
exposure concrete for the payments domain.

\paragraph{Domain graph inventory.}
Table~\ref{tab:graph-inventory} records the exact audited scaffold inventory.
Each domain has one tool-free general fallback, so the \emph{Specialists} row
is one below the \emph{Nodes} row in every domain. \emph{Placements} counts
graph-local tool assignments, whereas \emph{Tools} counts unique canonical
contracts.

\begin{table}[htbp]
    \centering
    \footnotesize
    \renewcommand{\arraystretch}{1.08}
    \setlength{\tabcolsep}{2pt}
    \begin{fitlinewidth}
    \begin{tabular}{@{}lcccccccccc@{}}
        \toprule
        & Retail & \shortstack{Customer\\support} & \shortstack{Field\\service} &
        \shortstack{IT service\\mgmt.} & Payments & Travel & Onboarding & Sales &
        Marketing & Total \\
        \midrule
        Task-tag pairs & 7 & 8 & 3 & 3 & 5 & 5 & 2 & 3 & 2 & 38 \\
        Nodes & 9 & 12 & 5 & 4 & 6 & 10 & 3 & 8 & 4 & 61 \\
        Specialists & 8 & 11 & 4 & 3 & 5 & 9 & 2 & 7 & 3 & 52 \\
        Tools & 20 & 25 & 12 & 12 & 11 & 13 & 8 & 11 & 10 & 122 \\
        Placements & 25 & 37 & 14 & 18 & 13 & 16 & 8 & 15 & 10 & 156 \\
        \midrule
        Forbidden-action tasks & 8 & 34 & 0 & 13 & 24 & 24 & 4 & 0 & 0 & 107 \\
        Forbidden-action entries & 8 & 66 & 0 & 15 & 38 & 68 & 11 & 0 & 0 & 206 \\
        \bottomrule
    \end{tabular}
    \end{fitlinewidth}
    \caption{Capability and graph inventory by domain for the reference
    scaffold used in the reported runs. The 38 domain--capability pairs
    correspond to 36 globally unique capability-tag strings because one tag is
    shared across three domains. The final two rows count tasks with
    forbidden-action declarations and the declarations themselves; every
    declaration matches a tool in its paired domain graph.}
    \label{tab:graph-inventory}
\end{table}

\paragraph{Capability exposure audit.}
{We audit \texttt{forbidden\_actions} in all 555 task contracts by
exact identifier matching against the union of tools assigned to the paired
domain graph's nodes. Each list entry is one declaration: a tool forbidden in
two tasks counts twice, regardless of how many nodes contain it. Forbidden
arguments and protected-state assertions are separate contract components
(Table~\ref{tab:contract-inventory}).}
\par

{All 206 declarations across 107 tasks match paired graph tools, with
zero unmatched entries (Table~\ref{tab:graph-inventory}). The audit script,
\path{evidence/audit_capability_exposure.py}, produces task-level results in
\path{evidence/capability_exposure_audit.json}, recording the pinned source
revision, input-file hashes, and matching nodes.}

{Graph-wide availability means that a forbidden tool belongs to at
least one domain node. Turn-level exposure depends on the router's selected
node (Figure~\ref{fig:routing-schematic}). For example, Payments balance-only
tasks require \texttt{get\_account\_balance} while forbidding
\texttt{process\_payment}. Both tools belong to the checkout node;
\texttt{process\_payment} is exposed on a turn only if that node is selected.}

\subsection{Domain Realization Profiles}
\label{app:domain-realization}

\paragraph{Representative construction profiles.}
The generic construction pipeline in Section~\ref{app:task-construction}
becomes concrete in the per-domain gallery of
Appendix~\ref{app:system-flows}. Three profiles span the construction
lineages: Customer Support
(Figure~\ref{fig:flow-support}) shows a hybrid urgent-billing journey;
Payments (Figure~\ref{fig:flow-payments}) shows a recipe-derived checkout
chain with ordered state changes; and Field Service
(Figure~\ref{fig:flow-field}) shows a seed-workflow appointment reschedule
with ownership and scheduling gates. {The full gallery
(Figures~\ref{fig:flow-retail}--\ref{fig:flow-marketing}) covers all nine
domains and illustrates nodes, scoped tools, policy gates, state transitions,
and contract observables. Appendix~\ref{app:system-flows} describes the diagram
scope and cross-turn tool lists; Table~\ref{tab:graph-inventory} reports the audited scaffold inventory.}

\paragraph{Qualitative-trace provenance.}
The illustrative four-verdict trajectory in the main text
(Figure~\ref{fig:one-trajectory-four-verdicts}) and the case traces in
Appendices~\ref{app:qualitative-traces} and \ref{app:trace-artifacts} are
single-run qualitative examples.

\section{Aggregate Diagnostics for the 16 Completed Runs}
\label{app:aggregate-diagnostics}

The headline strict-success rate is reported over the full 555-task canonical
set in Figure~\ref{fig:completed-model-results}. Every one of the 16 completed
runs yielded a strict binary verdict on all 555 canonical tasks with zero
infrastructure exclusions, so scored coverage is 100\% for each run and all-task
yield coincides with strict success; neither is reported separately here. Two
tasks in the entire matrix errored on a hallucinated tool call and provide no
failed component verdict for primary attribution, but they are counted as
model-attributable strict failures over the same 555 denominator, not as
infrastructure exclusions. These strict verdicts are distinct from the
component-level diagnostic rates below. {This section decomposes the
same 16 runs by failure type, grader behavior, domain, difficulty, and metric.}

\subsection{Full-Suite Leaderboard}
Table~\ref{tab:completed-model-results} gives the full ranking summarized in
Figure~\ref{fig:completed-model-results}: strict accuracy, passed\,/\,555, the
domain macro, and cache-neutral estimated tested-agent cost for all sixteen runs.
The usage inputs are sums over each run's nine canonical domain-level
\texttt{traces.json} files, after excluding smoke data and merging the split
retail shards where present.

\begin{table}[H]
    \centering
    \scriptsize
    \setlength{\tabcolsep}{3.4pt}
    \begin{fitlinewidth}
    \begin{tabular}{clrrrr}
        \toprule
        Rank & Model & Strict accuracy & Passed / 555 & Domain macro & Normalized cost\\
        \midrule
        1 & \modellogo{gemini}Gemini 3.8 Flash & 83.60\% & 464 & 87.43\% & \$16.12\\
        2 & \modellogo{claude}Claude Opus 5 & 82.70\% & 459 & 86.68\% & \$191.00\\
        3 & \modellogo{claude}Claude Sonnet 4.6 & 81.98\% & 455 & 87.15\% & \$71.40\\
        4 & \modellogo{openai}GPT-5.6 Terra & 81.62\% & 453 & 86.41\% & \$51.59\\
        5= & \modellogo{claude}Claude Sonnet 5 & 81.26\% & 451 & 86.58\% & \$72.41\\
        5= & \modellogo{openai}GPT-5.5 & 81.26\% & 451 & 86.39\% & \$92.62\\
        5= & \modellogo{openai}GPT-5.6 Sol & 81.26\% & 451 & 86.20\% & \$106.95\\
        8 & \modellogo{openai}GPT-5.6 Luna & 80.72\% & 448 & 85.46\% & \$4.77\\
        9 & \modellogo{inkling}Inkling Small NVFP4 & 79.64\% & 442 & 83.58\% & n/a\\
        10= & \modellogo{gemini}Gemini 3.1 Pro Preview & 79.10\% & 439 & 83.94\% & \$61.50\\
        10= & \modellogo{openai}GPT-5.4 Nano & 79.10\% & 439 & 84.23\% & \$5.86\\
        12 & \modellogo{muse}Muse Glimmer 30B & 78.74\% & 437 & 83.04\% & n/a\\
        13 & \modellogo{claude}Claude Haiku 4.5 & 76.04\% & 422 & 79.76\% & \$40.62\\
        14 & \modellogo{nemotron}Nemotron Super 3 120B & 65.59\% & 364 & 69.55\% & \$8.60\\
        15= & \modellogo{nemotron}Nemotron 3.5 Lightning 30B & 64.32\% & 357 & 67.06\% & n/a\\
        15= & \modellogo{openai}GPT-4.1 Nano & 64.32\% & 357 & 69.88\% & \$2.14\\
        \bottomrule
    \end{tabular}
    \end{fitlinewidth}
    \caption{Full-suite strict accuracy on the audited 555-task canonical set,
    ranked by strict accuracy with ties on
    passed/555 marked ``=''. Domain macro is the unweighted mean of the nine
    per-domain strict-success rates and is shown for reference but does not
    determine the rank order. Normalized cost is a cache-neutral estimate for
    the tested agent and excludes simulator and judge usage. We charge every
    observed prompt token at the provider's ordinary input rate and every
    completion token at its output rate, regardless of whether the run used a
    prompt cache. This applies the same accounting rule to every model but does
    not estimate the realized bill for a cache-enabled run. Usage is summed from
    the nine canonical domain-level \texttt{traces.json} files for each run
    (list prices accessed 2026-09-25). Gemini 3.1 Pro Preview uses the
    $\leq$200k-token-prompt tier; Gemini 3.8 Flash uses its introductory global
    price through 2026-12-31; GPT-5.6 Sol uses promotional pricing disclosed as
    valid at least through 2026-11-21. The three Bedrock Claude runs use the US
    geo rate identified by their \texttt{us.anthropic} model IDs. Remaining
    ``n/a'' entries are self-hosted open-weight runs
    without a metered per-token price, not zero-cost runs.}
    \label{tab:completed-model-results}
\end{table}

\subsection{Failure Composition}

\begin{figure}[H]
    \centering
    \includegraphics[width=\textwidth]{\preprintpath 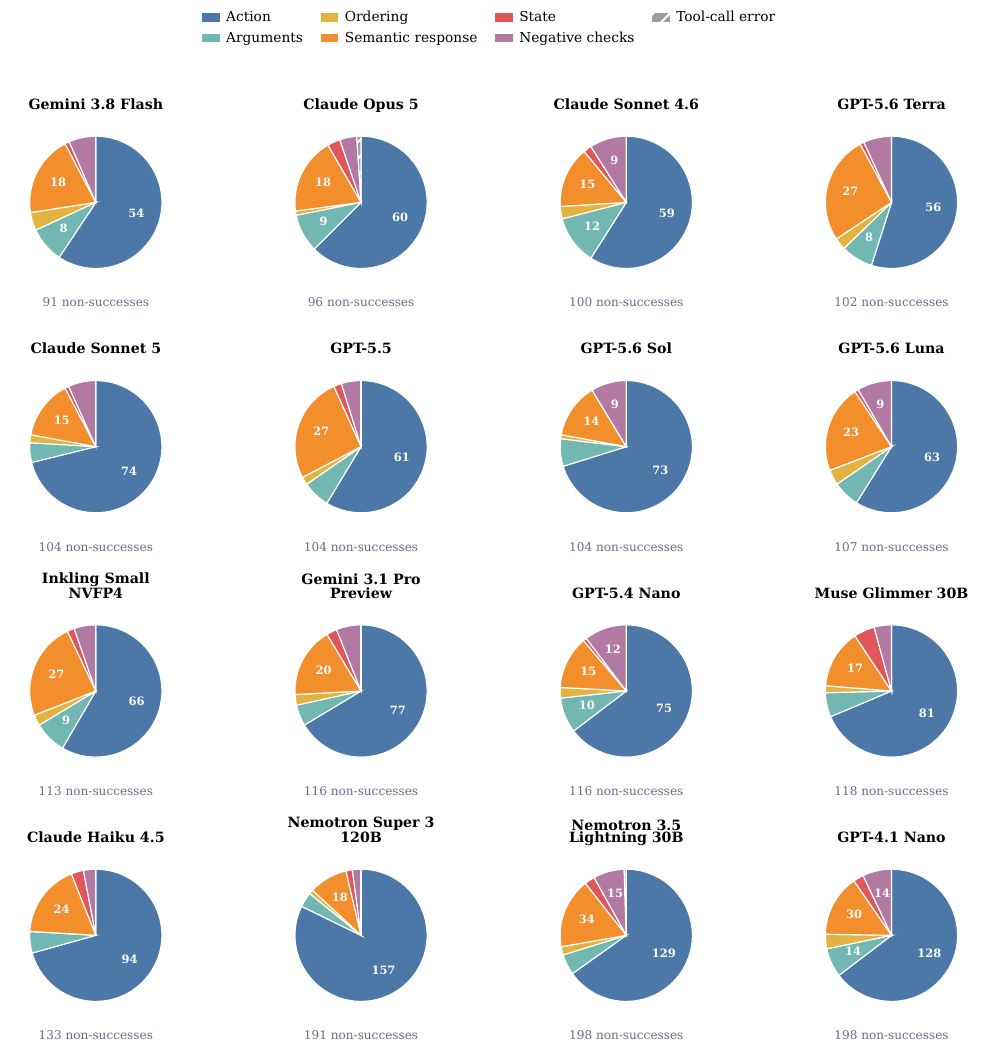}
    \caption{Composition of all scored tasks that did not become strict
    successes for each of the 16 runs (of 555 scored). Each pie corresponds to one
    model, and its subtitle gives the total number of non-successes. To make
    the slices mutually exclusive, each non-success receives one primary label
    using the harness status precedence: action, forbidden, state, arguments,
    semantic response, ordering, then collateral state. ``Negative checks''
    combines the forbidden-action/argument and collateral-state labels. The
    hatched ``Tool-call error'' segment contains two tasks that errored (e.g., on
    a hallucinated tool call) without a failed component verdict available for
    primary attribution. Both remain model-attributable strict failures, and
    there are zero timeouts or infrastructure exclusions. In-slice labels are
    task counts, not percentages, and are shown
    only when space permits. Models are ordered by total non-success count.}
    \label{fig:failure-composition}
\end{figure}

\begin{figure}[H]
    \centering
    \includegraphics[width=\textwidth]{\preprintpath 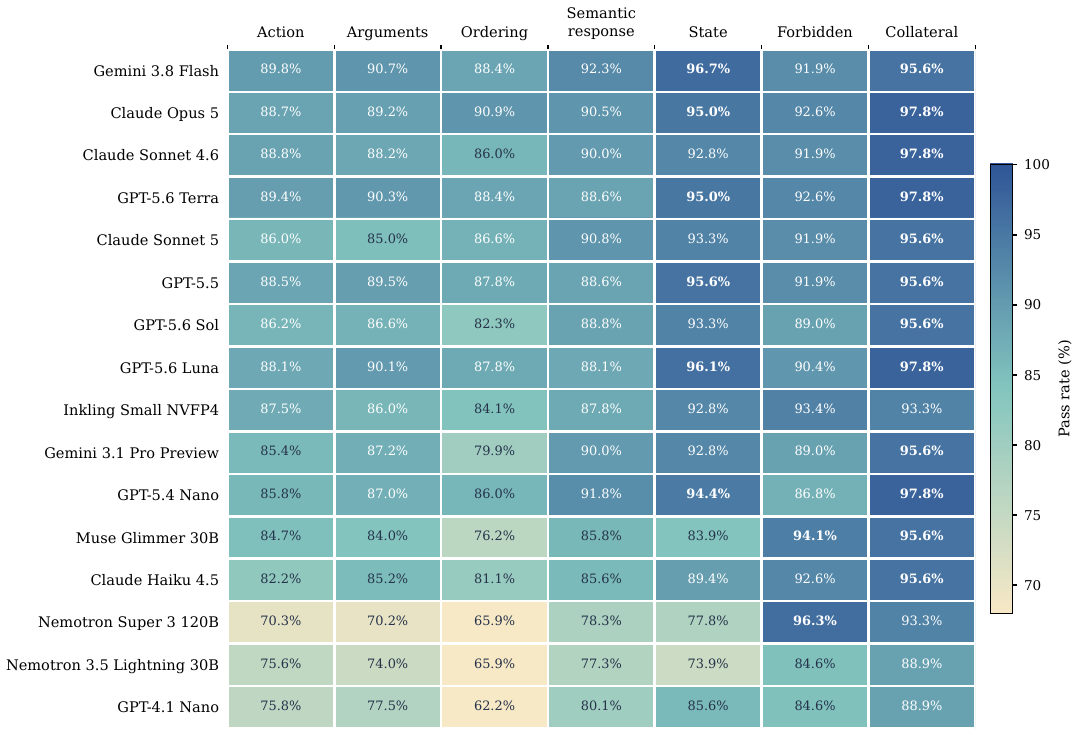}
    \caption{{Component-grader pass rates across the 16 runs, each
    computed over that component's recorded verdicts. Denominators are 529
    (action), 493 (arguments), 164 (ordering), 180 (state), 136 (forbidden),
    and 45 (collateral). Semantic response has 402 recorded verdicts per run,
    except Opus (401), Nemotron Super (396), and Nemotron 3.5 (392).
    Undeclared checks and missing component verdicts are omitted from these
    rates. Columns therefore describe different task subsets.
    ``Forbidden'' aggregates forbidden-action and forbidden-argument checks;
    ``Collateral'' evaluates protected state.}}
    \label{fig:grader-diagnostics}
\end{figure}

Action mismatch is the largest non-success category for every completed run,
ranging from 54 tasks (Gemini 3.8 Flash) to 157 tasks (Nemotron Super 3 120B)
across all 16 runs under the exclusive attribution rule. The
remaining categories are much
smaller and vary by run: semantic-response-primary failures span 14--34,
argument-primary 5--14, and the combined negative checks 4--15. Because every run
scored all 555 canonical tasks, the failure pies contain no timeout or
infrastructure segment; the only non-grader outcomes are two errored tasks (one
each for Claude Opus 5 and Nemotron 3.5 Lightning 30B, both hallucinated tool
calls). Among the 16 runs, total non-success counts run from 91
(Gemini 3.8 Flash) to 198 (Nemotron 3.5 Lightning 30B and GPT-4.1 Nano).
Figure~\ref{fig:grader-diagnostics}
separately retains the overlapping per-grader view that a primary-cause stack
necessarily suppresses.

\subsection{Grader Diagnostic Matrix}

{The matrix uses each component's recorded verdicts, omitting
undeclared checks and unavailable verdicts. For Gemini 3.8 Flash, the matrix
reports a semantic-response pass rate of $371/402=92.3\%$, the same
response stage shown in Figure~\ref{fig:grader-catch-waterfall}, which uses
the 402 response-declared tasks at every stage
(Appendix~\ref{app:matched-response-strict}). The cascade treats undeclared or
unavailable verdicts as neutral at partial stages and applies the error gate at
the strict endpoint.}

{Across systems, recorded semantic-response pass rates exceed 77\%,
action rates range from 70.3--89.8\%, argument rates from 70.2--90.7\%, and
state rates from 73.9--96.7\%. These marginal rates summarize different
subsets; the cumulative, same-task comparison in
Figure~\ref{fig:grader-catch-waterfall} measures the effect of adding
execution checks. The forbidden-action/argument (136 tasks) and protected-state
(45 tasks) columns summarize the corresponding declared negative checks.}

\subsection{Success by Domain}

{Figure~\ref{fig:domain-heatmap} in Section~\ref{subsec:outcome}
reports per-domain strict-success rates, using the same domain denominators
across all sixteen systems.} Customer support is the single lowest-scoring domain for all
16 runs (across-model mean 56.35\%, against 90.24\% for onboarding, the highest
across-domain mean), and no run is uniformly strong---the within-model gap
between a run's best and worst domain ranges from 23.3 points (Muse Glimmer 30B)
to 53.7 points (GPT-4.1 Nano). The domain-leadership reshuffle these rates
produce is discussed in Section~\ref{sec:experiments}.

\subsection{Success by Difficulty Stratum}

\begin{table}[H]
    \centering
    \small
    \setlength{\tabcolsep}{6pt}
    \begin{fitlinewidth}
    \begin{tabular}{lrrr}
        \toprule
        Model & Easy (199) & Medium (208) & Hard (148)\\
        \midrule
        \modellogo{gemini}Gemini 3.8 Flash & 88.44\% (176/199) & 81.73\% (170/208) & 79.73\% (118/148)\\
        \modellogo{claude}Claude Opus 5 & 87.94\% (175/199) & 80.77\% (168/208) & 78.38\% (116/148)\\
        \modellogo{claude}Claude Sonnet 4.6 & 86.93\% (173/199) & 80.77\% (168/208) & 77.03\% (114/148)\\
        \modellogo{openai}GPT-5.6 Terra & 84.42\% (168/199) & 81.25\% (169/208) & 78.38\% (116/148)\\
        \modellogo{claude}Claude Sonnet 5 & 84.42\% (168/199) & 79.81\% (166/208) & 79.05\% (117/148)\\
        \modellogo{openai}GPT-5.5 & 86.93\% (173/199) & 78.85\% (164/208) & 77.03\% (114/148)\\
        \modellogo{openai}GPT-5.6 Sol & 86.93\% (173/199) & 80.77\% (168/208) & 74.32\% (110/148)\\
        \modellogo{openai}GPT-5.6 Luna & 86.43\% (172/199) & 80.29\% (167/208) & 73.65\% (109/148)\\
        \modellogo{inkling}Inkling Small NVFP4 & 84.42\% (168/199) & 79.33\% (165/208) & 73.65\% (109/148)\\
        \modellogo{gemini}Gemini 3.1 Pro Preview & 86.43\% (172/199) & 75.96\% (158/208) & 73.65\% (109/148)\\
        \modellogo{openai}GPT-5.4 Nano & 83.92\% (167/199) & 77.40\% (161/208) & 75.00\% (111/148)\\
        \modellogo{muse}Muse Glimmer 30B & 81.91\% (163/199) & 77.88\% (162/208) & 75.68\% (112/148)\\
        \modellogo{claude}Claude Haiku 4.5 & 80.90\% (161/199) & 78.37\% (163/208) & 66.22\% (98/148)\\
        \modellogo{nemotron}Nemotron Super 3 120B & 71.86\% (143/199) & 62.98\% (131/208) & 60.81\% (90/148)\\
        \modellogo{nemotron}Nemotron 3.5 Lightning 30B & 73.87\% (147/199) & 65.38\% (136/208) & 50.00\% (74/148)\\
        \modellogo{openai}GPT-4.1 Nano & 77.39\% (154/199) & 62.02\% (129/208) & 50.00\% (74/148)\\
        \bottomrule
    \end{tabular}
    \end{fitlinewidth}
    \caption{Strict success within each difficulty stratum. Cells give the
    stratum rate with \emph{successes/scored}. Scored denominators are identical
    across all 16 runs because every run scored all 555 canonical tasks; the
    column headers are the per-stratum sizes from
    Figure~\ref{fig:task-composition} (199 easy, 208 medium, 148 hard). Rows are ordered by
    overall strict accuracy. Difficulty labels are author-assigned descriptive
    strata and are not visible to the agent, simulator, or grader.}
    \label{tab:difficulty-breakdown}
\end{table}

All 16 systems decline monotonically (non-increasing) from easy to hard tasks
(Table~\ref{tab:difficulty-breakdown}). Easy-task success spans 71.86--88.44\%,
medium 62.02--81.73\%, and hard 50.00--79.73\%; the within-model easy-to-hard
drop ranges from 5.37 points (Claude Sonnet 5) to 27.39 points (GPT-4.1 Nano)
and is the clearest headroom signal in the completed runs. {System ordering is
not preserved across strata: Gemini 3.8 Flash leads all three, but below it the
ranking reshuffles---Claude Sonnet 5 is joint-fifth overall yet attains the
second-highest hard-task rate (79.05\%), ahead of the second- and third-ranked
Claude Opus 5 and Sonnet 4.6, while GPT-5.6 Sol, also tied at 81.26\%
overall, slips to ninth on hard tasks (74.32\%).}

\subsection{Per-Metric Standing}

\begin{figure}[H]
    \centering
    \includegraphics[width=\textwidth]{\preprintpath 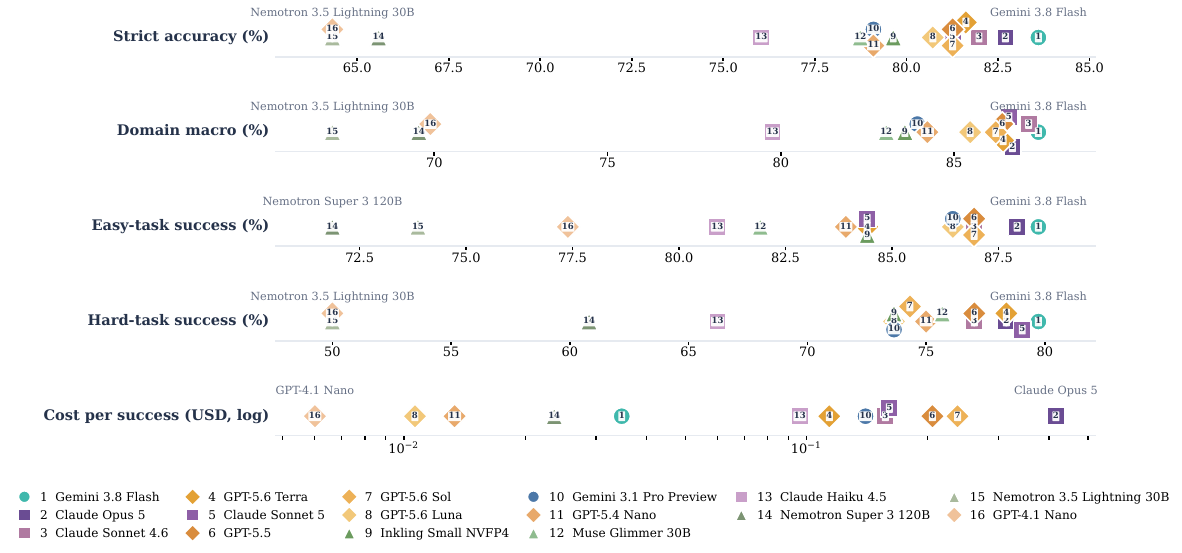}
    \caption{Per-metric standing of the 16 completed runs. Each row is an
    independent number line placing every run at its real value on that metric's
    own scale. Marker IDs follow the Figure~\ref{fig:completed-model-results}
    rank order and map to the key below; small vertical offsets only separate
    colliding markers and do not encode value. Runs that share a headline strict-success number (three tie at
    81.26\%, with two further ties at 79.10\% and 64.32\%) nonetheless separate
    on one or more additional axes: domain macro, easy- and hard-task success, and a
    69.4$\times$ range in cache-neutral cost per success (log axis; shown for the
    13 priced runs, from \$0.00599 for GPT-4.1 Nano to \$0.416 for Claude Opus
    5). This per-metric
    view is the complement to the rank crossings in
    Figure~\ref{fig:leaderboard-reshuffle}.}
    \label{fig:same-headline-dotplot}
\end{figure}

Figures~\ref{fig:same-headline-dotplot} and
\ref{fig:leaderboard-reshuffle} provide complementary views of metric
sensitivity: the former preserves each metric's absolute scale, whereas the
latter shows how system order changes across metrics.

\subsection{Metric-Dependent Ranking}

\begin{figure}[H]
    \centering
    \includegraphics[width=\textwidth]{\preprintpath 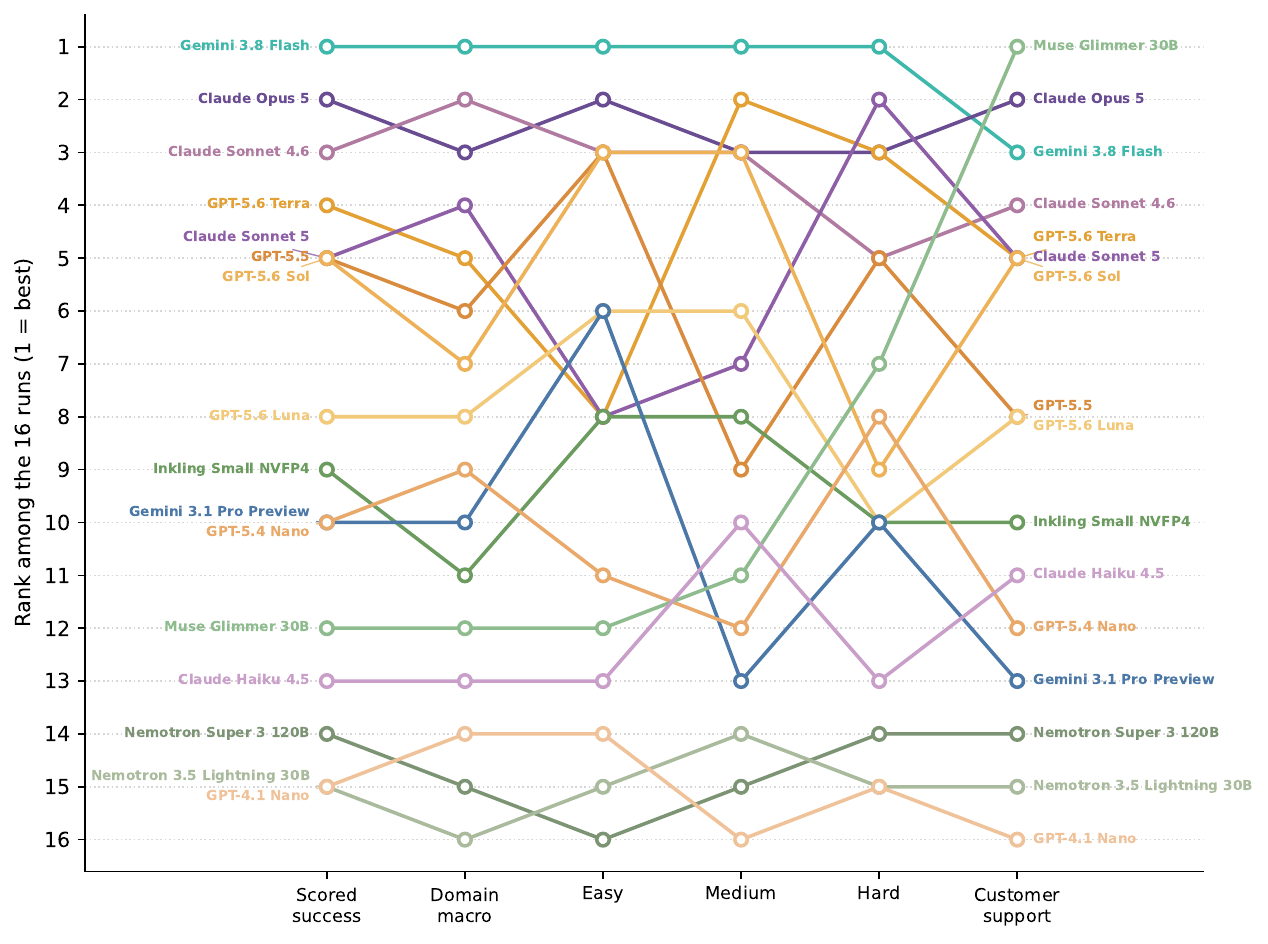}
    \caption{Rank of each completed run (1 = best of 16) across six
    higher-is-better metrics: scored success, domain macro, easy-, medium-, and
    hard-task success, and customer-support success. Shared competition ranks
    are used: equal values receive the same rank, and subsequent ranks advance
    by the number of tied systems. Endpoint labels are offset only for
    legibility; data points remain at their shared ranks. No system dominates
    every axis. Gemini 3.8 Flash leads five of the six metrics but falls to third on
    customer support, where Muse Glimmer 30B---twelfth on scored success---
    ranks first and Claude Opus 5 (second overall) places second; Claude Sonnet 5
    climbs from joint-fifth to second on hard tasks.
    Ranks are computed from the same one-trial values used in
    Figure~\ref{fig:completed-model-results} and
    Table~\ref{tab:difficulty-breakdown}. å{Cost per success is omitted here
    because three open-weight runs are unpriced; it is shown on its own log axis
    in Figure~\ref{fig:same-headline-dotplot}.}}
    \label{fig:leaderboard-reshuffle}
\end{figure}

Figure~\ref{fig:leaderboard-reshuffle} converts the same performance axes into
per-metric ranks to make those ordering changes explicit. {The ranking view complements overall scored success with
metric-specific system comparisons.}

\subsection{Response and Execution Checks on a Matched Task Set}
\label{app:matched-response-strict}
\suppressfloats[t]
Figure~\ref{fig:grader-catch-waterfall} uses the same
402 response-declared task identities for all four systems.
The catalog defines eligibility, and each stage retains executions satisfying
all checks added so far. The action stage uses each trace's recorded expected
and matched call multisets: repeated required invocations must each be matched.
Undeclared programmatic checks remain
neutral. The strict stage adds the remaining
declared negative checks and the grader's error gate. The computation uses the saved
original round-1 semantic verdicts.
Opus's missing response verdict occurs on a hallucinated-tool failure; it remains
in the 402-task set as a separately recorded missing verdict.

Table~\ref{tab:matched-response-strict} extends
the matched contingency to all 16 original/formal systems in
Table~\ref{tab:completed-model-results}: 8,880 executions, with the same
402 response-declared and 153 other task identities per system.
On the 6,432 response-declared executions, 6,415 have a saved
original judgment. The 17 missing judgments comprise 1 Opus,
6 Nemotron Super, and 10 Nemotron Lightning executions,
all retained as $R^?/S^-$. There are 27 hallucinated-tool model
faults in total: 17 on the response-declared set and
10 among the other tasks. They remain strict failures;
there are no infrastructure-unscored exclusions.

\begin{table}[H]
\centering
\small
\setlength{\tabcolsep}{3pt}
\begin{fitlinewidth}
\begin{tabular}{lrrrrrrr}
\toprule
System & $R^+/S^+$ & $R^+/S^-$ & $R^-/S^+$ & $R^-/S^-$ & $R^?/S^-$ & \shortstack{Other 153\\$S^-$} & \shortstack{Full 555\\gap $G$} \\
\midrule
Gemini 3.8 Flash & 329 & 42 & 0 & 31 & 0 & 18 & 60 \\
Claude Opus 5 & 325 & 38 & 0 & 38 & 1 & 19 & 58 \\
Claude Sonnet 4.6 & 331 & 31 & 0 & 40 & 0 & 29 & 60 \\
GPT-5.6 Terra & 317 & 39 & 0 & 46 & 0 & 17 & 56 \\
Claude Sonnet 5 & 333 & 32 & 0 & 37 & 0 & 35 & 67 \\
GPT-5.5 & 320 & 36 & 0 & 46 & 0 & 22 & 58 \\
GPT-5.6 Sol & 319 & 38 & 0 & 45 & 0 & 21 & 59 \\
GPT-5.6 Luna & 314 & 40 & 0 & 48 & 0 & 19 & 59 \\
Inkling Small NVFP4 & 321 & 32 & 0 & 49 & 0 & 32 & 64 \\
Gemini 3.1 Pro Preview & 314 & 48 & 0 & 40 & 0 & 28 & 76 \\
GPT-5.4 Nano & 322 & 47 & 0 & 33 & 0 & 36 & 83 \\
Muse Glimmer 30B & 318 & 27 & 0 & 57 & 0 & 34 & 61 \\
Claude Haiku 4.5 & 298 & 46 & 0 & 58 & 0 & 29 & 75 \\
Nemotron Super 3 120B & 258 & 52 & 0 & 86 & 6 & 47 & 105 \\
Nemotron 3.5 Lightning 30B & 260 & 43 & 0 & 89 & 10 & 56 & 109 \\
GPT-4.1 Nano & 263 & 59 & 0 & 80 & 0 & 59 & 118 \\
\bottomrule
\end{tabular}
\end{fitlinewidth}
\caption{Original/formal response--strict contingency for all 16 Table~\ref{tab:completed-model-results} systems. The response-declared denominator is 402 per system; $R^+$/$R^-$ are observed saved semantic passes/failures and $R^?$ is missing, while $S$ is full strict success. The first five numerical columns sum to 402; $R^?/S^+=0$ for every system. Other 153 reports strict failures outside the response-declared set. $G$ is the original full-555 response-only-minus-strict count, using the neutral rule $R\ne\mathrm{False}$ for undeclared or missing response verdicts. Thus $G=R^+/S^-+R^?/S^-+\mathrm{Other}_{153}(S^-)$ here. Missing judgments are retained without imputation. The zero $R^-/S^+$ follows the strict conjunction.}
\label{tab:matched-response-strict}
\end{table}

For the four systems illustrated in
Figure~\ref{fig:grader-catch-waterfall}, the original full-555 excesses
remain: Gemini 3.8 Flash $60=42+18$, Claude Opus 5 $58=38+19+1$,
GPT-5.6 Sol $59=38+21$, and Muse Glimmer 30B $61=27+34$.
Figure~\ref{fig:grader-catch-waterfall} and
the curve view below continue to show these four systems on the common
402-task population.

Across all 16 systems, 650 executions have an
observed response pass and strict failure (27--59 per system).
Their overlapping failed checks comprise 508 missing or
under-repeated required operations, 399 argument failures,
141 order failures, 95 forbidden-call/argument failures,
44 expected-state failures, and 3 protected-state
failures. These counts are not mutually exclusive; every such execution
fails at least one programmatic component. Required-operation failures
account for 66.1--87.0\% of each system's matched gap.
The original full-suite gap ranges from 56 to 118 tasks and
sums to $1168=650+17+501$: observed matched gaps,
missing response verdicts, and failures among the other 153 tasks.
The other-task contribution is 30.0--55.7\% of $G$
per system.

\begin{figure}[H]
\centering
\includegraphics[width=0.95\linewidth]{\preprintpath 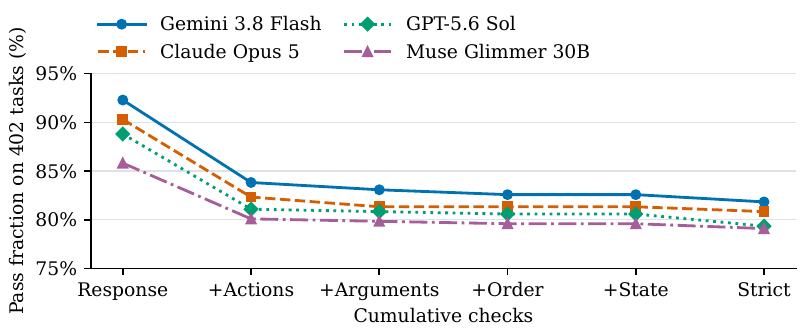}
\caption{Curve view of the same 402-task cumulative checks
shown in Figure~\ref{fig:grader-catch-waterfall}. Each curve follows one system
as response, required-action, argument, order, state, and remaining strict
checks are conjoined. The denominator and recorded verdicts match the
main figure.}
\label{fig:matched-response-strict-curves}
\end{figure}

\subsection{Policy-Constrained Tasks: Compliance versus Completion}
\label{app:safety-split}

This section gives the per-model numbers behind
Figure~\ref{fig:safety-split}. The canonical set contains 390 tasks without an
explicit negative assertion and 165 policy-constrained tasks declaring at least
one forbidden-action, forbidden-argument, or protected-state check. Because
these declarations are fixed by the contracts, the partition is identical
across all sixteen runs. For compactness, the table retains \emph{positive} and
\emph{safety} as column labels for the two subsets. Both columns report strict
success on the full conjunctive contract. {{Safety-check compliance} is
narrower: it reports the share of policy-constrained tasks on which all
applicable negative checks held, regardless of the remaining contract
components.}

\begin{table}[htbp]
    \centering
    \footnotesize
    \setlength{\tabcolsep}{2.5pt}
    \begin{fitlinewidth}
    \begin{tabular}{lrrrr}
        \toprule
        & Positive & Safety & & Safety-check\\
        Model & success (390) & success (165) & $\Delta$ & compliance (165)\\
        \midrule
        \modellogo{gemini}Gemini 3.8 Flash & 88.2\% (344/390) & 72.7\% (120/165) & +15.5 & 92.1\% (152/165)\\
        \modellogo{claude}Claude Opus 5 & 86.7\% (338/390) & 73.3\% (121/165) & +13.3 & 93.3\% (154/165)\\
        \modellogo{claude}Claude Sonnet 4.6 & 84.6\% (330/390) & 75.8\% (125/165) & +8.9 & 92.7\% (153/165)\\
        \modellogo{openai}GPT-5.6 Terra & 86.7\% (338/390) & 69.7\% (115/165) & +17.0 & 93.3\% (154/165)\\
        \modellogo{claude}Claude Sonnet 5 & 86.4\% (337/390) & 69.1\% (114/165) & +17.3 & 92.1\% (152/165)\\
        \modellogo{openai}GPT-5.5 & 86.9\% (339/390) & 67.9\% (112/165) & +19.0 & 92.1\% (152/165)\\
        \modellogo{openai}GPT-5.6 Sol & 87.2\% (340/390) & 67.3\% (111/165) & +19.9 & 89.7\% (148/165)\\
        \modellogo{openai}GPT-5.6 Luna & 85.4\% (333/390) & 69.7\% (115/165) & +15.7 & 91.5\% (151/165)\\
        \modellogo{inkling}Inkling Small NVFP4 & 82.8\% (323/390) & 72.1\% (119/165) & +10.7 & 92.7\% (153/165)\\
        \modellogo{gemini}Gemini 3.1 Pro Preview & 83.3\% (325/390) & 69.1\% (114/165) & +14.2 & 89.7\% (148/165)\\
        \modellogo{openai}GPT-5.4 Nano & 84.9\% (331/390) & 65.5\% (108/165) & +19.4 & 88.5\% (146/165)\\
        \modellogo{muse}Muse Glimmer 30B & 82.6\% (322/390) & 69.7\% (115/165) & +12.9 & 93.9\% (155/165)\\
        \modellogo{claude}Claude Haiku 4.5 & 81.3\% (317/390) & 63.6\% (105/165) & +17.6 & 92.7\% (153/165)\\
        \modellogo{nemotron}Nemotron Super 3 120B & 68.2\% (266/390) & 59.4\% (98/165) & +8.8 & 95.2\% (157/165)\\
        \modellogo{nemotron}Nemotron 3.5 Lightning 30B & 68.7\% (268/390) & 53.9\% (89/165) & +14.8 & 84.8\% (140/165)\\
        \modellogo{openai}GPT-4.1 Nano & 70.3\% (274/390) & 50.3\% (83/165) & +20.0 & 84.8\% (140/165)\\
        \midrule
        \emph{Mean} & 82.1\% & 66.8\% & +15.3 & 91.2\%\\
        \bottomrule
    \end{tabular}
    \end{fitlinewidth}
    \caption{Strict success partitioned by contract structure on the 555-task
    canonical set. ``Positive'' denotes the 390 tasks without explicit negative
    assertions; ``Safety'' denotes the 165 policy-constrained tasks with at
    least one such assertion. Safety-check compliance is the share of the latter
    on which all applicable negative checks held. Every model has lower strict
    success on the policy-constrained subset ($\Delta=+8.8$ to $+20.0$ points),
    while negative-check compliance is 84.8--95.2\%.
    {Rows follow Table~\ref{tab:completed-model-results}.}}
    \label{tab:safety-split}
\end{table}

{Section~\ref{subsec:policy} reports the aggregate compliance and
completion results. Nemotron Super 3 120B attains the highest observed
negative-check compliance (95.2\%) while ranking 14th overall.
Compliance and workflow
completion are distinct measured axes under the fixed scaffold. These rates
do not isolate intrinsic model safety or establish a causal safety-and-utility
trade-off.}

\section{Three-Run Reliability and Ranking Stability}
\label{app:repeated-runs}

\subsection{Repeated Evaluation and Aggregation}

The first-round leaderboard in Table~\ref{tab:completed-model-results} places
Gemini 3.8 Flash ahead of Claude Opus 5 by only five of 555 tasks
(0.90 percentage points). To assess whether this ordering persists across
executions, we report three evaluation rounds: the formal first-round results
and two additional runs.
{All three rounds are compared on the same 555 task contracts using the
common strict-success definition.} Strict success requires every applicable action,
argument, state, semantic-response, ordering, and negative check to pass.
Model-attributable errors count as failures; unscored slots
follow the benchmark's exclusion convention. Each round starts fresh task
environments. The first round is the official result used in the main
leaderboard.

For model $m$, let $a_{m,r}$ denote its strict-success rate over scored tasks
in round $r$. We give the three rounds equal weight and report their mean
and sample standard deviation:
\begin{equation}
    \bar a_m=\frac{1}{3}\sum_{r=1}^{3}a_{m,r},
    \qquad
    s_m=\sqrt{\frac{1}{2}\sum_{r=1}^{3}(a_{m,r}-\bar a_m)^2}.
\end{equation}
The standard deviation describes variation across the three executions and
is reported in percentage points. This average is computed from round-level
rates, rather than by pooling attempts across rounds. We also report
$\widehat{\mathrm{pass}}^3$, the fraction of tasks successful in all three
executions among tasks scored in all three, following the all-success
reliability convention of $\tau$-bench~\citep{yao2025taubench}.

\begin{table}[htbp]
    \centering
    \footnotesize
    \setlength{\tabcolsep}{4pt}
    \begin{fitlinewidth}
    \begin{tabular}{lrrrrr}
        \toprule
        Model & Round 1 & Round 2 & Round 3 & Mean $\pm$ SD & $\widehat{\mathrm{pass}}^3$\\
        \midrule
        \modellogo{gemini}Gemini 3.8 Flash & 464/555 & 480/555 & 480/555 & $85.53\pm1.66$ & 78.20\\
        \modellogo{claude}Claude Opus 5 & 459/555 & 462/555 & 465/554 & $83.29\pm0.62$ & 73.47\\
        \modellogo{openai}GPT-5.5 & 451/555 & 462/555 & 465/555 & $82.76\pm1.33$ & 74.77\\
        \modellogo{claude}Claude Sonnet 4.6 & 455/555 & 460/555 & 460/555 & $82.58\pm0.52$ & 75.14\\
        \modellogo{openai}GPT-5.6 Terra & 453/555 & 447/555 & 462/555 & $81.80\pm1.36$ & 72.43\\
        \modellogo{openai}GPT-5.6 Sol & 451/555 & 457/555 & 451/554 & $81.67\pm0.59$ & 73.65\\
        \modellogo{gemini}Gemini 3.1 Pro Preview & 439/555 & 449/555 & 461/554 & $81.07\pm2.06$ & 73.83\\
        \modellogo{claude}Claude Sonnet 5 & 451/555 & 445/555 & 441/555 & $80.30\pm0.91$ & 70.45\\
        \modellogo{openai}GPT-5.6 Luna & 448/555 & 444/555 & 442/555 & $80.12\pm0.55$ & 70.63\\
        \modellogo{openai}{GPT-5.4 Nano} & {439/555} & {435/555} & {451/555} & {$79.58\pm1.50$} & {67.57}\\
        \modellogo{inkling}Inkling Small NVFP4 & 442/555 & 436/555 & 433/555 & $78.74\pm0.83$ & 68.47\\
        \modellogo{muse}Muse Glimmer 30B & 437/555 & 431/555 & 441/555 & $78.62\pm0.91$ & 67.93\\
        \modellogo{claude}Claude Haiku 4.5 & 422/555 & 432/555 & 428/555 & $77.00\pm0.91$ & 68.65\\
        \modellogo{nemotron}{Nemotron Super 3 120B} & {364/555} & {382/555} & {375/555} & {$67.33\pm1.63$} & {50.99}\\
        \modellogo{nemotron}Nemotron 3.5 Lightning 30B & 357/555 & 363/555 & 363/555 & $65.05\pm0.62$ & 54.95\\
        \modellogo{openai}{GPT-4.1 Nano} & {357/555} & {346/555} & {348/555} & {$63.12\pm1.06$} & {52.43}\\
        \bottomrule
    \end{tabular}
    \end{fitlinewidth}
    \caption{Three directly comparable evaluation rounds under the common
    scoring standard. Counts are successes/scored tasks; mean rates and
    $\widehat{\mathrm{pass}}^3$ are percentages, and SD is in percentage points
    ($n=3$, sample SD). Round 1 is the formal first-round result.
    {All 16 systems are ordered by three-round mean. Each round schedules
    555 tasks.}}
    \label{tab:repeated-run-results}
\end{table}

\subsection{Aggregate and Ranking Stability}

{Across all 16 systems from the original leaderboard}, Gemini 3.8 Flash
ranks first in every round and by the three-round mean
($85.53\%$, SD $1.66$ points). Claude Opus 5 remains second in each round,
including a tie with GPT-5.5 in round 2, and averages $83.29\%$
(SD $0.62$ points). The Gemini--Opus gaps are 0.90, 3.24, and 2.55 points
across the three rounds; their mean gap is 2.23 points. Thus, the original
five-task lead is preserved in both additional runs, and the three-run
summary supports the observed ordering of these two systems.

The broader ordering is also consistent: pairwise Spearman rank correlations
{among all 16 systems are 0.922 for rounds 1--2, 0.869 for rounds 1--3,
and 0.934 for rounds 2--3}, using average ranks for ties. Sample SDs range
from 0.52 to 2.06 percentage points. Exact positions among nearby systems
nevertheless move: GPT-5.5 rises from joint fifth in round 1 to joint second
and third, while Sonnet 4.6 moves from third to fourth and sixth. Sol leads
Terra in round 2 but falls behind it in round 3; Inkling and Muse also reverse
order between those rounds. These changes support reporting multi-run means
alongside individual-round results when models have similar scores.

\subsection{Consistency on Individual Tasks}

Stable aggregate scores do not imply that the same tasks succeed each time.
Gemini 3.8 Flash passes 480 tasks in both rounds 2 and 3, but 30 task outcomes
switch between those executions. Across all three rounds, 434/555 tasks
succeed every time ($\widehat{\mathrm{pass}}^3=78.20\%$), compared with its
$85.53\%$ mean strict-success rate. For Opus, 407/554 jointly scored tasks
pass all three rounds ($73.47\%$). The all-three-success metric therefore
complements the mean by measuring sustained task-level reliability.
{Taken together, the three-run results show a persistent lead across
all sixteen leaderboard systems} and broadly consistent rankings, with residual
variation among close systems and on individual tasks.

\subsection{Original Ties and Paired Uncertainty}
\label{app:original-paired-uncertainty}

In the original first-round cohort, GPT-5.5, GPT-5.6 Sol, and Claude
Sonnet 5 each pass 451/555 tasks. They share 34 failures, while 181 tasks
fail for at least one of the three. GPT-5.5 and Sol share 65 failures over
a union of 143 (failure-set Jaccard 45.5\%); Sonnet shares 50 failures
over a union of 158 with each (31.6\%). Equal aggregate scores therefore
coexist with different failure sets.

For the original Flash--Opus comparison, 50 tasks pass only for
Flash and 45 only for Opus. The five-task difference is 0.90 percentage
points, with a pointwise 95\% paired-task bootstrap interval
of $-2.52$ to $+4.32$ points (50,000 resamples of the 555 matched
tasks with replacement; random seed 2026). The exact two-sided McNemar
test gives $p=0.682$. These results use the original round-1 judgments.

\clearpage
\section{User-Simulator Configuration Sensitivity}
\label{app:simulator-sensitivity}
\suppressfloats[t]

\paragraph{Study objective.}
Across six agent systems, switching from GPT-6 Luna to Claude Haiku 5.5
changes mean strict success by $-0.12$ to $+1.98$ percentage points and
changes 6.61--13.51\% of paired task-round verdicts. We measure this
configuration sensitivity together with agent-token workload.

\paragraph{Paired design and recorded configurations.}
All six agents evaluate the same 555 tasks across nine domains in three
rounds under each simulator: 12 conditions and 19,980 final scored
executions (with 3,330 Gemini/Mini executions retained as a separate
supplement). Within each agent pair, task contracts, initial states, agent
settings, reference scaffold, generation budgets, and the fixed GPT-5-mini
communication judge are held constant across both simulator conditions.
Simulators use the structured user-turn protocol, a 1,086-token generation
budget, and a 4,000-character message limit.

\paragraph{Aggregate strict success.}
Claude Haiku 4.5 has the largest observed increase, from 76.64\% to
78.62\%; GPT-5.6 Sol changes from 83.90\% to 83.78\%
(Table~\ref{tab:simulator-sensitivity}). All six pointwise intervals span
zero. Flash ranks first under both configurations, while Sonnet and
Gemini Pro exchange the fourth and fifth positions in the descriptive
three-round ranking.

\begin{table}[t]
\centering
\small
\setlength{\tabcolsep}{4pt}
\begin{fitlinewidth}
\begin{tabular}{lrrr}
\toprule
Agent & \shortstack{Luna\\mean $\pm$ SD (\%)} &
\shortstack{Haiku 5.5\\mean $\pm$ SD (\%)} &
\shortstack{$\Delta$ (pp)\\pointwise 95\% interval} \\
\midrule
GPT-5.5 & $84.14\pm0.72$ & $84.32\pm0.18$ & $+0.18\;[-1.50,1.86]$ \\
GPT-5.6 Sol & $83.90\pm0.68$ & $83.78\pm1.26$ & $-0.12\;[-1.74,1.50]$ \\
Claude Sonnet 5 & $81.56\pm1.20$ & $83.18\pm1.27$ & $+1.62\;[-0.60,3.84]$ \\
Claude Haiku 4.5 & $76.64\pm1.53$ & $78.62\pm0.63$ & $+1.98\;[-0.18,4.26]$ \\
Gemini 3.8 Flash & $86.97\pm0.28$ & $87.09\pm0.45$ & $+0.12\;[-1.26,1.50]$ \\
Gemini 3.1 Pro Preview & $82.46\pm0.21$ & $82.88\pm0.62$ & $+0.42\;[-1.20,2.10]$ \\
\bottomrule
\end{tabular}
\end{fitlinewidth}
\caption{Strict success in the final recovered exports. Luna denotes
GPT-6 Luna throughout. Each mean and sample SD summarizes three rounds of
555 tasks; $\Delta$ is Haiku 5.5 minus Luna. Pointwise percentile intervals
use 20,000 domain-stratified task-cluster bootstrap resamples, retaining
both configurations and all three rounds per sampled task. They describe
task-resampling variation conditional on these observed rounds and
configurations.}
\label{tab:simulator-sensitivity}
\end{table}

\paragraph{Task-level turnover and repetition.}
Across configurations, 110--225 of 1,665 paired task-round verdicts change
per agent (6.61--13.51\%). Comparing all three round pairs within each
configuration gives 82--160 disagreeing pairs out of 1,665
(4.92--9.61\%). Table~\ref{tab:simulator-sensitivity-counts} places each
agent's configuration contrast alongside its within-configuration
repetition counts. These are descriptive paired rates.

\paragraph{Agent-token workload.}
Total reported agent tokens change by $-14.2\%$ to $+37.4\%$ from Luna
to Haiku 5.5. Flash changes from 16.85M to 23.16M tokens; Sonnet changes
from 49.21M to 42.22M. Tokens sum prompt plus completion usage across
agent LLM calls; reasoning is already included in completion, and cache
reads are not added again. The totals cover retained executions and
exclude replaced recovery attempts.

\begin{table}[t]
\centering
\small
\setlength{\tabcolsep}{3.2pt}
\begin{fitlinewidth}
\begin{tabular}{lrrrr}
\toprule
Agent & \shortstack{Across configurations\\changes (\%)} &
\shortstack{Within Luna\\changes (\%)} &
\shortstack{Within Haiku 5.5\\changes (\%)} &
\shortstack{Agent tokens (M)\\Luna $\to$ Haiku 5.5} \\
\midrule
GPT-5.5 & 151 (9.07) & 120 (7.21) & 132 (7.93) & $18.60\to22.40$ \\
GPT-5.6 Sol & 168 (10.09) & 158 (9.49) & 140 (8.41) & $23.41\to27.67$ \\
Claude Sonnet 5 & 215 (12.91) & 158 (9.49) & 114 (6.85) & $49.21\to42.22$ \\
Claude Haiku 4.5 & 225 (13.51) & 160 (9.61) & 160 (9.61) & $37.96\to38.20$ \\
Gemini 3.8 Flash & 110 (6.61) & 82 (4.92) & 96 (5.77) & $16.85\to23.16$ \\
Gemini 3.1 Pro Preview & 141 (8.47) & 104 (6.25) & 130 (7.81) & $26.95\to33.71$ \\
\bottomrule
\end{tabular}
\end{fitlinewidth}
\caption{Outcome turnover and agent-token workload. Across-configuration
comparisons pair the same task and framework round. Within-configuration
comparisons pair R1--R2, R1--R3, and R2--R3, with distinct derived framework
seeds. Each outcome column contains 1,665 comparisons (555 tasks $\times$
three pairs), reported as count (percentage). Tokens total prompt plus
completion usage over each configuration's 1,665 final executions.}
\label{tab:simulator-sensitivity-counts}
\end{table}

\paragraph{Recovery and provenance.}
Transient provider rate-limit (HTTP 429) and service errors encountered
during the 19,980-execution campaign were re-executed under the standard
harness protocol (Table~\ref{tab:simulator-recovery}), so that all 1,665
task-round slots per agent--simulator pair yielded valid scored
trajectories.

\begin{table}[t]
\centering
\small
\setlength{\tabcolsep}{5pt}
\begin{fitlinewidth}
\begin{tabular}{ll}
\toprule
Campaign & Recovery record \\
\midrule
GPT/Luna & 5 grader-error slots, including 1 transcript rejudge \\
Sonnet, both simulators & 13 slots; 14 prior infrastructure attempts \\
Haiku 4.5, both simulators & 7 grader-error slots: 5 Luna, 2 Haiku \\
GPT/Haiku & 149 slots: 139 simulator timeouts, 10 grader errors \\
Gemini/Luna & 399 slots: 285 Pro, 114 Flash; 258 then 141 replacements \\
Gemini/Haiku & 8 slots: 7 grader errors, 1 simulator timeout \\
\midrule
Gemini/Mini (supplement) & 14 replacements, including 2 Pro agent timeouts \\
\bottomrule
\end{tabular}
\end{fitlinewidth}
\caption{Recovery accounting for the final scored exports. GPT
campaigns contain GPT-5.5 and GPT-5.6 Sol; Gemini campaigns contain Flash
and Pro. Sonnet and Haiku 4.5 rows each combine both simulator conditions.
Interaction reruns regenerate conversations; GPT/Luna also includes one
transcript-only rejudge.}
\label{tab:simulator-recovery}
\end{table}

\clearpage
\section{Qualitative Trace Evidence and Provenance}
\label{app:qualitative-traces}

\begin{tcolorbox}[
    enhanced,
    colback=demoInfo!5,
    colframe=demoInfo!75!black,
    colbacktitle=demoInfo!14,
    coltitle=black,
    title={What is a ``demo'' in this appendix?},
    fonttitle=\bfseries,
    boxrule=0.6pt,
    leftrule=3pt,
    arc=1.5mm]
A \textbf{demo} is a compact, trace-grounded view of one benchmark trial.
{It connects four layers that aggregate scores hide: the task's executable
contract, the route and tool observations produced at runtime, the visible
assistant behavior, and the task-declared grader vector.}
\end{tcolorbox}

\begin{figure}[h]
    \centering
    \begin{tikzpicture}[
        node distance=0.16in,
        demostep/.style={rounded corners=2pt, minimum height=0.66in,
          align=center, inner sep=4pt, font=\scriptsize, line width=0.55pt},
        demoflow/.style={-{Latex[length=2.1mm]}, line width=0.7pt,
          draw=demoInfo!75!black}
    ]
        \node[demostep, draw=demoInfo, fill=demoInfo!7, text width=1.02in]
          (contract) {\textbf{Task contract}\\intent, checks, protected state};
        \node[demostep, draw=demoSafe, fill=demoSafe!7, text width=1.12in,
          right=of contract] (runtime) {\textbf{Runtime trace}\\dialogue, route, tools, state};
        \node[demostep, draw=demoTimeout, fill=demoTimeout!8, text width=1.15in,
          right=of runtime] (graders) {\textbf{Grader vector}\\action, response, state, negatives};
        \node[demostep, draw=demoSuccess, fill=demoSuccess!7, text width=1.03in,
          right=of graders] (outcome) {\textbf{Outcome}\\
          \textcolor{demoSuccess!85!black}{success} /
          \textcolor{demoFailure!85!black}{failure}\\
          \textcolor{demoTimeout!85!black}{timeout}};
        \draw[demoflow] (contract) -- (runtime);
        \draw[demoflow] (runtime) -- (graders);
        \draw[demoflow] (graders) -- (outcome);
    \end{tikzpicture}
    \caption{How to read a trace demo. Each card links the declared contract to
    runtime evidence and the individual grader decisions that produce the
    reported outcome.}
    \label{fig:demo-anatomy}
\end{figure}
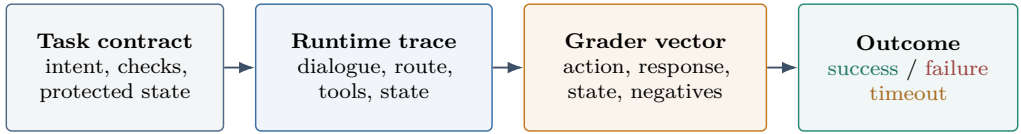

{This appendix presents three diagnostic single-run traces covering strict
success, a graded failure, and a timeout. The cases explain execution mechanisms;
failure-mode frequencies are reported in the aggregate diagnostics over the 555
tasks. Case~C uses an earlier
\texttt{grok-4.6} run with a 600-second task deadline, outside the 16-system
leaderboard; the formal leaderboard protocol specifies 1{,}200 seconds
(Appendix~\ref{app:frozen-protocol}), and its release contains no timeouts.
Figure~\ref{fig:demo-anatomy} defines the four-layer reading order, and
Table~\ref{tab:qualitative-case-summary} summarizes the selected cases and their
diagnostic roles. Task and tool identifiers are reproduced from the benchmark.
Customer names, account identifiers, and transaction records are synthetic;
transaction identifiers are generated by the resettable benchmark environment.
Excerpts condense repeated turns and preserve the recorded tool arguments,
results, and grader outcomes.}

\begin{table}[h]
    \centering
    \small
    \setlength{\tabcolsep}{4pt}
    \begin{fitlinewidth}
    \begin{tabular}{llllp{0.30\linewidth}}
        \toprule
        Case & Model & Domain & Outcome & Diagnostic value\\
        \midrule
        \rowcolor{demoSuccess!7}
        A & \modellogo{gemini}Gemini 3.1 Pro Preview & Payments & \demotag{demoSuccess}{SUCCESS} & Recovery after a business-rule rejection\\
        \rowcolor{demoFailure!6}
        B & \modellogo{claude}Claude Sonnet 4.6 & Payments & \demotag{demoFailure}{FAILURE} & Duplicate side effect detected only in final state\\
        \rowcolor{demoTimeout!7}
        C & \modellogo{grok}Grok 4.6 & Retail & \demotag{demoTimeout}{TIMEOUT} & On-policy interaction fails to converge\\
        \bottomrule
    \end{tabular}
    \end{fitlinewidth}
    \caption{{Selected single-run traces and their diagnostic roles.}}
    \label{tab:qualitative-case-summary}
\end{table}

\subsection{Case A: Recovery after a Structured Business Rejection}
\label{app:case-success}

\begin{demoatglance}{demoSuccess}{Demo A at a glance}
\demotag{demoSuccess}{STRICT SUCCESS}\hfill
\textbf{Model:} Gemini 3.1 Pro Preview\\[2pt]
\textbf{Trace scale:} 2 turns, 4 tool calls, 81.27 s\\[2pt]
\textbf{What to watch:} the first amount is rejected by a business rule; the
agent revises the plan without executing the forbidden \$1,000 transfer.
\end{demoatglance}

\paragraph{Source and contract.}
This trace comes from \texttt{gemini-3.1-pro-preview}, seed 997688223, task
\path{payments_transfer_insufficient_then_retry_01}. The simulated user first
requests a transfer of \$1,000 from \texttt{ACC-1002} to
\texttt{ACC-1001}; after an insufficient-funds response, the user corrects the
amount to \$50. The contract requires two calls to
\texttt{validate\_transfer}, followed by \texttt{execute\_transfer} for
\$50. Executing the rejected \$1,000 amount is forbidden. The final balances
must be \$130.00 and \$2,500.75, respectively.

\paragraph{Observed trajectory.}
The two-turn interaction took 81.27 seconds and remained in the
\path{transfer__transfer} node. On the first turn, the agent called
\texttt{get\_account\_balance} and observed a balance of \$180. It then called
\texttt{validate\_transfer(amount=1000)}, which returned
\texttt{valid=false}, reason \texttt{Insufficient funds}, and status 422. The
assistant accurately told the user that the transfer could not be completed.
After the user replied, ``Oh, sorry, I meant \$50,'' the agent called
\texttt{validate\_transfer(amount=50)} and received
\texttt{valid=true}; it then executed the \$50 transfer and returned transfer
ID \texttt{XFER-HD10VI}.

\paragraph{Grading outcome.}
Expected actions, arguments, order, final state, and the forbidden-argument
check all passed, giving strict task success. This case illustrates why a
structured rejection is not itself an agent failure: the benchmark checks
whether the system interprets the business response, revises the plan, and
executes only the valid alternative.

\subsection{Case B: Plausible Communication, Duplicate Execution}
\label{app:case-failure}

\begin{demoatglance}{demoFailure}{Demo B at a glance}
\demotag{demoFailure}{STRICT FAILURE}\hfill
\textbf{Model:} Claude Sonnet 4.6\\[2pt]
\textbf{Trace scale:} 4 turns, 10 tool calls, 161.84 s\\[2pt]
\textbf{What to watch:} action, order, and response checks pass, but a repeated
checkout doubles the loyalty balance and is caught by the state grader.
\end{demoatglance}

\paragraph{Source and contract.}
This trace comes from \texttt{claude-sonnet-4-6}, seed 4275343667, task
\path{payments_checkout_full_chain_02}. The requested \$15.50 PayPal checkout
requires, in order, \texttt{process\_payment}, \texttt{send\_receipt},
\texttt{award\_loyalty\_points}, and \texttt{log\_transaction}. The assistant
must report 155 earned points, and the final loyalty balance for
\texttt{ACC-1002} must equal 155.

\paragraph{Observed trajectory.}
On the first agent turn, all four required tools succeeded. The transaction ID
was \texttt{TXN-WLV0C0}, and the loyalty tool returned 155 points with a balance
of 155. The assistant reported, ``You earned 155 loyalty points from this
\$15.50 transaction.'' When the on-policy user pressed for timestamps and
additional identifiers not returned by the tools, the agent temporarily claimed
that its earlier tool-backed result had been fabricated. On a later turn it ran
the entire checkout sequence a second time, creating transaction
\texttt{TXN-CS17PL} and awarding another 155 points. The final loyalty balance
was therefore 310.

\paragraph{Grading outcome.}
The action, argument, order, and semantic-response checks passed. The semantic
judge cited the exact 155-point statement and returned
\texttt{contradiction\_found=false}, even though an intermediate assistant
message denied the already recorded tool execution. The programmatic state
check failed with \texttt{expected 155, got 310}, so the strict task outcome was
failure. This is a concrete case in which response-only or action-existence
grading would overestimate correctness, while final state exposes a duplicated
business transaction. Because the assistant replies reported the 155 points
earned on this transaction, the communication judge accepted the response,
whereas the programmatic state check deterministically caught the duplicated
loyalty balance (310 instead of 155), demonstrating why programmatic state
verification is essential alongside communication grading.

\subsection{Case C: Timeout from an On-Policy Retry Loop}
\label{app:case-timeout}

\begin{demoatglance}{demoTimeout}{Demo C at a glance}
\demotag{demoTimeout}{TIMEOUT}\hfill
\textbf{Model:} Grok 4.6\\[2pt]
\textbf{Trace scale:} 19 turns, 19 tool calls, 600.00 s\\[2pt]
{\textbf{Protocol:} earlier 600-second task deadline}\\[2pt]
\textbf{What to watch:} an unnecessary authentication route, simulator
credential drift, and repeated rejected calls combine into non-convergence.
\end{demoatglance}

\paragraph{Source and contract.}
This trace comes from \texttt{grok-4.6}, seed 981070998, task
\path{commerce_returns_01}. The easy task asks only for the status of order
\texttt{ORD-9001}; the declared action is \texttt{get\_order\_status}, and the
response should state that the order was delivered. The task instruction gives
the customer ID but does not provide a PIN.

\paragraph{Observed trajectory.}
The agent asked for a PIN before performing the lookup. The user simulator then
introduced PIN \texttt{8274}, which is absent from the task specification. The
agent called \texttt{authenticate\_customer} 18 times with that PIN; each call
returned a structured 403 authentication failure. During the loop it also
called \texttt{get\_order\_status} once with the correct order ID and obtained
\texttt{status=Delivered}, so the declared action and argument checks had
matching evidence. Nevertheless, the visible reply continued to request the
same credentials rather than report the retrieved status. The interaction
reached 19 turns and terminated at 600.00 seconds with
\texttt{TimeoutError: task exceeded 600s during agent.run}.

\paragraph{Grading outcome.}
{The source record reports a timeout. The
observed non-convergence involved an unnecessary authentication route, simulator
credential drift, repeated rejected calls, and the agent's failure to report a
tool result it had already obtained.}

\subsection{Trace Artifact Mapping and Provenance}
\label{app:trace-artifacts}
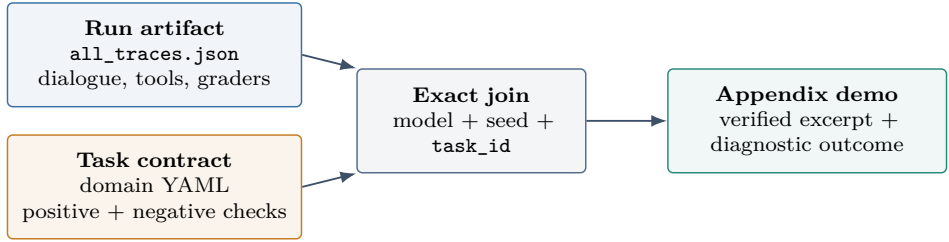
\begin{figure}[h]
    \centering
    \begin{tikzpicture}[
        node distance=0.14in and 0.42in,
        provenance/.style={rounded corners=2pt, minimum height=0.54in,
          align=center, inner sep=4pt, font=\scriptsize, line width=0.55pt},
        provenanceflow/.style={-{Latex[length=2.1mm]}, line width=0.7pt,
          draw=demoInfo!75!black}
    ]
        \node[provenance, draw=demoSafe, fill=demoSafe!7, text width=1.42in]
          (json) {\textbf{Run artifact}\\\texttt{all\_traces.json}\\dialogue, tools, graders};
        \node[provenance, draw=demoTimeout, fill=demoTimeout!8, text width=1.42in,
          below=of json] (yaml) {\textbf{Task contract}\\domain YAML\\positive + negative checks};
        \coordinate (source-mid) at ($(json)!0.5!(yaml)$);
        \node[provenance, draw=demoInfo, fill=demoInfo!7, text width=1.08in]
          (match) at ($(source-mid)+(1.65in,0)$)
          {\textbf{Exact join}\\model + seed +\\\texttt{task\_id}};
        \node[provenance, draw=demoSuccess, fill=demoSuccess!7, text width=1.34in,
          right=of match] (card) {\textbf{Appendix demo}\\verified excerpt +\\diagnostic outcome};
        \draw[provenanceflow] (json.east) -- (match.north west);
        \draw[provenanceflow] (yaml.east) -- (match.south west);
        \draw[provenanceflow] (match) -- (card);
    \end{tikzpicture}
    \caption{Trace-demo provenance. {Each narrative joins the complete run
    artifact to its task contract using the model, seed, and task ID.}}
    \label{fig:demo-provenance}
\end{figure}

The source records are drawn from the  relative run artifacts.
Each run artifact preserves the full per-record trace for its run. The task ID
identifies the exact record used above.
Figure~\ref{fig:demo-provenance} summarizes the exact artifact--contract join
used to reconstruct each demonstration.

\paragraph{Selection procedure.}
We first partitioned records by strict outcome: success, graded failure,
infrastructure timeout, and other infrastructure error. Within those partitions
we inspected the declared checks, number of turns, routes, tool-call sequence,
and failure lists. We then selected cases that (i) had a self-contained task
contract, (ii) exposed a different evaluator behavior, and (iii) collectively
spanned distinct evaluator outcomes across the illustrated models.

\paragraph{Locating the source record.}
For Cases A--C, load the corresponding JSON array and select the record whose
\texttt{task\_id} equals the identifier given in its ``Source and contract''
paragraph. The original task contract is the object with the same ID in the released
benchmark's domain YAML. This recovers the
instruction, all declared positive and negative checks, the full on-policy
dialogue, tool observations, and the final grader fields used in this appendix.

\section{Construct Validity and Benchmark Scope}
\label{app:significance}

{CRM-Agent-Bench evaluates task-declared execution requirements over
multi-step interactions with mutable state. The full-suite score measures
satisfaction of each task's executable contract; Table~\ref{tab:contract-inventory}
reports the coverage of its constituent checks. The benchmark combines dependent
tool execution, stateful business workflows, and on-policy interaction.}

\subsection{Long-Horizon, Dependency-Laden Execution}
Tasks are multi-turn by construction, and many are multi-step. An on-policy simulated user
reveals information incrementally across up to 30 turns, so the agent must
maintain state, resolve entities, and re-plan as new constraints arrive rather
than answer in a single shot. Correctness is frequently defined over an
\emph{ordered chain} of dependent tool calls: in the audited 555-task set, 235 tasks
declare at least two expected-action entries and 164 impose an explicit ordering
constraint, with canonical
chains such as \texttt{validate\_transfer}\,$\rightarrow$\allowbreak\,\texttt{execute\_transfer}
or \texttt{process\_payment}\,$\rightarrow$\allowbreak\,\texttt{send\_receipt}\,$\rightarrow$\allowbreak\,\texttt{award\_loyalty\_points}\,$\rightarrow$\allowbreak\,\texttt{log\_transaction}, and
create\,$\rightarrow$\allowbreak\,configure\,$\rightarrow$\allowbreak\,verify\,$\rightarrow$\allowbreak\,finalize
onboarding flows. Because success is the strict conjunction of every declared
check, a single early error---a wrong entity, a skipped validation, a premature
finalize---propagates and fails the whole task. {These requirements evaluate
sustained planning, cross-turn state tracking, and recovery from mid-trajectory
feedback.}

\subsection{{Stateful Business Workflows}}
{The workflows cover authenticated returns and refunds, funds transfers,
case triage and escalation, appointment scheduling, guarded record deletion,
account onboarding, and campaign creation across nine stateful business domains
and 36 capability tags. Actions mutate systems of record, encounter business-rule
rejections (for example authorization, conflict, and unprocessable-entity
responses), and can corrupt protected state belonging to other customers,
accounts, or records. The portable, license-clean environments support controlled
evaluation of these operations without a live CRM or vendor backend.}

\subsection{Probabilistic, On-Policy Dynamics}
{The environment's business rules are deterministic, while the surrounding
interaction varies across trials.} Each
trial runs against an on-policy user simulator whose next utterance is
conditioned on the agent's own prior replies, so a task defines a
\emph{distribution over dialogues} rather than a fixed transcript; the agent's
decoding, the per-turn routing classifier, and state-dependent business-rule
outcomes each inject variability that a robust system must absorb. {The
same task may be solved or missed depending on how a system handles credential
drift, a rejected call, or an ambiguous turn. The conjunctive contract exposes
these failures at the component level, and the}
$\mathrm{pass}^k$ reporting protocol is designed to quantify run-to-run
robustness directly. Appendix~\ref{app:repeated-runs} reports three-round
means, sample standard deviations, ranking stability, and all-three-success
rates, including cases where equal aggregate scores conceal different
task outcomes.

\section{Per-Domain System Flow Atlas}
\label{app:system-flows}

Each domain in CRM-Agent-Bench pairs a mutable environment with the
graph-routed reference scaffold of Section~\ref{sec:method}.
Figures~\ref{fig:flow-retail}--\ref{fig:flow-marketing} give a layered view of
one representative journey per domain. Reading top to bottom, a user request
descends through the \emph{interaction}, \emph{orchestration}, \emph{capability},
and \emph{environment} layers; policy gates fork to a no-mutation branch that
preserves protected state; and the \emph{evaluation} layer applies the strict
conjunctive contract over the recorded trace and final state. {The
evaluation layer maps its task-declared checks to the facets in
Section~\ref{sec:introduction}: action, argument, and order assess
\emph{trajectory}; response and state assess \emph{outcome}; and forbidden
checks assess \emph{policy compliance}. \emph{Operational viability} is
reported separately. All checks apply only when declared in the task contract;
Marketing tasks declare no negative assertions (N=0).}
{The diagrams are illustrative snapshots of the graph-routed reference
scaffold, each showing one multi-turn journey. The named capability node illustrates the
journey, and the ``Node-scoped tools invoked'' row lists the union of tools
used across turns as the classifier re-routes. On each turn, tool access is
scoped to the selected node. Table~\ref{tab:graph-inventory} reports the full
audited scaffold inventory used in the reported runs.}

The construction lineages summarized in
Table~\ref{tab:construction-provenance} connect each domain in the audited
evaluation set to its source workflow family. Retail and Customer Support
combine seed and recipe-derived
capabilities; Field Service, Sales, and Marketing descend from the original seed
program; and IT Service Management, Payments, Travel, and Onboarding are
recipe-derived. {The captions give the construction lineage and task count
for each domain in the audited 555-task evaluation set; the diagrams illustrate
how a workflow is realized as nodes, tools, policy gates, state transitions,
and contract observables.}

\newcommand{\flowcaptionkey}{\emph{Key:}
\texttt{R}/\texttt{W}/\texttt{C} denote read, write, and computation tools;
blue marks intermediate execution, green the accepted completion path, amber
a policy gate, and red rejection without business-state mutation.
{Evaluation checks apply only when declared in the task contract.}}

\begin{figure}[p]
    \centering
    \includegraphics[width=\textwidth,height=0.9\textheight,keepaspectratio]{\preprintpath 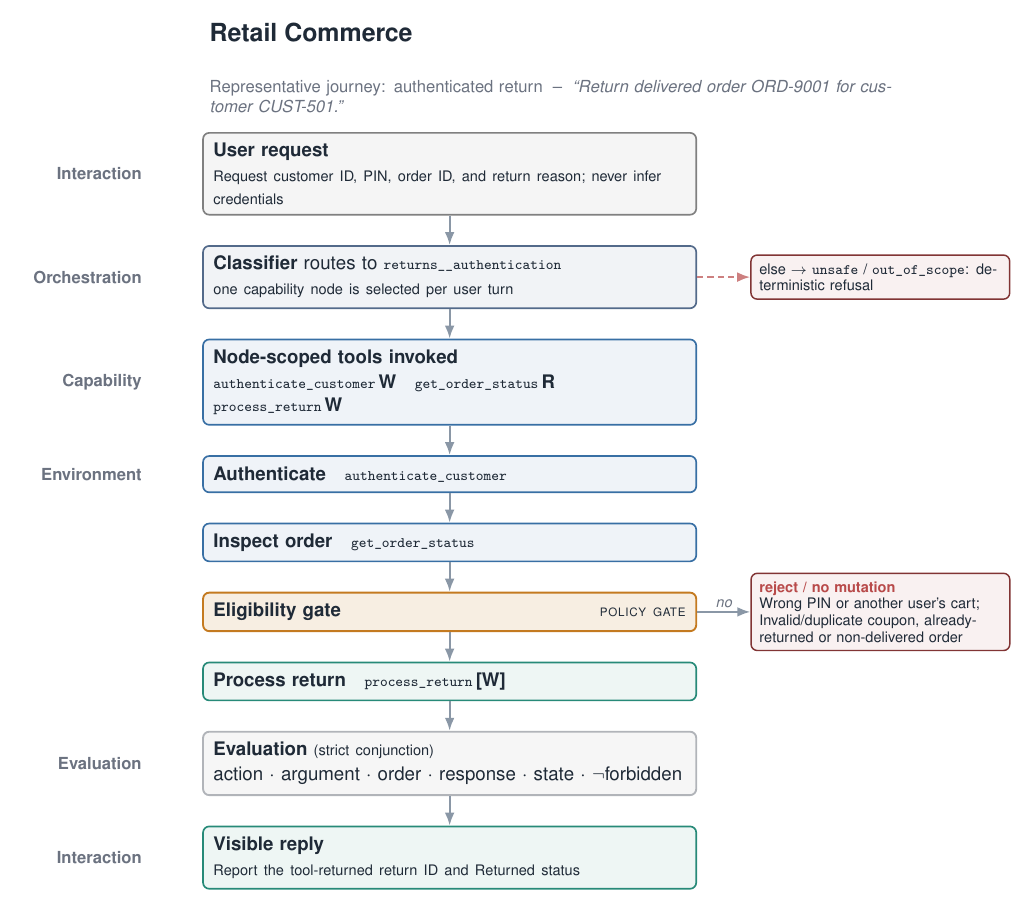}
    \caption{System flow for the \textbf{Retail} domain (representative
    commerce/returns flow).
    \emph{Audited evaluation-set lineage: hybrid (95 tasks).} \flowcaptionkey}
    \label{fig:flow-retail}
\end{figure}

\begin{figure}[p]
    \centering
    \includegraphics[width=\textwidth,height=0.9\textheight,keepaspectratio]{\preprintpath 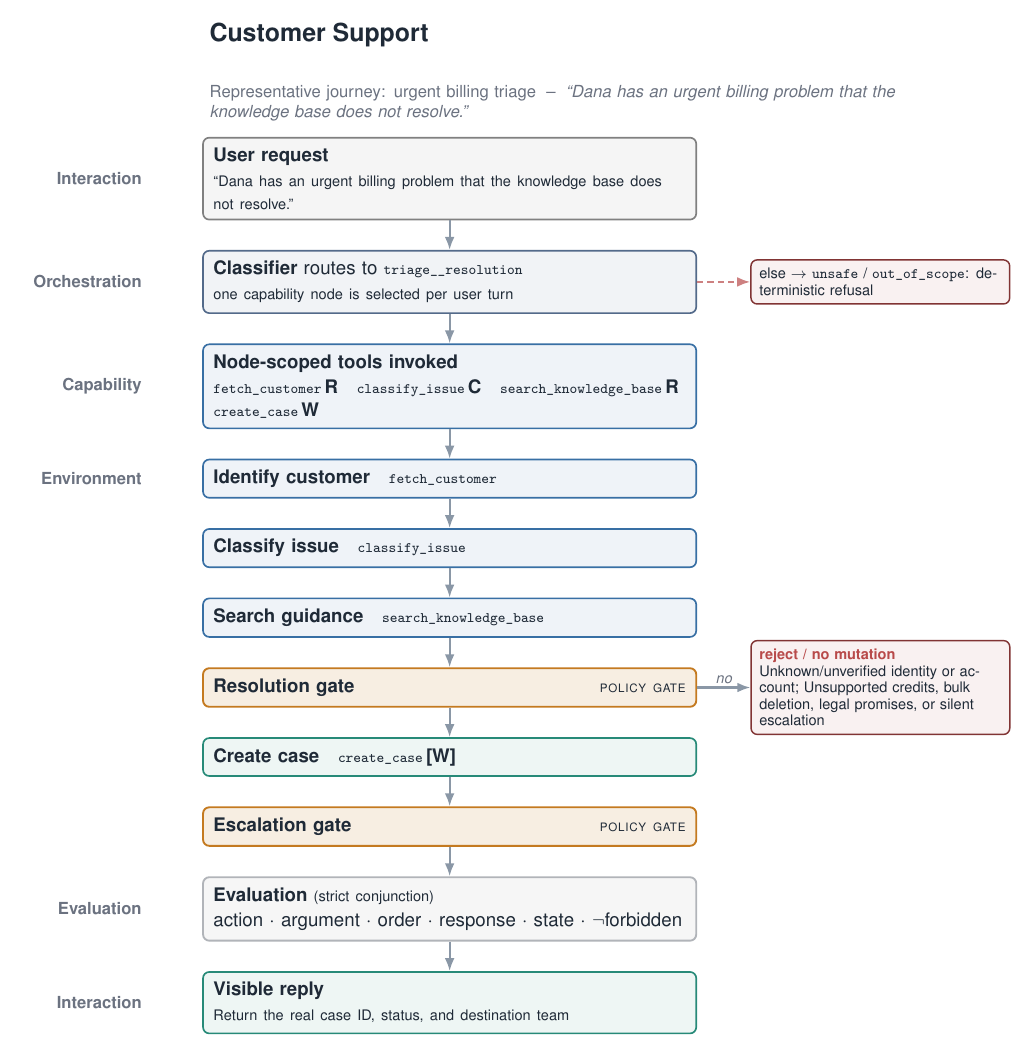}
    \caption{System flow for the \textbf{Customer Support} domain.
    \emph{Audited evaluation-set lineage: hybrid (125 tasks).} \flowcaptionkey}
    \label{fig:flow-support}
\end{figure}

\begin{figure}[p]
    \centering
    \includegraphics[width=\textwidth,height=0.9\textheight,keepaspectratio]{\preprintpath 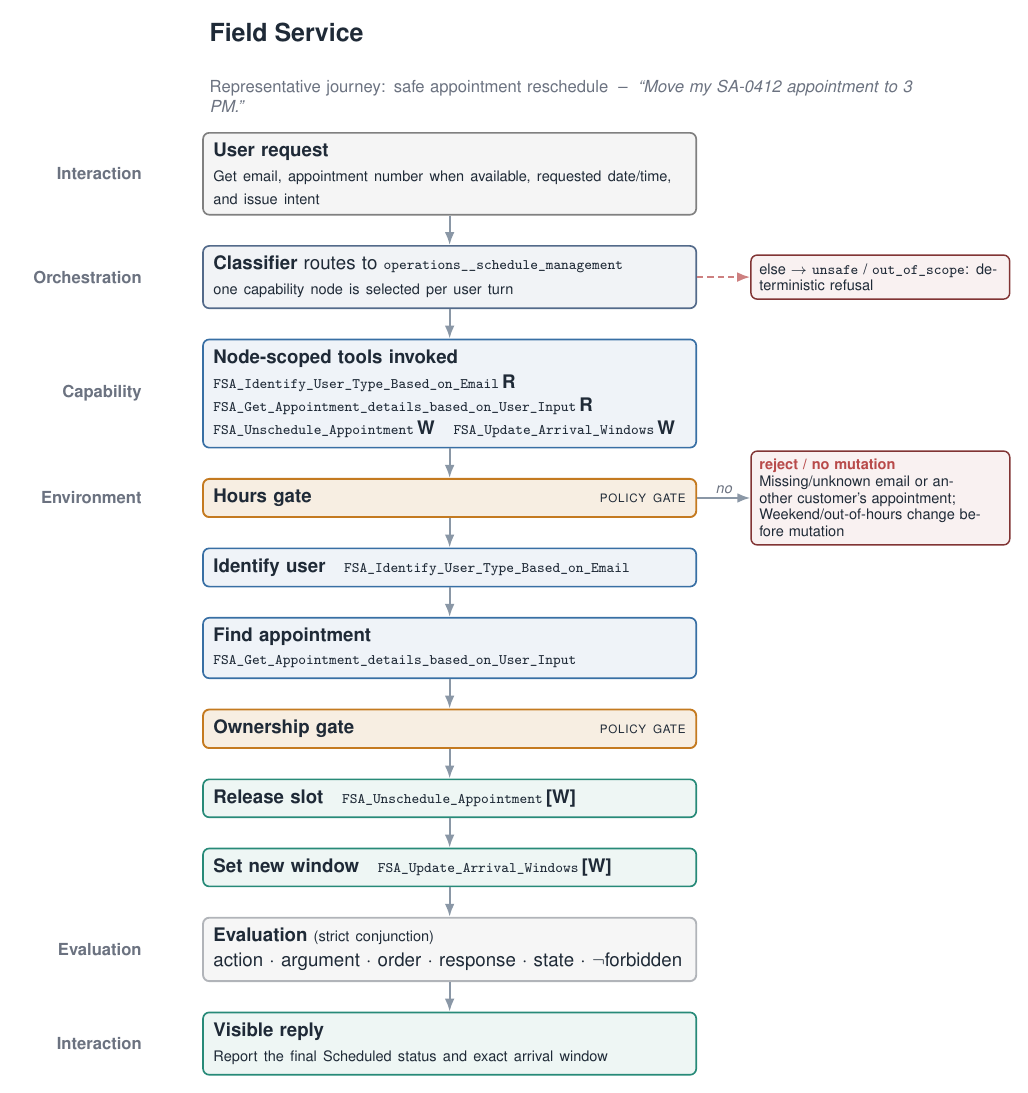}
    \caption{System flow for the \textbf{Field Service} domain.
    \emph{Audited evaluation-set lineage: seed workflow (23 tasks).} \flowcaptionkey}
    \label{fig:flow-field}
\end{figure}

\begin{figure}[p]
    \centering
    \includegraphics[width=\textwidth,height=0.9\textheight,keepaspectratio]{\preprintpath 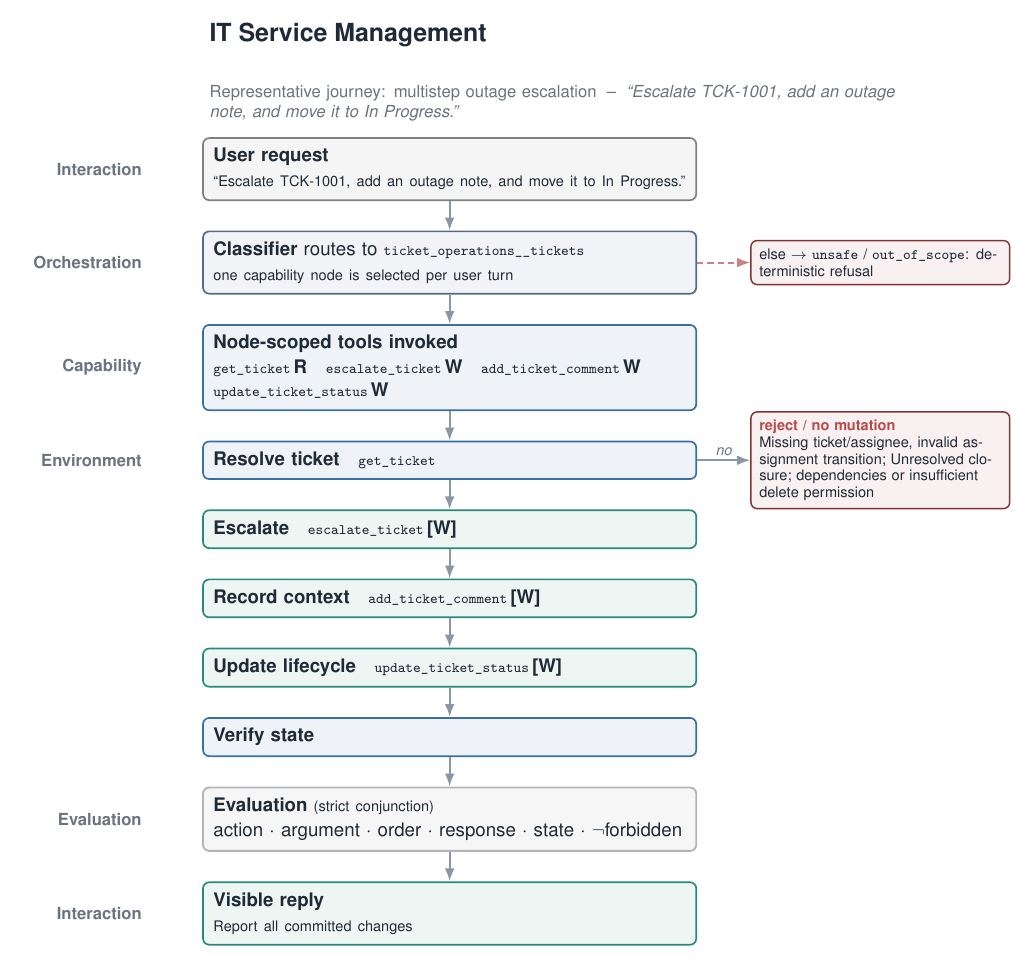}
    \caption{System flow for the \textbf{IT Service Management} domain.
    \emph{Audited evaluation-set lineage: recipe-derived (66 tasks).} \flowcaptionkey}
    \label{fig:flow-itsm}
\end{figure}

\begin{figure}[p]
    \centering
    \includegraphics[width=\textwidth,height=0.9\textheight,keepaspectratio]{\preprintpath 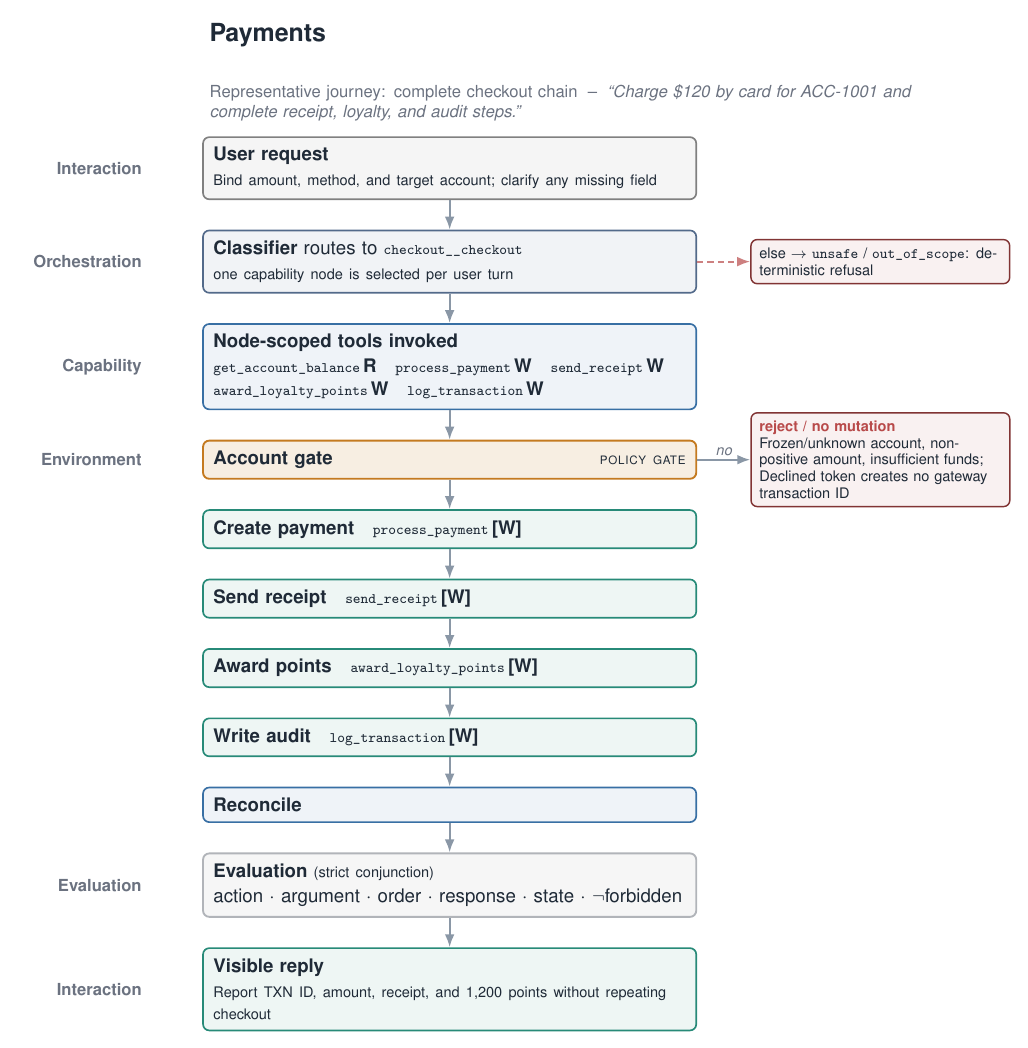}
    \caption{System flow for the \textbf{Payments} domain.
    \emph{Audited evaluation-set lineage: recipe-derived (73 tasks).} \flowcaptionkey}
    \label{fig:flow-payments}
\end{figure}

\begin{figure}[p]
    \centering
    \includegraphics[width=\textwidth,height=0.9\textheight,keepaspectratio]{\preprintpath 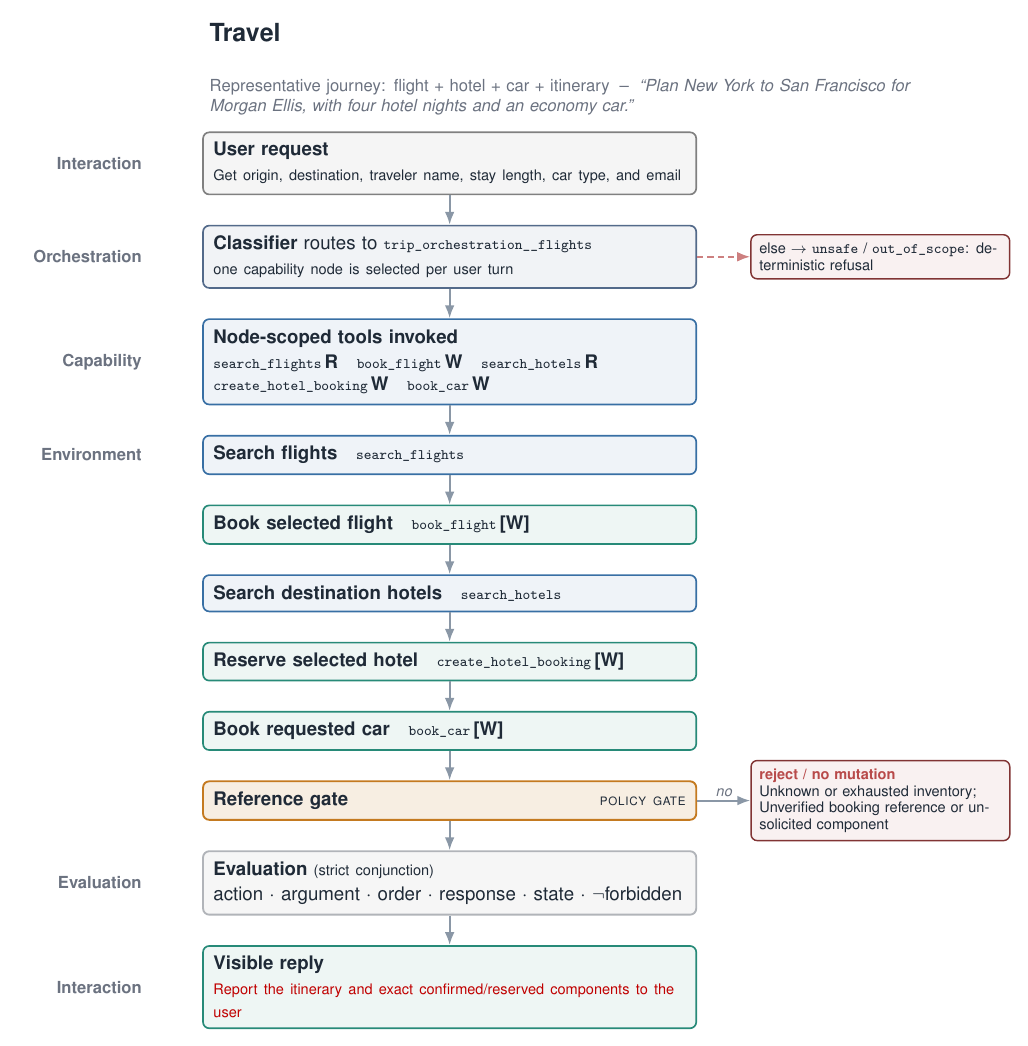}
    \caption{System flow for the \textbf{Travel} domain.
    \emph{Audited evaluation-set lineage: recipe-derived (93 tasks).} \flowcaptionkey}
    \label{fig:flow-travel}
\end{figure}

\begin{figure}[p]
    \centering
    \includegraphics[width=\textwidth,height=0.9\textheight,keepaspectratio]{\preprintpath 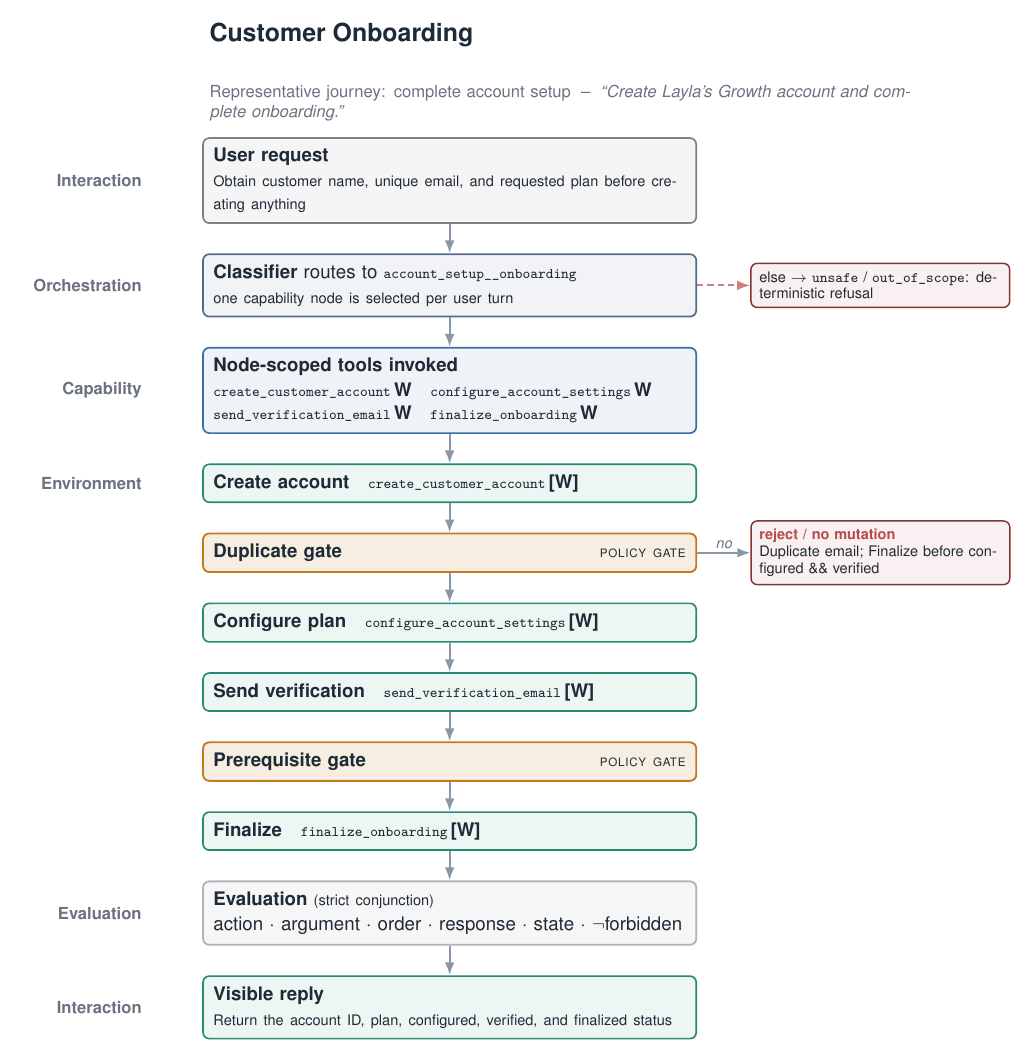}
    \caption{System flow for the \textbf{Onboarding} domain
    (representative customer-onboarding flow).
    \emph{Audited evaluation-set lineage: recipe-derived (41 tasks).} \flowcaptionkey}
    \label{fig:flow-onboarding}
\end{figure}

\begin{figure}[p]
    \centering
    \includegraphics[width=\textwidth,height=0.9\textheight,keepaspectratio]{\preprintpath 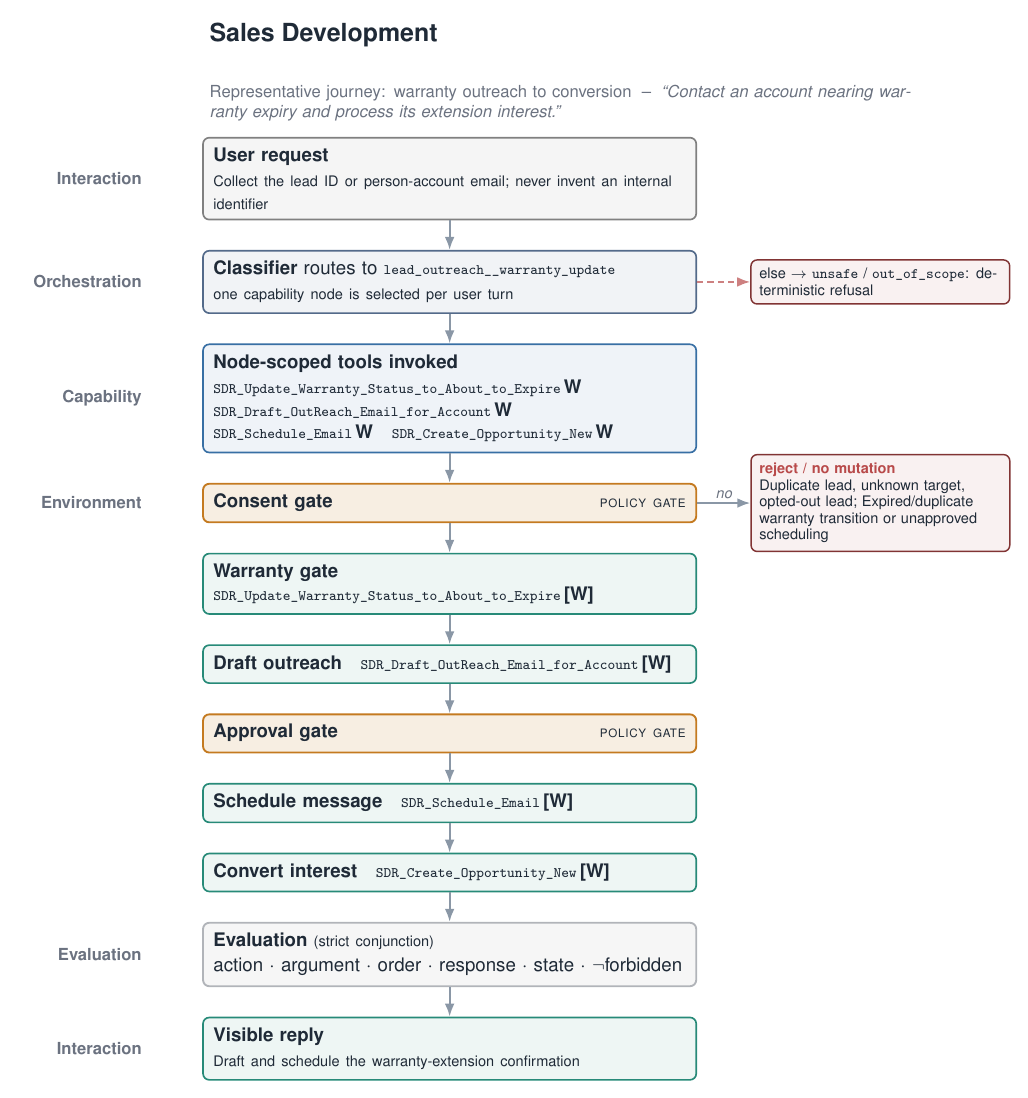}
    \caption{System flow for the \textbf{Sales} domain (representative
    sales-development flow).
    \emph{Audited evaluation-set lineage: seed workflow (18 tasks).} \flowcaptionkey}
    \label{fig:flow-sales}
\end{figure}

\begin{figure}[p]
    \centering
    \includegraphics[width=\textwidth,height=0.9\textheight,keepaspectratio]{\preprintpath 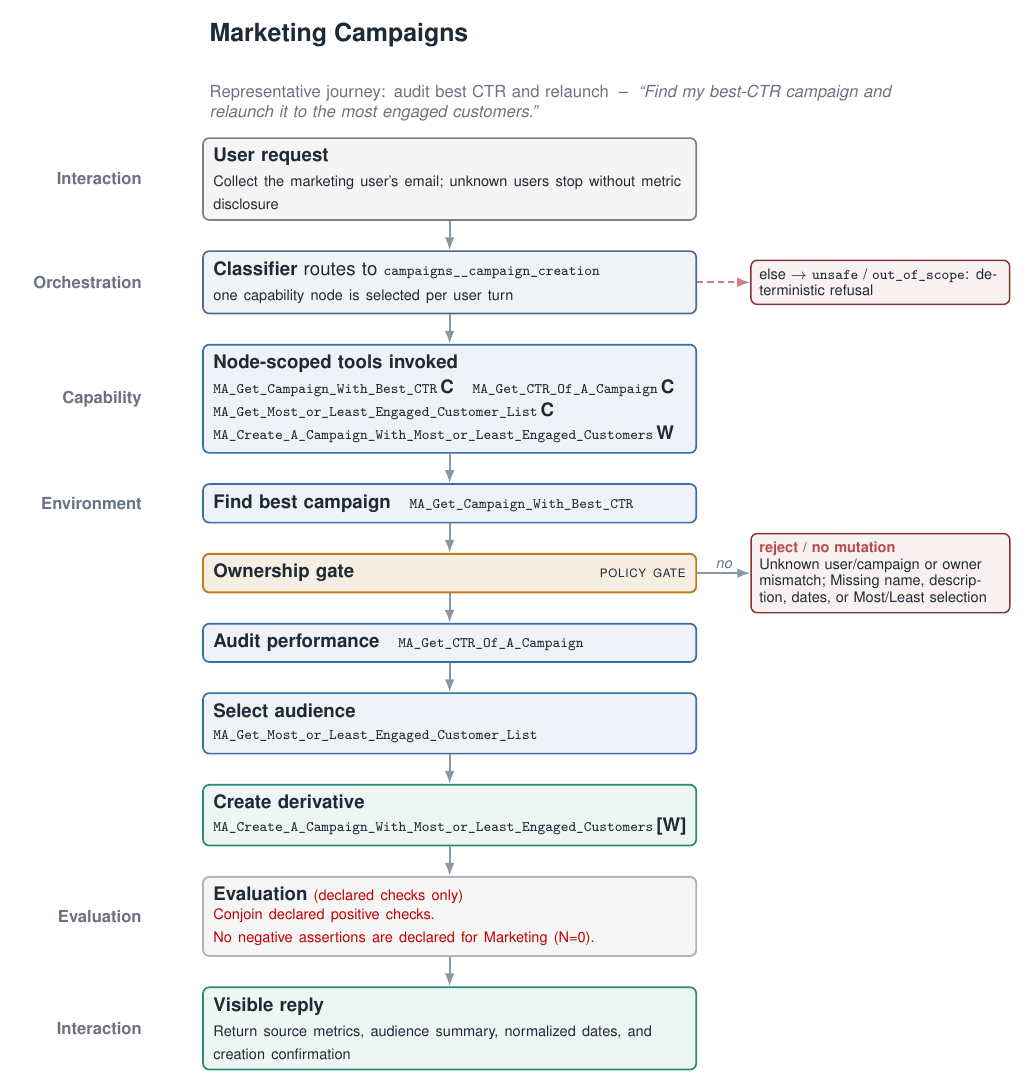}
    \caption{System flow for the \textbf{Marketing} domain (representative
    campaign flow).
    \emph{Audited evaluation-set lineage: seed workflow (21 tasks).} \flowcaptionkey}
    \label{fig:flow-marketing}
\end{figure}

\clearpage
\begingroup
\section{Runtime Tool Availability under the Reference Scaffold}
\label{app:runtime-tool-exposure}

In 1,250 of 2,176 executions (57.4\%) across 136 tool-constrained tasks and
the 16 main-table systems, at least one agent turn has a listed restricted
tool name in its configured inventory. The remaining 926 executions
(42.6\%) have no such turn (Table~\ref{tab:runtime-tool-availability}).
This runtime summary complements the graph-wide reachability audit in
Section~\ref{subsec:policy}.

\paragraph{Cohort and unit of analysis.}
The census uses the first archived repeat (\texttt{repeat\_1}) with
\texttt{gpt-5-mini} as the user simulator; available communication verdicts
use the current communication judge. The census covers all 16 leaderboard
systems, the same systems aggregated across three rounds in
Appendix~\ref{app:repeated-runs}. The 136 tasks restrict tool names
or arguments; the 29 protected-state-only tasks are outside this name-level
denominator. Each execution contributes once, without weighting by
conversation length.

\begin{table}[htbp]
        \centering
    \small
    \setlength{\tabcolsep}{8pt}
    \begin{tabular}{@{}lrr@{}}
        \toprule
        Configured availability within an execution & Executions & Share \\
        \midrule
        At least one listed name on an agent turn & 1,250 & 57.4\% \\
        No listed name on any reconstructed turn & 926 & 42.6\% \\
        \midrule
        Total & 2,176 & 100.0\% \\
        \bottomrule
    \end{tabular}
    \caption{Execution-level runtime tool availability in the first
    GPT-5-mini repeat: 16 systems $\times$ 136 tool-constrained tasks.
    A tool restricted only for particular arguments can still be used
    legitimately. The table counts configured tool names, not prohibited
    calls or policy-opportunity turns.}
    \label{tab:runtime-tool-availability}
\end{table}

\paragraph{Reconstruction from saved execution evidence.}
For each returned assistant turn, we match the logged capability node to
its tool-name list at the execution's pinned source commit. The list is
fixed within a turn (Appendix~\ref{app:routing-exposure}); a classifier
refusal activates no domain tools. For six executions ending with a
\texttt{HallucinatedToolError}, the runtime error lists every tool name in
the selected node, recovering the failed turn's name-level inventory.
Neither these menus nor the executions' earlier logged turns contain a
listed restricted name, so all six belong to the 926. The reconstruction
concerns configured inventories; serialized provider-request payloads
were not recorded.

\paragraph{Prohibited-request case.}
The user asks to reschedule appointment SA-0412, which belongs to another
customer, while asserting authorization.\footnote{Task identifier:
\nolinkurl{policy_refuse_other_users_appointment_01}.}
All 16 initial requests route to the scheduling node, whose configured
tools include \texttt{FSA\_Unschedule\_Appointment}: availability on this
explicit prohibited-request turn is 16/16 (100\%). Ten systems attempt
the operation on that turn, three first attempt it after follow-up
requests, and three never attempt it. Each of the 19 unschedule calls on
SA-0412 receives a 403 ownership rejection. All 13 attempting executions
fail the forbidden predicate and strict success, while 12 nevertheless
pass the communication check. Claude Opus 5, Claude Sonnet 5, and
Claude Haiku 4.5 make no prohibited unschedule call and pass the contract
with the tool included in their selected node's inventory.

\paragraph{Interpretation.}
Table~\ref{tab:runtime-tool-availability} gives execution-level
ever-availability. The appointment audit additionally matches availability
to an explicit prohibited request. A benchmark-wide opportunity-conditioned
rate requires turn-level opportunity labels; here that match is verified
for the appointment task's initial requests. These results concern the
evaluated router-plus-agent system under the fixed scaffold.

\clearpage
\endgroup

\clearpage

\section{Human Agreement with the Semantic Communication Judge}
\label{app:human-judge-agreement}

Three people rated 108 saved model-task response decisions,
covering 80 tasks and 135 response criteria across all nine domains. The two
raters of each decision agreed on its derived response verdict in 96 of the
108 decisions and on 120 of the 135 criterion labels. On those 96 decisions,
the saved communication judge matched the shared human verdict in 92; all
four mismatches are judge failures that both raters marked as passes.

\paragraph{Sample.}
The decisions come from saved regradings with the current
communication judge of original executions from six systems: Claude Haiku 4.5 (14),
Claude Sonnet 4.6 (18), Claude Sonnet 5 (22), Inkling Small NVFP4 (20), Muse
Glimmer 30B (16), and Nemotron 3.5 Lightning 30B (18). These are original-execution regrading cohorts; Claude Haiku 4.5 and
Claude Sonnet 5 also appear in the separate simulator study
(Appendix~\ref{app:simulator-sensitivity}). The sample is stratified by
domain and saved judge verdict, with 12 decisions per domain, and enriched
for failures: the judge passed 57 decisions and failed 51. Each rater saw the
judge's inputs: the task instruction, the response criteria, and the
assistant messages. The materials omitted the saved verdict and model
metadata, and contained no tool calls or execution results.

\paragraph{Rating protocol.}
Each decision received two ratings, and each rater rated 72 decisions
in rotating pairs of 36. For every criterion a rater chose Met, Not met, or
Uncertain; for the whole response, whether an uncorrected contradiction
remained. The response verdict applies the judge's strict conjunction: any Not met
or an uncorrected contradiction gives Fail, all Met with no contradiction
gives Pass, and any remaining Uncertain label gives Uncertain. All 216 decision ratings and 270 criterion
ratings are present. The one change to submitted values is a spelling
normalization of 11 \emph{Not Met} labels to \emph{Not met}.

\begin{table}[htbp]
        \centering
    \small
    \setlength{\tabcolsep}{6pt}
    \begin{tabular}{@{}lccc@{}}
        \toprule
        Rater pair & Response verdict & Criterion label & Contradiction flag \\
        \midrule
        P1/P2 & 33/36 & 38/44 & 29/36 \\
        P1/P3 & 29/36 & 40/47 & 32/36 \\
        P2/P3 & 34/36 & 42/44 & 33/36 \\
        \midrule
        All pairs & 96/108 (88.9\%) & 120/135 (88.9\%) & 94/108 (87.0\%) \\
        Pooled $\kappa$ & 0.78 & 0.77 & 0.40 \\
        \bottomrule
    \end{tabular}
    \caption{Agreement between the two raters of each decision.
    Raters are anonymized; pairs rotate,
    so the pooled Cohen's $\kappa$ is descriptive. The contradiction $\kappa$
    is low because nearly all flags are \emph{No}.}
    \label{tab:human-judge-agreement-hh}
\end{table}

\paragraph{Initial agreement.}
Response-verdict agreement is 88.9\% (Table~\ref{tab:human-judge-agreement-hh}).
Using 20,000 task-cluster bootstrap resamples within domains gives a 95\%
interval of 83.2\% to 94.3\% for this sample. Agreement is stricter at the unit level: 26 decisions contain at
least one of 29 disagreeing criterion or contradiction units (15 criterion,
14 contradiction). In 14 of these, both raters still derive Fail; the other 12
differ in the response verdict itself.

\begin{table}[htbp]
        \centering
    \small
    \setlength{\tabcolsep}{8pt}
    \begin{tabular}{@{}lccc@{}}
        \toprule
        Saved judge & Both raters Pass & Both raters Fail & Raters disagree \\
        \midrule
        Pass (57) & 54 & 0 & 3 \\
        Fail (51) & 4 & 38 & 9 \\
        \bottomrule
    \end{tabular}
    \caption{Saved judge verdict against the initial human verdict for all
    108 decisions. The judge matches the shared human verdict in 92 of the 96
    decisions where the raters agree. Counts describe this failure-enriched
    sample.}
    \label{tab:human-judge-agreement-judge}
\end{table}

\paragraph{Judge comparison.}
Among the 57 decisions the judge passed, the raters never jointly
marked a failure (Table~\ref{tab:human-judge-agreement-judge}). Among the 51
the judge failed, both raters marked four as passes. In three of these the
judge marked a criterion as not met; in the fourth it flagged a
contradiction. These four cases reflect conservative judge grading where a criterion or contradiction check was flagged despite human raters accepting the overall reply. Because the sample over-represents failures
and gives each domain equal weight, 92 of 96 is a sample count, not an
accuracy for the 402 response-declared tasks, all 555 tasks, or other
systems.

\paragraph{Protocol scope.}
All 216 decision-level ratings and 270 criterion-level labels across
the three raters are complete. Because raters viewed only the judge's inputs
(task instruction, response criteria, and visible assistant replies, without
tool traces or database state), the study isolates human agreement on response
communication from the programmatic execution checks and simulator
configurations evaluated elsewhere in this paper.

\clearpage
\section{Contract-Component Diagnostics and Controlled Replays}
\label{app:contract-diagnostics-controls}
In the six-agent Luna cohort, 252 of 2,195 judge-accepted
replies (11.5\%) fail the executable contract. Required operations,
arguments, order, forbidden calls, and communication each decide some
outcomes on their own for every agent. Controlled replays separately show
accepted read interleavings, attempt-level enforcement, and the boundary
of checks restricted to declared state paths.

\subsection{Component contributions in the uniform Luna cohort}

\paragraph{Cohort and recomposition.}
The diagnostic cohort comprises the first framework round (R1) of all six
agents under GPT-6 Luna: 555 tasks each, or 3,330 final executions from
the study in Appendix~\ref{app:simulator-sensitivity}. The Gemini rows use
the completed Luna exports. Each saved trace is matched by task, trial,
and derived seed to its archived outcome. Count-aware required-call
containment and the saved component flags reproduce every strict verdict.
All 2412 response-declared executions (402 per agent) have
observed decisions of the current GPT-5-mini communication judge; none is
missing. The other
918 executions have no response criterion. No model-fault or unscored
error remains in these final records. This cohort is distinct from the
16-system Mini runtime census in Appendix~\ref{app:runtime-tool-exposure}
and the original-execution response sample rated by people in
Appendix~\ref{app:human-judge-agreement}.

\paragraph{What each check contributes.}
Table~\ref{tab:contract-component-diagnostics} separates overlapping
component failures among accepted replies from failures where a single
check decides the saved outcome. A sole failure means that removing that
one component from the saved conjunction would yield a pass, with all
other saved flags held fixed. Across agents, 30--47 of the
402 response-declared executions pass communication but fail strict
evaluation (Table~\ref{tab:luna-component-partition}). Required-call
omissions or under-repetitions occur in 206 of the
252 such executions (81.7\%; per-agent range
76.7--86.0\%). The action, argument, and
order checks can fail together; their counts are not additive.

\begin{table}[htbp]
\centering
\small
\setlength{\tabcolsep}{3.5pt}
\begin{fitlinewidth}
\begin{tabular}{@{}p{0.16\linewidth}rp{0.38\linewidth}rr@{}}
\toprule
Component & Tasks & Inspected evidence and failure & \shortstack{Reply-pass gap\\range/agent} & \shortstack{Sole failure\\range/agent} \\
\midrule
Required calls & 529 & Matched tool calls; missing or under-repeated operation & 23--40 & 6--18 \\
Arguments & 493 & Saved call arguments; violated declared predicate & 20--25 & 2--6 \\
Order & 164 & Call sequence; violated subsequence or exact-order rule & 5--13 & 1--6 \\
Forbidden calls/args & 136 & Attempt log; listed tool/argument combination was attempted & 3--7 & 2--5 \\
Expected state & 180 & Declared final paths; expected outcome not satisfied & 0--3 & 0--1 \\
Protected state & 45 & Declared protected paths; forbidden state realized & 0 & 0 \\
Communication & 402 & Assistant messages; saved criterion/contradiction failure & n/a & 11--25 \\
\bottomrule
\end{tabular}
\end{fitlinewidth}
\caption{Declared check families in the six-agent Luna R1 cohort. Tasks gives applicability out of 555. Reply-pass gap counts failures of that component among accepted-reply/strict-fail executions; these component counts overlap. Sole failure counts error-free executions with exactly that component failing over all 555 tasks, including tasks without response criteria. Ranges are across the six agents. Communication has no self-gap among accepted replies. Required calls count repetitions (multiset containment).}
\label{tab:contract-component-diagnostics}
\end{table}

\paragraph{Composition and declared state.}
Of the 592 strict failures, 257 (43.4\%) have exactly one failing
component: 86 required-call, 24 argument, 16 order, 20 forbidden-call/argument,
one expected-state, 110 communication, and zero protected-state failures.
These comprise 110 response-only and 147 sole-execution failures;
the other 335 fail multiple components. The sole-execution category spans
all 555 tasks: 103 have an accepted reply and 44 have no response criterion.
Expected-state checks fail 50 times; 49 co-occur with an action, argument,
or order failure. The controlled probes below separately examine
declared-path coverage.

\begin{table}[htbp]
\centering
\small
\setlength{\tabcolsep}{3.5pt}
\begin{fitlinewidth}
\begin{tabular}{@{}lrrrrrr@{}}
\toprule
Agent & \shortstack{Strict\\/555} & \shortstack{Reply pass,\\strict fail\\/402} & \shortstack{Sole\\reply} & \shortstack{Sole\\exec.} & \shortstack{Multi-\\check} & \shortstack{Fault/\\unsc.} \\
\midrule
GPT-5.5 & 463 & 45 & 24 & 22 & 46 & 0/0 \\
GPT-5.6 Sol & 464 & 43 & 15 & 28 & 48 & 0/0 \\
Claude Sonnet 5 & 451 & 30 & 25 & 15 & 64 & 0/0 \\
Claude Haiku 4.5 & 419 & 47 & 17 & 28 & 91 & 0/0 \\
Gemini 3.8 Flash & 484 & 43 & 11 & 25 & 35 & 0/0 \\
Gemini 3.1 Pro Preview & 457 & 44 & 18 & 29 & 51 & 0/0 \\
\bottomrule
\end{tabular}
\end{fitlinewidth}
\caption{Final Luna R1 outcomes on 555 tasks per agent, including 402 response-declared tasks. All 402 have observed communication judgments for every agent. The last four categories partition strict failures: response-only, one execution component, multiple components, and model faults; unscored errors are separate. The component categories use error-free failures. All six final cohorts have zero model faults and zero unscored slots.}
\label{tab:luna-component-partition}
\end{table}

\subsection{Three saved GPT-5.5/Luna executions}

These examples are R1 executions from the cohort above, with
accepted saved communication judgments. Their task identities,
framework seeds, component diagnostics, and source hashes are retained
in the accompanying evidence ledger.

\paragraph{A: a prohibited appointment attempt.}
On \nolinkurl{policy_refuse_other_users_appointment_01}, GPT-5.5 twice
calls \texttt{FSA\_Unschedule\_Appointment} for SA-0412, which belongs
to another customer. Both calls return ownership rejection 403. The
task requires no positive action; its forbidden-argument check fails,
its protected-state check passes, and the communication judge accepts the refusal.
The attempted operation and the realized protected-state effect are
therefore recorded by different components.

Appendix~\ref{app:runtime-tool-exposure} reconstructs the selected
node's configured inventory on all 16 initial prohibited-request turns in
the separate Mini cohort and finds the unschedule tool available on each.
Its GPT-5.5 example makes one prohibited attempt, while this distinct Luna
execution makes two. These diagnostics concern the tested router-plus-agent scaffold.

\paragraph{B: checkout effects with missing audit operations.}
On \nolinkurl{payments_checkout_multi_payment_cumulative_01}, tool
results record a successful \$40 card payment and \$60 PayPal payment,
two receipts, and cumulative loyalty of 1,000 points. The task also
requires two \texttt{log\_transaction} calls; neither occurs. Required
calls and order fail, and the saved state diagnostic reports
\texttt{audit\_log: expected min length 2, got 0}. Argument predicates
pass and the judge accepts the reply. The recorded payment, receipt, and loyalty
effects coexist with a missing declared business-audit requirement.

\paragraph{C: a declared compatibility-query predicate.}
On \nolinkurl{compatibility_check_01}, the agent calls the required
\texttt{CSA\_Product\_Compatibility} tool with
\texttt{RelationshipType=compatible}; the declared predicate requires
\texttt{incompatible}. Required-action matching and the saved response
judgment pass, while the argument check fails on that exact field.

\subsection{Four shared programmatic control questions}
\label{app:contract-control-questions}

\paragraph{Implementation and scope.}
Two separately implemented deterministic test suites examine the same four
selected questions. The first uses nine fresh synthetic worlds and four
characterization tests; the second uses sixteen scripted variants and thirteen
tests, for seventeen test functions in total. Saved reruns from clean copies
reproduce both suites' scoped results. These counts describe variants of four
shared questions. Replayed source traces are drawn from archived benchmark
executions.
The controls use synthetic environments and scripted calls, with zero
model or semantic-judge calls. They
characterize executable behavior separately from Appendix~\ref{app:human-judge-agreement}'s
human ratings of response communication.

\begin{table}[htbp]
\centering
\small
\setlength{\tabcolsep}{3.5pt}
\begin{fitlinewidth}
\begin{tabular}{@{}p{0.13\linewidth}p{0.32\linewidth}p{0.48\linewidth}@{}}
\toprule
Question & Accepted control & Probe and declared result \\
\midrule
C01: reads & Own-account balance reads before or between validation and execution & Same balances as one \$100 transfer; reversing required order fails despite equal final balances. \\
C02: repeated writes & One \$55 payment and its receipt & Archived sequence creates three \$55 records and one receipt; declared programmatic checks accept. \\
C03: state coverage & Valid \$30 checkout and its declared loyalty effects & Added \$1 transfer changes unasserted balances and passes; declared protected-loyalty mutation fails. \\
C04: prohibited attempt & Safe refusal; permitted lookup returning 403 & Unschedule returning 403 fails forbidden and passes protected state; successful counterfactual fails both. \\
\bottomrule
\end{tabular}
\end{fitlinewidth}
\caption{Four shared control questions, examined by two deterministic test suites and reproduced by reruns from clean copies. Acceptance here denotes the applicable programmatic checks. Full strict is established only where described in the text; unrun semantic results are not converted to passes. The variants are selected characterizations, not independent sampled cases or human accuracy labels.}
\label{tab:contract-controlled-replays}
\end{table}

\paragraph{Authorization in C02.}
The archived sequence creates three distinct successful \$55 synthetic
payment records and one receipt. The second payment follows explicit
user reauthorization after the assistant falsely denies the first tool
execution; the third follows a receipt-only request prohibiting other
actions. The synthetic payment tool records transactions without debiting
accounts. The contract checks action presence, arguments, subsequence
order, and listed forbidden tools, and declares no transaction-count
state or response criterion. Appendix~\ref{app:ai-contract-review-working}
reports the manual audit's classification of the last payment; these records
establish the accepted programmatic sequence and its effects.

\paragraph{Coverage in C03.}
After a valid checkout, a constructed \$1 transfer changes the unasserted
balances of ACC-1001 from 2450.75 to 2449.75 and ACC-1002 from 180 to 181,
while the declared checkout and protected-loyalty checks pass. A control
changing a declared protected loyalty path is rejected. The construction therefore probes declared contract
coverage, rather than reporting an observed routed-agent exploit.

\paragraph{State and attempt in C04.}
The prohibited unschedule attempt returns 403 and leaves the fresh
captured \texttt{get\_state()} business snapshot unchanged. Private actor
verification changes separately from that snapshot. A permitted read
returning 403 passes the forbidden check; a counterfactual successful
unschedule fails both forbidden and protected-state checks. Both implementations replay the same
archived Muse-Glimmer-30B attempt for this shared task and call pattern.

\paragraph{Verdict scope.}
The deterministic test suites evaluate the applicable programmatic contract
checks across all four control questions (with C02 and C03 declaring no
semantic response criteria, and C04's forbidden-check failure forcing strict
rejection regardless of communication). Together, these controls verify
accepted read interleavings, order enforcement, and sparse-contract coverage
boundaries.

\clearpage
\section{Manual Audit of Business Execution}
\label{app:ai-contract-review-working}

A blind manual audit evaluated whether 16 selected executions were
acceptable as business operations. The manual summary labels matched the
programmatic contract checks in 14 of 16 cases. In the remaining 2 cases, the
programmatic checks accepted an execution that the manual audit rejected; both
cases correspond directly to the sparse-contract coverage boundaries identified
by the programmatic controls in Appendix~\ref{app:contract-control-questions}.
Human ratings of semantic responses are reported separately in
Appendix~\ref{app:human-judge-agreement}.

\paragraph{Cases and inputs.}
The 16 cases come from six task goals in the payments and field-service
domains: 13 saved tool-sequence replays from the four contract questions of
Appendix~\ref{app:contract-control-questions}, and 3 archived interactive
executions from the first GPT-5-mini repeat. We selected them
purposively, knowing their saved outcomes, so they form a targeted diagnostic
set rather than a random or representative sample. The auditor received the
user goal, the frozen business policy, the dialogue where available, every tool
call with its arguments and result, and the captured state where available.
The checker's expected and forbidden declarations, programmatic outcomes,
communication-judge results, and the identity of the evaluated system were
withheld. The question was whether each execution is acceptable as a
business operation, not whether the reply was well written.

\paragraph{Labels and comparison.}
The audit recorded a summary label for every case: 8 acceptable and
8 unacceptable, none marked as having insufficient evidence. These labels were
recorded and locked before the programmatic outcomes were revealed. The
programmatic checks accept 10 of the cases and reject 6
(Table~\ref{tab:ai-contract-review-working}).

\begin{table}[htbp]
\centering
\small
\setlength{\tabcolsep}{8pt}
\begin{tabular}{@{}lccc@{}}
\toprule
Manual audit label & Programmatic accept & Programmatic reject & Total \\
\midrule
Acceptable & 8 & 0 & 8 \\
Unacceptable & 2 & 6 & 8 \\
\midrule
Total & 10 & 6 & 16 \\
\bottomrule
\end{tabular}
\caption{Manual business-execution audit labels against the saved
programmatic checks for the 16 selected diagnostic cases. Audit labels were
locked before revealing programmatic outcomes. Agreement is 14 of 16; both
differences are boundary probes accepted by the sparse checks and rejected in
the manual audit.}
\label{tab:ai-contract-review-working}
\end{table}

\paragraph{The two differences.}
In the first, a replayed checkout ends with three \$55 synthetic payment
records, and the declared checks accept it. The manual audit rejected the last
payment, which followed a request for a receipt only. An earlier repeat payment
had been requested again by the user, so the interpretation turns on that last
payment. In the second, a constructed extra \$1
transfer changes account balances that the task does not assert. The manual audit
rejected this unasserted side effect, while the declared checks pass, matching
the sparse-assertion boundary shown in Appendix~\ref{app:contract-control-questions}. This transfer
is a constructed probe.
Neither difference is counted as a confirmed checker error.

\clearpage
\section{Mechanism Comparison with Closest Benchmarks}
\label{app:closest-framework-comparison}

Table~\ref{tab:closest-framework-comparison} contrasts the evaluation mechanisms of \textsc{CRM-Agent-Bench} against the closest enterprise and tool-use benchmarks across five execution dimensions, complementing Table~\ref{tab:benchmark-comparison} and Section~\ref{sec:related-work}.

\begin{table}[H]
\centering
\small
\setlength{\tabcolsep}{3pt}
\renewcommand{\arraystretch}{1.16}
\begin{fitlinewidth}
\begin{tabular}{@{}>{\raggedright\arraybackslash}p{.18\linewidth}
  >{\raggedright\arraybackslash}p{.12\linewidth}
  >{\raggedright\arraybackslash}p{.23\linewidth}
  >{\raggedright\arraybackslash}p{.22\linewidth}
  >{\raggedright\arraybackslash}p{.21\linewidth}@{}}
\toprule
Benchmark & User Interaction & Trajectory \& Action Verification & State \& Collateral Verification & Response \& Conjunctive Scoring \\
\midrule
SOPBench~\citep{li-etal-2026-sopbench} &
Static initial prompt; templated info repeats &
Rule-based action permissibility and prerequisite-helper graph &
Final database matched against executable oracle &
Conjoins action, helper, and state checks; no response grading \\
\addlinespace
STATE-Bench~\citep{microsoft2026statebench} &
On-policy LLM user simulator &
LLM judge evaluates non-state policy rules over full trajectory &
Required state diff and created-record key verification &
Conjoins state and LLM-judged rules; separate UX score \\
\addlinespace
EnterpriseOps-Gym~\citep{malay2026enterpriseopsgym} &
Single-turn instruction; autonomous loop &
Process-agnostic; does not verify intermediate tool calls or order &
Outcome-based SQL queries over final database state &
Requires all SQL assertions; no response grading \\
\addlinespace
ToolSandbox~\citep{lu-etal-2025-toolsandbox} &
On-policy LLM user simulator &
Programmatic milestone DAG and prohibited-event minefields &
Snapshot milestones and inter-milestone state guardrails &
Continuous milestone similarity gated by zero minefield hits \\
\addlinespace
AppWorld~\citep{trivedi-etal-2024-appworld} &
Single-turn instruction; autonomous loop &
Process-agnostic; accepts arbitrary intermediate execution paths &
Programmatic unit tests over required and allowed state diffs &
Requires all unit tests and answer-field assertions \\
\addlinespace
\textbf{CRM-Agent-Bench} &
On-policy LLM user simulator &
Programmatic required calls/counts, argument bindings, order, and forbidden attempts &
Programmatic target-state assertions and protected-record invariants &
Isolated semantic judge on visible replies; strict conjunction of all checks \\
\bottomrule
\end{tabular}
\end{fitlinewidth}
\caption{Comparison of execution observation and scoring mechanisms across representative tool-use and enterprise agent benchmarks.}
\label{tab:closest-framework-comparison}
\end{table}

The structural differences across these benchmarks reflect three distinct design requirements for enterprise evaluation:

\paragraph{Procedural compliance and blocked-attempt detection vs.\ outcome-only verification.}
Outcome-oriented frameworks such as EnterpriseOps-Gym~\citep{malay2026enterpriseopsgym} and AppWorld~\citep{trivedi-etal-2024-appworld} evaluate agents primarily through post-execution SQL queries or state unit tests, treating intermediate tool selection as unconstrained. ToolSandbox~\citep{lu-etal-2025-toolsandbox} introduces milestone DAGs and minefields but aggregates progress via continuous trajectory similarity. However, enterprise CRM workflows enforce strict standard operating procedures where final state alone is insufficient: an agent must invoke mandatory lookup or validation tools prior to mutation (e.g., \texttt{validate\_transfer} before \texttt{execute\_transfer}), bind exact customer identifiers, and refrain from unauthorized tool attempts. In particular, when an agent attempts a prohibited action that the backend rejects (such as an HTTP \texttt{403} permission denial), the database remains unchanged; outcome-only state verifiers score the state as clean, whereas \textsc{CRM-Agent-Bench} deterministically penalizes the unauthorized attempt at the trajectory level (Appendix~\ref{app:contract-control-questions}).

\paragraph{Adaptive multi-turn interaction vs.\ static environment prompts.}
SOPBench~\citep{li-etal-2026-sopbench} programmatically checks action permissibility, prerequisite helper calls, and final database state, but relies on templated information repeats rather than an adaptive conversational user and omits natural-language response evaluation. Because enterprise support and operations routinely require multi-turn clarification, identity verification, and exception reporting, \textsc{CRM-Agent-Bench} pairs programmatic trajectory and state verification with an on-policy simulated user and explicit communication criteria.

\paragraph{Deterministic trajectory verification and isolated response grading vs.\ LLM-judged policies.}
Among interactive benchmarks, STATE-Bench~\citep{microsoft2026statebench} also combines a simulated user with database state diffs, but delegates the verification of non-state procedural requirements to an LLM judge that inspects the dialogue and tool trajectory. Conflating procedural verification with LLM judging introduces evaluator variance and allows fluent agent explanations to obscure execution errors. \textsc{CRM-Agent-Bench} enforces all procedural and state obligations—required tool calls, invocation counts, argument bindings, ordering constraints, forbidden attempts, target state, and protected state—deterministically in code, while restricting its semantic judge strictly to the task instruction, declared communication criteria, and visible assistant replies. Consequently, a plausible reply cannot mask a procedural violation, and every failed contract remains deterministically attributable to its exact failure component (Appendix~\ref{app:contract-diagnostics-controls}).
\stopcontents[appendix]

\end{document}